\documentstyle[12pt]{article}

\newcommand{\tr}{\mbox{tr}}
\newcommand{\eqn}[1]{(\ref{#1})}
\newcommand{\del}{\partial}
\newcommand{\complex}{{\bb C}} 
\newcommand{\zed}{{\bb Z}} 
\newcommand{\real}{{\bb R}} 
\newcommand{\reals}{{\bbs R}} 
\newcommand{\zeds}{{\bbs Z}} 
\newcommand{\id}{{\bb I}} 
\newcommand{\alg}{{\cal A}} 
\def\cstars{$C^*$-algebras }
\def\cstar{$C^*$-algebra }
\newcommand{\quater}{{\bb H}} 
\newcommand{\unit}{\id}
\def\norm#1{{\Vert#1\Vert}}

\font\mybb=msbm10 at 12pt
\def\bb#1{\hbox{\mybb#1}}
\font\mybbs=msbm10 at 9pt
\def\bbs#1{\hbox{\mybbs#1}}

\def\e{{\rm e}}
\def\slash{\!\!\!\!/}
\def\Dirac{{D\slash\,}}
\def\beq{\begin{equation}}
\def\eeq{\end{equation}}
\def\bea{\begin{eqnarray}}
\def\eea{\end{eqnarray}}
\def\bd{\begin{displaymath}}
\def\ed{\end{displaymath}}

\renewcommand{\theequation}{\thesection.\arabic{equation}}

\makeatletter
\newdimen\normalarrayskip              
\newdimen\minarrayskip                 
\normalarrayskip\baselineskip
\minarrayskip\jot
\newif\ifold             \oldtrue            \def\new{\oldfalse}
\def\arraymode{\ifold\relax\else\displaystyle\fi} 
\def\@arrayskip{\ifold\baselineskip\z@\lineskip\z@
     \else
     \baselineskip\minarrayskip\lineskip2\minarrayskip\fi}
\def\@arrayclassz{\ifcase \@lastchclass \@acolampacol \or
\@ampacol \or \or \or \@addamp \or
   \@acolampacol \or \@firstampfalse \@acol \fi
\edef\@preamble{\@preamble
  \ifcase \@chnum
     \hfil$\relax\arraymode\@sharp$\hfil
     \or $\relax\arraymode\@sharp$\hfil
     \or \hfil$\relax\arraymode\@sharp$\fi}}
\def\@array[#1]#2{\setbox\@arstrutbox=\hbox{\vrule
     height\arraystretch \ht\strutbox
     depth\arraystretch \dp\strutbox
     width\z@}\@mkpream{#2}\edef\@preamble{\halign \noexpand\@halignto
\bgroup \tabskip\z@ \@arstrut \@preamble \tabskip\z@ \cr}%
\let\@startpbox\@@startpbox \let\@endpbox\@@endpbox
  \if #1t\vtop \else \if#1b\vbox \else \vcenter \fi\fi
  \bgroup \let\par\relax
  \let\@sharp##\let\protect\relax
  \@arrayskip\@preamble}
\makeatother

\newcommand{\newsection}[1]
{\vspace{5mm}
\pagebreak[3]
\addtocounter{section}{1}
\setcounter{equation}{0}
\setcounter{subsection}{0}
\setcounter{footnote}{0}
\begin{flushleft}
{\large\bf \thesection. #1}
\end{flushleft}
\nopagebreak
\medskip
\nopagebreak}

\newlength{\extraspace}
\newlength{\extraspaces}
\begin{document}

\renewcommand{\footnotesize}{\small}

\addtolength{\baselineskip}{.8mm}

\thispagestyle{empty}

\begin{flushright}
\baselineskip=12pt
{\sc OUTP}-97-22P\\
DSF/22-97\\
hep-th/9707202\\
\hfill{  }\\ July 1997
~~
\end{flushright}
\vspace{.2cm}

\begin{center}
{\Large\bf{Duality Symmetries and Noncommutative Geometry of
String Spacetimes}}\\[12mm]

{\sc Fedele Lizzi}\footnote{Permanent Address: Dipartimento di Scienze Fisiche,
Universit\`a di Napoli Federico II and INFN, Sezione di Napoli,
Italy.\\
E-mail: {\tt lizzi@na.infn.it}} {\sc
and Richard J.\ Szabo\footnote{E-mail: {\tt r.szabo1@physics.oxford.ac.uk}}}
\\[2mm]
{\it Department of Physics -- Theoretical Physics\\ University of Oxford\\ 1
Keble Road, Oxford OX1 3NP, U.K.} \\[15mm]

\vskip 0.2 in

{\sc Abstract}

\begin{center}
\begin{minipage}{15cm}

We examine the structure of spacetime symmetries of toroidally
compactified string theory within the framework of noncommutative
geometry. Following a proposal of Fr\"ohlich and Gaw\c edzki, we
describe the noncommutative string spacetime using a detailed
algebraic construction of the vertex operator algebra. We show that
the spacetime duality and discrete worldsheet symmetries of the string
theory are a consequence of the existence of two independent Dirac
operators, arising from the chiral structure of the conformal field
theory. We demonstrate that these Dirac operators are also responsible
for the emergence of ordinary classical spacetime as a low-energy
limit of the string spacetime, and from this we establish a
relationship between T-duality and changes of spin structure of the
target space manifold. We study the automorphism group of the vertex
operator algebra and show that spacetime duality is naturally a gauge
symmetry in this formalism. We show that classical general covariance
also becomes a gauge symmetry of the string spacetime. We explore some
larger symmetries of the algebra in the context of a universal gauge
group for string theory, and connect these symmetry groups with some
of the algebraic structures which arise in the mathematical theory of
vertex operator algebras, such as the Monster group. We also briefly
describe how the classical topology of spacetime is modified by the
string theory, and calculate the cohomology groups of the
noncommutative spacetime.  A self-contained, pedagogical introduction
to the techniques of noncommmutative geometry is also included.

\end{minipage}
\end{center}

\end{center}

\centerline{To appear in Communications in Mathematical Physics}
\vfill
\newpage
\pagestyle{plain}
\setcounter{page}{1}
\stepcounter{subsection}

\newsection{Introduction}

Duality has emerged as an important non-perturbative tool for the understanding
of the spacetime structure of string theory and certain aspects of confinement
in supersymmetric gauge theories (see \cite{Tdualrev,Alvarez} for respective
reviews). Recently, its principal applications have been in string theory
within the unified framework of M theory \cite{Mtheory}, in which all five
consistent superstring theories in ten-dimensions are related to one another by
duality transformations. Target space duality, in its simplest toroidal
version, i.e. T-duality, relates large and small compactification radius
circles to one another. It therefore relates two different spacetimes in which
the strings live and, implicitly, large and small distances. The quantum string
theory is invariant under such a transformation of the target space. The
symmetry between these inequivalent string backgrounds leads to the notion of a
{\it stringy} or {\it quantum} spacetime which forms the moduli space of string
vacua and describes the appropriate stringy modification of classical general
relativity.

T-duality naturally leads to a fundamental length scale in string theory, which
is customarily identified as the Planck length $l_P$. A common idea is that at
distances smaller than $l_P$ the conventional notion of a spacetime geometry is
inadequate to describe its structure. As strings are extended objects, the
notion of a `point' in the spacetime may not make sense, just as the notion of
a point in a quantum phase space is meaningless. In fact, it has been
conjectured that the string configurations themselves obey an uncertainty
principle, so that at small distances they become smeared out. A recent
candidate theory for this picture is the effective matrix field theory for
D-branes \cite{Dfield,Matrix} in which the spacetime coordinates are described
by noncommuting matrices. In this paper we shall describe a natural algebraic
framework for studying the geometry of spacetime implied by string theory.

Duality and string theory seem to point to a description of spacetime which
goes beyond the one given by ordinary geometry with its concept of manifolds,
points, dimensions etc. In this respect, a tool ideally suited to generalize
the concept of ordinary differential geometry is {\it Noncommutative Geometry}
\cite{Book}. Indeed, the foundations of noncommutative geometry were developed
by von Neumann in an attempt to understand the geometry of a quantum phase
space. The central idea of Noncommutative Geometry is that, since a generic
(separable) topological space (for example a manifold) is completely
characterized by the commutative
 $C^*$-algebra of continuous complex-valued functions
defined on it, it may be useful to regard this algebra as the algebra
generated by the coordinates of the space. Conversely, given a {\em
commutative} $C^*$-algebra, it is possible to construct, with purely algebraic
methods, a topological space. The terminology {\em noncommutative} geometry
refers to the possibility of generalizing these concepts to the
case where the algebra is noncommutative. This would be the case of some sort
of space in which the coordinates do not commute. The structure of the theory
is, however, even more powerful that this first generalization.

The commutative algebra gives not only the possibility to construct the points
of a space, but also to give them a topology, again using only algebraic
methods. Metric aspects (distances, etc.) can also be constructed by
representing the algebra as bounded operators on a Hilbert space, and then
defining on it a generalization of the Dirac operator. Differential forms are
also represented as operators on the Hilbert space and gauge transformations
act as conjugation by elements of the group of unitary operators of the
algebra. This set of three objects, an algebra, a Hilbert space on which the
algebra is represented, and the generalized Dirac operator, goes under the name
of a {\em Spectral Triple}.

One is now led to ask if the generalization of spacetime hinted to by string
theory can find a place in the framework of noncommutative geometry. Namely, if
there is an algebra which can provide (very loosely speaking) the
noncommutative coordinates of string theory. The structure should be such that
in the low-energy limit, in which strings are effectively just point particles
described by ordinary quantum field theory, one recovers the usual
(commutative) spacetime manifolds. In the following we shall elaborate on a
program initiated by Fr\"ohlich and Gaw\c edzki \cite{FG} (see also
\cite{ChamFro}) for understanding the geometry of string spacetimes using
noncommutative geometry. This program has also been pursued recently by
Fr\"ohlich, Grandjean and Recknagel \cite{fgrleshouches}. In \cite{FG} it was
proposed that the algebra representing the stringy generalization of
spacetime is the {\em vertex operator algebra} of the underlying conformal
field theory for the strings. We shall analyse some of the properties of this
spacetime through a detailed algebraic study of the properties of vertex
operator algebras.

A conformal field theory has a natural chiral structure, from which we
show how to naturally construct two Dirac operators. These operators
are crucial to the construction of the low-energy limit of the
noncommutative spacetime which gives the conventional spacetimes of
classical general relativity at large distance scales. We will
explicitly construct this limit, using the tools of noncommutative
geometry, and show further how our Dirac operators are related to the
more conventional ones that arise from $N=1$ superconformal field
theories. Chamseddine \cite{Chamseddine} has recently used these Dirac
operators in the spectral action principle \cite{ChamsConnes} of
noncommutative geometry and shown that they lead to the desired
effective superstring action.  However, the Dirac operators that we
incorporate into the spectral triple are more general and as such they
illuminate the full structure of the duality symmetries of the string
spacetime.  All of the information concerning the target space
dualities and discrete worldsheet symmetries of the string theory lies
in the relationships between these two Dirac operators. They define
isometric noncommutative spacetimes at the level of their spectral
triples, and as such lead naturally to equivalences between their
low-energy projective subspaces which imply the duality symmetries
between classical spacetimes and the quantum string theory.  {}From
this we also deduce a non-trivial relation between T-duality
transformations and changes of spin structure of the spacetime.

The main focus of this paper will be on the applications of noncommutative
geometry to a systematic analysis of the symmetries of a string spacetime using
the algebraic formulation of the theory of vertex operator algebras (see for
example \cite{flm,huang,gebert}). We will show (following \cite{strsymsdual})
that target space duality is, at the level of the string theory spectral
triple, just a very simple inner automorphism of the vertex operator algebra,
i.e. a gauge transformation (this was anticipated in part in \cite{ls}). This
transformation leaves the algebra (which represents the noncommutative topology
of the stringy spacetime) invariant, but changes the Dirac operator, and hence
the metric properties. We shall also describe other automorphisms of the vertex
operator algebra. For example, discrete worldsheet symmetries of the conformal
field theory appear as outer automorphisms of the noncommutative geometry. For
the commutative algebra of functions on a manifold there are no inner
automorphisms (gauge symmetries), and the group of (outer) automorphisms
coincides with the diffeomorphism group of the manifold. We will show that the
outer automorphisms of the low-energy projective subalgebras of the vertex
operator algebra, which define diffeomorphisms of the classical spacetime, are
induced via the projections from {\it inner} automorphisms of the full string
theory spectral triple. This implies that, in the framework of noncommutative
geometry, general covariance appears naturally
as a gauge symmetry of the quantum
spacetime, and general relativity is therefore formulated as a gauge theory.

We shall also analyse briefly the problem of computing the full automorphism
group of the noncommutative string spacetime. This ties in with the problem of
finding a universal gauge group of string theory which overlies all of its
dynamical symmetries. We are not aware of a full classification of the 
automorphims of a
vertex operator algebra, nor will we attempt it here. Some very important
mathematical aspects of this group are known, such as its relation to the
Monster group \cite{flm}, and we examine these properties within the context of
the noncommutative geometry of the string spacetime. We also briefly
discuss some properties of the noncommutative differential topology of the
string spacetime and compare them with the classical topologies. From this
we shall see the natural emergence of spacetime topology change.

For definitiveness and clarity we restrict ourselves in this paper to
a detailed analysis of essentially the simplest string theory, the
linear sigma-model, i.e. closed strings compactified on a flat
$n$-dimensional torus. The advantage of studying this class of models
is that everything can be represented in more or less explicit
form. Already at this level we will find a very rich noncommutative
geometrical structure, which illustrates how the geometry and topology
relevant for general relativity must be embedded into a larger
noncommutative structure in string theory. Our analysis indicates that
noncommutative geometry holds one of the best promises of studying
physics at the Planck scale, which at the same time incorporates the
dynamical features governed by string theory. Our results also apply
to $N=1$ superconformal field theories (whose target spaces are
effectively restricted to tori), and we also indicate along the way
how the analysis generalizes to other conformal field
theories. However, the most interesting generalization would be in the
context of {\it open} strings, where the two chiral sectors of the
theory merge and T-duality connects strings with D-branes, solitonic
states of the theory that appear if one allows the endpoints of an
open string to have Dirichlet rather than the customary Neumann
boundary conditions. D-branes have been discussed in the context of
Noncommutative Geometry recently in \cite{Dbranes} (see also
\cite{fgrleshouches}).  Duality now relates spectral triples of {\it
different} string spacetimes, which could have remarkable implications
in the context of M theory, the membrane theory which purports to
contain (as different low energy regimes) all consistent superstring
theories. A complete dynamical description of M theory is yet unknown,
but in the context of this paper this could be achieved by some large
vertex operator algebra. In this respect the conjecture of
\cite{Matrix}, which relates M-theory to a matrix model, could be
interpreted as a particular truncation of a set of vertex operators to
finite-dimensional $N\times N$ matrices. The large-$N$ limit then
recovers the aspects of the full theory.

\subsubsection*{Structure and Outline of the Paper}

This paper has been written with the hope of being accessible to both
string theorists with no prior knowledge of noncommutative geometry,
zzand also to mathematicians and mathematical physicists with no detailed
knowledge of string
theory. It also merges two very modern branches of mathematics (both
inspired in large part by physics), the algebraic theory of vertex
operator algebras and noncommutative geometry.  It therefore contains
some review sections in the hope of keeping the presentation
relatively self-contained. We start in section 2 with a brief,
self-contained review of the basic techniques of noncommutative
geometry in general, where the aspects relevant for our constructions
are described, often in a heuristic way. In section 3 we then begin a
systematic construction (following \cite{FG}) of the
string spacetime. We begin this construction in a way that could be
generalized to more general worldsheet geometries than the ones which
we consider, and also to more complicated target spaces such as
toroidal orbifolds. The Dirac operators of the theory are introduced
in section 4, along with their alternative spin structures, and are
related to similar objects which appear in supersymmetric
sigma-models. The relevant algebra (the vertex operator algebra) is
introduced in section 5. Here we aim at giving a more algebraic
description of this space of operators than is usually presented, so
that this section is somewhat foundational to the general study of
noncommutative string spacetimes, in the sense that the full algebraic
properties of it are not fully exploited in the remainder of the
paper.  A full analysis of the string spacetime that captures all of
these rich algebraic aspects of the vertex operator algebra could lead
to remarkable insights into the spacetime structure of string
theory. Sections 6 and 7 then contain the main results of this paper.
In section 6 the spectral triples are analysed and the low-energy
sectors are constructed from the two Dirac operators. The duality
symmetries are also presented as isomorphisms between the spectral
triples. Section 7 discusses the symmetries and topology of the
noncommutative string spacetime in a general setting, and some aspects
of a universal string theory gauge group, Borcherds algebras and
relations with the Monster group are presented. Some more formal
algebraic aspects of vertex operator algebras, along with some
features about how they construct the string spacetime, are included
in an appendix at the end of the paper. This expands the descriptions
already presented in section 5 with emphasis on how the analysis and
results of this paper extend to more general string theories.

\newsection{Elements of Noncommutative Geometry}

In this section we will give, for the sake of completeness, a very brief
introduction to the basic tools of noncommutative geometry. We shall work mostly
with examples of commutative algebras, corresponding to ordinary (commutative)
geometry. We will not be concerned with mathematical details, but aim mostly at
giving an overview. The classical reference is Connes' book \cite{Book}. In the
following we shall for the most part follow the introduction to noncommutative
geometry given in the excellent forthcoming book by Landi \cite{Landi}, to
which we refer the reader for details. Another introduction can be found in
\cite{madore}. The reader already conversant in noncommutative geometry may
wish to skip this section.

\subsubsection*{Topological Spaces from Algebras}

The main idea behind noncommutative geometry is to study the topology (and
geometrical properties) of a space by not seeing it as a set of points, but
rather by investigating the set of fields defined on it. In this sense the
tools of noncommutative geometry resemble the methods of modern theoretical
physics. The main mathematical result which enables such a study lies in a
series of theorems due to Gel'fand and Naimark (for a review see for example
\cite{FellDoran,Di}). They established a complete equivalence between {\em
Hausdorff Topological Spaces} and {\em Abelian $C^*$-algebras}.

It is worthwhile to recall these two concepts. A topological space is
a set with a topology (a collection of open subsets obeying certain
conditions) defined on it. A Hausdorff topology is one that makes the
space separable, i.e.  given two points it is always possible to find
two disjoint open sets each containing one of the two points. The
common topological spaces encountered in physics (for example
manifolds) are separable\footnote{For an example of a {\em
non-Hausdorff} space in the context of noncommutative geometry see
\cite{NCL}.}.  A $C^*$-algebra $\alg$ is, first of all, an
associative algebra over the complex numbers $\complex$, i.e. a set on
which two operations, sum (commutative) and product (not necessarily
commutative), are defined with the following properties:
\begin{itemize}
\item[1)] $\alg$ is a vector space over $\complex$, i.e. $\alpha a + \beta b
\in \alg$ for $a, b \in \alg$ and $\alpha,\beta \in \complex$.
\item[2)] It is distributive over addition with respect to left and right
multiplication, i.e. $a(b+c) = ab + ac$ and $(a+b)c=ac+bc$, $\forall~a,b,c
\in\alg$.
\end{itemize}
\noindent
$\alg$ is further required to be a Banach algebra:
\begin{itemize}
\item[3)] It is complete with respect to a norm
$\norm{\cdot}:\alg\to\real$ with the usual properties
\begin{itemize}
\item[a)] $\norm{a} \geq 0~, ~~~ \norm{a} = 0 ~\iff ~ a = 0~$
\item[b)] $\norm{\alpha a} = |\alpha|\,\norm{a}$
\item[c)] $\norm{a+b} \leq \norm{a} + \norm{b}$
\item[d)] $\norm{ab} \leq \norm{a}\, \norm{b}$
\end{itemize}
\end{itemize}
The Banach algebra $\alg$ is called a \cstar if, in addition to the properties
above, it also has defined on it a conjugation operation $*$ (analogous to the
one usually defined for complex numbers) with the properties
\begin{itemize}
\item[4)] $a^{**} = a$
\item[5)] $(ab)^* = b^* a^*$
\item[6)] $(\alpha a + \beta b)^* = \bar{\alpha} a^* + \bar{\beta}b^*$
\item[7)] $\norm{a^*}=\norm{a}$
\item[8)] $\norm{a^*a}=\norm{a}^2$
\end{itemize}
for any $a,b \in \alg$ and $\alpha, \beta \in \complex$, where $\bar\alpha$
denotes the usual complex conjugate of $\alpha\in\complex$.

Note that the existence of an identity $\unit$ is not an axiom. If an identity
with the property $a\,\unit=\unit\,a=a~~\forall a\in\alg$ exists then the
algebra is called {\em unital}. The simplest example of a $C^*$-algebra is the
algebra of complex numbers $\complex$ itself. Matrix algebras, as well as
algebras of bounded (or compact) operators on a separable Hilbert space, are
also examples of $C^*$-algebras, in general noncommutative. The $*$ conjugation
in these cases is the usual Hermitian conjugation $\dagger$, while the norm is
the  operator norm, i.e. $\norm{T}^2$ is the largest eigenvalue
of $T^\dagger T$. Another example of an abelian \cstar is the algebra $C(M)$ of
{\em continuous} complex-valued functions on a topological set $M$ \footnote{In
the case where the space is non-compact one has to further restrict to
functions which vanish on the frontier of the space.}. The norm in this case is
the $L^\infty$-norm, i.e. the maximum value attained by the function:
\beq
\norm{f}_\infty= \sup_{x\in M}|f(x)|~.
\eeq

The Gel'fand-Naimark theorem states that {\em there is a one-to one
correspondence between {\em Hausdorff} topological spaces and {\em
commutative} $C^*$-algebras}. The correspondence in one direction is
rather simple. Given a topological space $M$ it is possible to
naturally construct the algebra $C(M)$ of continuous complex-valued
functions on it. This is a commutative $C^*$-algebra, and moreover it
depends (through the continuity criterion) on the topology of the
original space. The correspondence in the other direction is much more
sophisticated and can be seen in a constructive way. Given an abelian
\cstar $\alg$, i.e. a set of elements with two commutative operations,
a norm and a conjugation, one can {\em reconstruct} the topological
space $M$ for which $\alg\cong C(M)$ is the algebra of continuous
complex-valued functions on $M$.

\subsubsection*{Topological Space as a Set of Ideals}

Let us now briefly describe this reconstruction process. An {\em ideal}
$I\subset\alg$ of an algebra $\alg$ is a subalgebra with the property:
\beq
ab\in I\ \mbox{and}\ ba\in I\ ~~~ \forall a\in I,b\in\alg
\eeq
That is, not only is the ideal closed under summation and multiplication (as it
is a subalgebra), but the product of any element of the ideal with any element
of the whole algebra is also in the ideal. A {\em maximal ideal} is an ideal
which is not contained in any other ideal (apart from the trivial ideal which
is the whole algebra $\alg$).

The relevant example of an ideal is the collection of continuous
functions which vanish on some subset of a topological space $M$. It
is easy to check that they form an ideal, and moreover that the sets
$I_x$ of functions which vanish at a single point $x\in M$ are maximal
ideals. Note that if $I_{x,y}$ denotes the ideal of functions which
vanish at two points $x,y\in M$, then $I_{x,y}=I_x\cap I_y\subseteq
I_x,I_y$. In this sense a reduction in the number of points is
equivalent to an increase in the number of functions in the ideal.

We will therefore identify the points of the space under
reconstruction with the maximal ideals of the given commutative
algebra $\cal A$ \footnote{If the algebra has no unit then a further
requirement (modularity) must be imposed \cite{FellDoran}.}. We must
at this point reconstruct the topology of the space, i.e. the
relations between the points. This can be done using the {\em
Hull kernel} or {\em Jacobson} topology. As is well-known the topology
of a space can be given in various equivalent ways, for example in
terms of open sets, or their complement closed sets. Another
equivalent way is to give the closure of every set. Closed subsets are
then defined as the ones which coincide with their closure. The
closure operation must satisfy some axioms due to Kuratowski
\cite{Landi}:
\begin{enumerate}
\item $\overline{\emptyset} = \emptyset$ \ ;
\item $W \subseteq \overline{W}~, ~~~ \forall ~W$ \ ;
\item $\overline{\overline{W}} = \overline{W}~, ~~~ \forall ~W$\ ;
\item $\overline{W_1 \cup W_2} = \overline{W}_1 \cup \overline{W}_2~,
{}~~~\forall ~W_1, W_2$ \ .
\end{enumerate}

In an abelian \cstar we can define the closure of any subset $W$ of the set of
maximal ideals. Being the union of ideals, $W$ is itself an ideal, although not
a maximal one in general. We define the closure $\overline{W}$ of $W$ as the
set of maximal ideals $I_m$ with
\beq
\overline{W}\equiv\left\{ I_m\subseteq\alg:\bigcap_{I\in W}I\subseteq
I_m\right\}
\eeq
That is, the point corresponding to the ideal $I_m$ belongs to the closure of
$W$ if $I_m$ contains the intersection of all ideals $I$ of $W$. All
maximal ideals which comprise $W$ belong to $\overline W$.
It is possible to prove that the above definition of closure satisfies
the Kuratowski axioms \cite{Landi}.

As an example let us find the closure of the open interval. In terms
of the above construction, $W$ is the ideal of all functions which vanish at
some point in the open interval. The intersection of all of these ideals is the
ideal of functions vanishing on the open interval. An ideal $I_m$ therefore
belongs to the closure of $W$ if it contains all functions which vanish in the
open interval. Recall now that the functions we are considering are {\em
continuous} (this carries the information about the topology), and, therefore,
if they vanish in the open interval, they vanish at the endpoints of the
interval as well. Thus the functions which vanish at any point of the {\em
closed} interval belong to the closure of $W$.

\subsubsection*{Topological Space as a Set of Irreducible Representations}

There is an alternative construction of the space and its topology. In this
construction the points are the irreducibile representations of the algebra (or
rather their equivalence classes modulo algebra-isomorphism). This space is
called the {\em structure space} of $\alg$ and is denoted $\hat\alg$. In the
case of commutative algebras the irreducible representations are all
one-dimensional. They are thus the (non-zero) $*$-linear functionals
$\phi:\alg\to\complex$ which are multiplicative, i.e. 
$\phi(ab)=\phi(a)\phi(b)$
for any $a,b\in\alg$. It follows that for unital algebras
$\phi(\unit)=1,~\forall\phi\in\hat\alg$. For commutative algebras the set of
irreducible representations and the set of maximal ideals coincide
\cite{FellDoran}. We can consider a representation $\phi_x$ as the evaluation
map which gives the values of the elements of the algebra at a point $x$:
\beq
\phi_x(f)\equiv f(x)~~~~,~~~~ f\in\alg \label{deltafun}
\eeq
The topology in the structure space is given using the concept of pointwise
convergence. A sequence $\phi_{x_n}$ of representations is said to converge to
$\phi_x$ if and only if for any $a\in\alg$ the sequence
$\phi_{x_n}(a)\to\phi_x(a)$ in the usual topology of $\complex$. It is not at
all trivial to show that the two topologies we described above are equivalent
\cite{Landi,FellDoran}, and in the proof all of the ingredients that make up a
\cstar play a role.

Yet another way to reconstruct the topological space, which is very similar to
this latter method, is to consider the space of {\em characters} of the
algebra. A character is a multiplicative $*$-linear  functional
\beq
\phi:\alg\to\complex~~~~;~~~~ \phi(ab) = \phi(a)\phi(b)~~\ \forall a,b\in\alg
\eeq
For abelian algebras the space of characters is the same as the space of
irreducible representations. The
space of characters of the algebra is therefore also called the structure space
of the algebra. The connection with the space of maximal ideals is easily done
by considering the space of {\em primitive ideals}. An ideal $I$ is primitive
if given an irreducible representation $\pi$,
\beq
I = \ker\pi
\eeq
For commutative algebras a primitive ideal is also maximal, and vice-versa, so
that the two spaces coincide. For a noncommutative algebra, for which
not all irreducible representations are one-dimensional, this is no longer
true.

The important aspect that we want to stress is that the investigation
of the topology of a space $M$ in terms of the relations among its points is
completely equivalent to the analysis of the algebra of functions defined on
it. Therefore the study of topological spaces (manifolds etc.) can be
substituted by a study of $C^*$-algebras. In the case where $M$ is a manifold,
in addition to the topology the differentiable structure is determined by the
algebra $C^\infty(M)$ of smooth functions on $M$. The algebra $C^\infty(M)$ is
not a
\cstar (but only a $*$-algebra) because it is not complete with respect
to the $L^\infty$-norm. The fact that with it we reconstruct the same space as
$C(M)$ is therefore not a violation of the Gel'fand-Naimark theorem, since in
fact the completion of $C^\infty(M)$ is $C(M)$. Manifolds are
characterized by the presence of tangent bundles. In this respect an
important theorem due to Serre and Swan \cite{SerreSwan} provides a one-to-one
correspondence between smooth sections of a vector bundle over a
manifold $M$ and finitely-generated projective modules over the algebra
$C^\infty(M)$, i.e. vector spaces on which the algebra acts and which can be
generated using a projector acting on finitely-many tensor products of the
algebra. Thus the study of the differentiable structures of
manifolds is equivalent to the study of finitely-generated projective modules.

\subsubsection*{Metric Spaces from Dirac Operators}

A physical space has much more structure to it than just topology,
which is in fact the most basic aspect of it, having to do mostly with global
properties. Connes \cite{Book} has shown that a metric and other local aspects
can be encoded at the level of algebras, and how fermionic and bosonic actions
can be written. The key property is another important result due to Gel'fand,
the fact that any \cstar $\alg$ (commutative or otherwise) can be represented
faithfully as a subalgebra of the algebra ${\cal B}({\cal H})$ of 
bounded operators 
on an infinite-dimensional separable Hilbert space ${\cal H}$. 
In the following we
will not distinguish between the algebra and this representation, and we take
$\alg\subset{\cal B}({\cal H})$. The norm of $\alg$ is represented by the
operator norm on ${\cal B}({\cal H})$,
\beq
\norm{a}=\sup_{\langle|\phi,\phi\rangle\leq 1}|\langle\phi|a|\phi\rangle|
\label{opnorm}\eeq
The metric structure, as well as the noncommutative generalization of
differential and integral calculus, is obtained via an operator which is the
generalization of the usual Dirac operator. We shall call it a Dirac operator
even in this generalized context.

{}From the point of view of noncommutative geometry the Dirac operator is a
(not necessarily bounded) operator $D$ on $\cal H$ with the following
properties:
\begin{enumerate}
\item $D$ is self-adjoint, i.e. $D=D^\dagger$ on the common domain of the
two operators.
\item The commutator $[D,a]$ ($a\in\alg$) is bounded on a dense
subalgebra of $\alg$.
\item $D$ has compact resolvent, i.e. $(D-\lambda)^{-1}~{\rm for}~\lambda
\not\in\real$ is a compact operator on ${\cal H}$.
\end{enumerate}
If the algebra is commutative, and therefore it has a structure space which
determines a corresponding algebra of continuous functions on a topological
space, then the Dirac operator enables one to give the topological space a
metric structure via the definition of the {\em distance}. We define the
distance between two points of the structure space as
\beq
d(x,y)=\sup_{a\in\alg}\left\{|a(x) -
a(y)|~:~\norm{[D,a]}\leq1\right\}\label{dist1} \ \ .
\eeq
This notion of distance coincides with the usual definition of geodesic
distance for a Riemannian manifold with Euclidean-signature metric
$g_{\mu\nu}$:
\beq
\bar d(x,y) = \inf_\gamma~\ell_\gamma(x,y)\label{dist2}
\eeq
where $\ell_\gamma(x,y)$ is the length of the path $\gamma$ from $x$ to $y$ with
respect to $g_{\mu\nu}$.

It is easy to convince oneself of the equality of these two definitions in a
simple one-dimensional example. In this case the algebra is the set of
functions on a line (or interval or circle) and the Dirac operator is the
derivative $D=i\frac d{dx}$. The distance between $x$ and $y$ is simply $|x-y|$
if we use the usual metric. The definition (\ref{dist1}) says that we must take
the supremum over those functions whose derivatives have norm $\leq 1$. This
means that $\frac{da}{dx}$ must be nowhere greater than 1. It follows that then
$|a(x)-a(y)|\leq|x-y|$, the inequality being saturated by any function with the
property $a(t)=t$ for $t\in[x,y]$.

To see the relation between the two definitions in a higher-dimensional
example, consider the usual Dirac operator $D= i\gamma^\mu \partial_\mu$ acting
on some dense domain of the Hilbert space ${\cal H}=L^2({\rm spin}(\real^d))$
of square-integrable spinors $\psi(x)$ defined on $\real^d$. The real-valued
gamma-matrices generate the Euclidean Dirac algebra
$\{\gamma^\mu,\gamma^\nu\}=2\delta^{\mu\nu}$. We take $\alg$ to be the algebra
of continuous complex-valued functions on $\real^d$ which vanish at infinity,
and consider them as operators on ${\cal H}$ acting by pointwise
multiplication. The commutator $[D,a]$, acting on the spinor field $\psi$, is
\beq
[D,a]\psi=(i\gamma^\mu\partial_\mu a)\psi + ia\gamma^\mu\partial_\mu\psi -
ia\gamma^\mu\partial_\mu\psi=i(\gamma^\mu\partial_\mu a)\psi
\eeq
The $L^\infty$-norm of the commutator is thus the maximum value of the
$\complex^d$ norm $\sqrt{\del^\mu a^* \del_\mu a}$, which is also equal
\cite{VG} to the Lipschitz norm of $a$:
\beq
\norm{[D,a]}_\infty = \sup_{x\in\reals^d}~\sqrt{\del^\mu a^*(x) \del_\mu
a(x)}= \norm{a}_{\rm Lip}\equiv\sup_{x\neq y}~{|a(x) - a(y)|\over \bar
d(x,y)}\label{Lip}
\eeq
To check that the definitions of distance in \eqn{dist1} and \eqn{dist2} agree,
note that the condition on the norm of $[D,a]$ in \eqn{dist1} along with
\eqn{Lip} imply
\beq
d(x,y)\leq\bar d(x,y) \ \ .
\eeq
But $\bar a(t)\equiv\bar d(x,t)$ as a function of $t$ (suitably regularized)
satisfies
\beq
\norm{[D,\bar a(t)]}_\infty = \norm{\bar d(x,t)}_{\rm Lip} = 1
\eeq
which gives
\beq
d(x,y) = \bar d(x,y) \ \ .
\eeq

Given the notion of distance between pairs of points, the other properties of a
metric space can be obtained by the traditional techniques. The set of data
$(\alg,{\cal H},D)$, i.e. a $C^*$-algebra $\alg$ of bounded operators on a
Hilbert space $\cal H$ and a Dirac operator $D$ on $\cal H$, is called a {\em
spectral triple}, and it encodes the geometry and topology of a space under
consideration. The pair $({\cal H},D)$ is called a {\em Dirac K-cycle} for
$\alg$ and it can be used to describe the cohomology of a space using K-theory
\cite{Book}. A different choice of Dirac operator will alter the metric
properties of the space. Let us go back to the simple one-dimensional example
above and suppose that we choose as Dirac operator
\beq
D'=i\frac{d}{dF(x)}=i\left(\frac{dF(x)}{dx}\right)^{-1} \frac{d}{dx}
\eeq
for a monotonic real-valued function $F$. It is then easy to see that
\beq
[D',a]\leq 1 ~~~ \Rightarrow ~~~ {da(x)\over dx}\leq   {dF(x)\over
dx}~~~~\forall x
\eeq
and that therefore
\beq
|a(x)-a(y)|\leq|F(x)-F(y)| \ \ ,
\eeq
again the equality being attained for $a=F$.

The change in Dirac operator in this simple example corresponds of course to
the use of the measure $dF(x)$ instead of $dx$. But we want to stress the point
that a change of measure is, in this context, a change of Dirac operator. Later
on we will see how different choices of the Dirac operator for the case of
string theory corresponds to spacetimes whose metric structures are related by
T-duality and other geometric transformations.

\subsubsection*{Distance between States of an Algebra}

The notion of distance between points does not generalize to the noncommutative
case, where it is in general impossible to speak of points in a topological
space, let alone of distances between them. It is, however, possible to
speak of distances between the {\em states} of the algebra $\alg$. A state is a
positive definite unit norm map,
\beq
\Psi:\alg\to\complex~~~~;~~~~\Psi (a^*a)\geq0\ ,\forall a\in
\alg,~~~\norm{\Psi}_\infty=1\ .
\label{statedef}\eeq

If the algebra is represented on a Hilbert space $\cal H$, then vectors
$|\psi\rangle\in{\cal H}$ define states via the expectation
values\footnote{Even if one starts with an abstract algebra $\alg$ it is always
possible to associate
a representation of $\alg$ to a state via a construction due to
Gel'fand, Naimark and Segal. The space of states of an algebra (after
quotienting out the states for which $\Psi(a^*a)=0$) can be made into
a Hilbert space. In the cases under consideration in this paper, however, a
Hilbert space is provided from the onset by the quantum theory of the given
physical problem.}:
\beq
\Psi(a)=\langle\psi|a|\psi\rangle \ \ .
\eeq
States are, however, a more general concept, as the `delta-functions' defined
in (\ref{deltafun}) are states in the sense of \eqn{statedef}
as well although, as is well-known,
they do not correspond to the expectation value of any vector of the Hilbert
space. We denote the space of states by ${\cal S}(\alg)$. Since
\beq
\lambda \Psi + (1-\lambda) \Phi \in {\cal S}(\alg)\ ,~~~~\forall  \Psi, \Phi
\in
{\cal S}(\alg),\ \lambda\in[0,1]
\eeq
the set of all states of an algebra $\alg$ is a convex space. Being a convex
space ${\cal S}(\alg)$ has a boundary whose elements are called {\it pure
states}. The delta-functions \eqn{deltafun} are examples of pure states.
Namely, a state is called pure if it cannot be written as the convex
combination of (two) other states. Another characterization of the structure
space is as the space of pure states, which in the commutative case coincides
with the set of irreducible representations, and hence the space of
characters, and moreover coincides with the space of maximal ideals of $\alg$.

The distance between states (pure or mixed) is then defined as
\beq
d(\Psi,\Phi) = \sup_{a\in\alg}\left\{|\Psi(a) - \Phi(a)|~:~\norm{[D,a]}\leq
1\right\}\label{diststa} \ \ .
\eeq
This distance is much more general than the distance between
the points (pure states) as it could be applied to calculate the
distance between any two quantum states of a physical system.

\subsubsection*{Differential Forms on Noncommutative Spaces}

Another important role played by the Dirac operator $D$ is in the construction
of the algebra of differential forms in the context of noncommutative geometry.
The key idea is to also represent differential forms as operators on ${\cal
H}$, on a par with $D$ and $\alg$. We first define the (abstract) {\em
universal differential algebra of forms} as the $\zed$-graded algebra
\beq
\Omega^*\alg = \bigoplus_{p\geq0}\Omega^p\alg
\eeq
which is generated as follows:
\beq
\Omega^0\alg=\alg
\eeq
and $\Omega^1\alg$ is generated by a set of abstract symbols $da$ which
satisfy:
\bea
d(ab) &=& (d a)b + a d b~,~~~ \forall ~a,b \in \alg\ ~~~\mbox{(Leibnitz
Rule)}  \label{leib} \\
d(\alpha a + \beta b) &=& \alpha d a + \beta d b~,~~~ \forall ~a,b \in
\alg~, ~~\alpha , \beta \in \complex\ ~~~\mbox{(Linearity)}  \label{leibbis}
\eea
Elements of $\Omega^p\alg$ are linear combinations of elements of the form
\beq
\omega=a_0da_1\cdots da_p
\label{pforms}\eeq
Because of (\ref{leib}) a generic $p$-form can be written as a linear
combination of forms of the kind (\ref{pforms}), with $a_0$ possibly a multiple
of $\unit$ in the unital case. This makes $\Omega^p\alg$ a $\zed_2$-graded
$\alg$-module. The graded exterior derivative operator is the nilpotent linear
map $d:\Omega^p\alg\to\Omega^{p+1}\alg$ defined by
\beq
d(a_0da_1\cdots da_p)=da_0da_1\cdots da_p
\eeq

We define a linear representation $\pi_D:\Omega^*\alg\to{\cal B}({\cal H})$ of
the universal algebra of abstract forms by
\beq
\pi_D(a_0da_1\cdots da_p )= a_0[D,a_1]\cdots [D,a_p]
\eeq
Notice, however, that $\pi_D(\omega)=0$ does not necessarily imply
$\pi_D(d\omega)=0$. Forms $\omega$ for which this happens are called {\em junk
forms}. They generate a $\zed$-graded ideal in $\Omega^*\alg$ and have to be
quotiented out \cite{Book,Landi}. Then the noncommutative differential algebra
is represented by the quotient space
\beq
\Omega_D^*\alg=\pi_D\left[\Omega^*\alg/(\ker\pi_D\oplus d\ker\pi_D)\right]
\label{OmegaD}\eeq
which we note depends explicitly on the particular choice of Dirac operator $D$
on the Hilbert space $\cal H$.

The algebra $\Omega_D^*\alg$ determines a DeRham complex whose cohomology
groups can be computed using the conventional methods. With this machinery, it
is also possible to naturally define (formally at this level) a
vector bundle $E$ over $\alg$ as a finitely-generated projective left
$\alg$-module, and along with it the usual definitions of connection,
curvature, and so on \cite{gravityNCG}. However, in what follows we shall for
the most part use only the trivial bundle over a unital $C^*$-algebra $\alg$.
For this we define a gauge group ${\cal U}(\alg)$ as the group of unitary
elements of $\alg$,
\beq
{\cal U}(\alg)=\left\{u\in\alg~|~u^\dagger u=uu^\dagger=\id\right\}
\label{unitarygp}\eeq
Alternatively, the gauge group can be defined as the subgroup of unimodular
elements $u\in{\cal U}(\alg)$, i.e. the unitary operators of unit determinant
$\det u=1$. The presence of one-forms is
then tantamount to the possibility of defining a connection, which is a generic
Hermitian one-form $\rho=\sum_ia_i[D,b_i]$, and with it 
a covariant Dirac operator
$D_\rho=D+\rho$.  The curvature of a connection $\rho$ is defined to be
\beq
\theta=[D,\rho]+\rho^2 \label{curv}
\eeq

\subsubsection*{Integration in Noncommutative Geometry}

The final ingredient of differential geometry we discuss is the integral. Since
we are representing all of the objects as operators on ${\cal H}$, it is
natural to define the integral as a trace. It is in fact \cite{Book} a
regularized trace called the {\it Dixmier trace} which is defined as follows.
Consider a generic bounded operator $L$ on $\cal H$ of discrete spectrum with
eigenvalues $\lambda_n$ ordered according to modulus and counted with the
appropriate multiplicities. We then define its Dixmier trace $\tr_\omega L$ to
be
\beq
\tr_\omega L=\lim_{N\to\infty}{1\over \log N}\sum_{n=1}^N \lambda_n
\eeq
For the algebra of continuous functions on a $p$-dimensional compact manifold
$M$, this definition then yields \cite{Book}
\beq
\int_M f(x)~d^px =\tr_\omega~ f |D|^{-p}
\label{intdirac}\eeq
and we see that in a loose sense the Dirac operator $D$ is the ``inverse'' of
the infinitesimal $dx$. The right-hand side of \eqn{intdirac} can be calculated
using heat-kernel methods.

\subsubsection*{Example of a Manifold}

Let us work out the example\footnote{The beginning of this subsection is a very
simplified version of section 6.2.1 of \cite{Landi}.} of a $p$-dimensional
spin-manifold $M$ with metric $g_{\mu\nu}$ and the usual Dirac operator
$D=i\gamma^\mu\nabla_\mu$. The real-valued gamma-matrices generate the
corresponding Clifford algebra $\{\gamma^\mu,\gamma^\nu\}=2g^{\mu\nu}$ of the
spin bundle ${\rm spin}(M)$ of $M$, and $\nabla_\mu$ is the usual covariant
derivative constructed from the Levi-Civita spin-connection of $M$.
Notice how these latter two dependences of $D$ characterize the
Riemannian geometry of $M$, and that, in this example, $D$ acts
densely on the Hilbert space, i.e. it maps a dense subspace of ${\cal
H}=L^2({\rm spin}(M))$ into itself. Zero-forms
are just complex-valued functions, while we can represent an exact one-form as
\beq
\pi_D(\del_\mu f dx^\mu)=i\del_\mu f\gamma^\mu
\eeq
so that a generic one-form is represented as $f_\mu\gamma^\mu$, i.e.
$\Omega_D^1\alg$ is a free $\alg$-module with basis $\{\gamma^\mu\}$. As we
build two-forms we run into the problem of junk forms. Consider the form
$\alpha=fdf-(df)f$. As a form in the universal algebra it is non-zero. Its
representation does, however, vanish:
\beq
\pi_D(\alpha)=\pi_D(fdf-(df)f)=i\gamma^\mu(f\del_\mu f -(\del_\mu f)f)=0
\eeq
while the representation of its differential does not vanish:
\beq
\pi_D(d\alpha)=-\gamma^\mu\gamma^\nu\del_\mu f\del_\nu f=
-2g^{\mu\nu}\del_\mu f\del_\nu f
\eeq
We can therefore identify junk forms as the symmetric part of the product of
two one-forms. This has to be quotiented out leaving only the antisymmetric
part, so that a generic two-form is represented as $f_{\mu\nu}\gamma^{\mu\nu}$
where $\gamma^{\mu\nu}=(\gamma^\mu\gamma^\nu-\gamma^\nu
\gamma^\mu)/2$, i.e. $\Omega_D^2\alg$ is a free $\alg$-module with basis
$\{\gamma^{\mu\nu}\}$.
Analogously one constructs higher-degree forms by antisymmetrizations of the
$\gamma$'s. Forms of degree $p+1$ or higher are all junk forms.

A connection is a generic one-form $\rho=A_\mu\gamma^\mu$
and the curvature defined in \eqn{curv} is the familiar Maxwell
tensor:
\beq
\theta=\mbox{$1\over 2$}F_{\mu\nu}\gamma^{\mu\nu}
\eeq
with $F_{\mu\nu}=\partial_\mu A_\nu-\partial_\nu A_\mu$. The Dixmier trace
reduces to the usual spacetime integral along with a trace over spinor indices.

Note that in this case arbitrary iterated commutators of the form
$[D,[D,[\cdots,[D,a]\cdots]$ are bounded only on the dense subalgebra
$C^\infty(M)\subset C(M)$. Thus to ensure boundedness of all operators under
consideration, one should restrict attention to the $*$-algebra $C^\infty(M)$,
rather than the $C^*$-algebra $C(M)$. As mentioned before, nothing is lost in
such a restriction, and therefore in the following we shall for the most part
not be concerned with the completeness of the algebra in the spectral triple.
Notice also that $C^\infty(M)$ acts densely on $\cal H$.

\subsubsection*{Fermionic and Bosonic Actions; The Standard Model}

It is also possible to define the action of a `noncommutative gauge
theory'. The fermionic action is defined as
\beq
S_F=\tr_\omega~\psi^\dagger D_\rho\psi=\langle\psi|D_\rho|\psi\rangle
\eeq
where the last equality gives the usual inner product on the Hilbert space of
square-integrable spinors, while the bosonic action is
\beq
S_B=\tr_\omega~\theta^2
\eeq
Other forms of the bosonic action have been introduced more recently in
\cite{ChamsConnes,CostaRica}. There is also another version of gauge theories
in noncommutative geometry which employs Lie superalgebras
\cite{Dubois}.

The fermionic and bosonic actions for the abelian case described in the example
above combine into the Dirac-Maxwell action for quantum electrodynamics. But
the most interesting perspective is of course the generalization to the case of
noncommutative algebras. Connes and Lott \cite{ConnesLott} have shown how the
application of this machinery gives a simple generalization of spacetime to a
two-sheeted spacetime, described by the algebra of functions with values in
the space of diagonal $2\times2$ matrices, that has a bosonic action which
naturally gives not only a Yang-Mills theory, but also the Higgs potential with
its biquadratic form. Connes \cite{RealNCG} (for a review including
more recent developments see \cite{cordelia}) has generalized the model to the
full standard model with a noncommutative algebra
\beq
\alg_{SM}=C^\infty(M)\otimes\left[\complex\oplus\quater\oplus
M(3,\complex)\right]
\label{standardalg}\eeq
where $\quater$ is the algebra of quaternions and $M(3,\complex)$ is the
algebra of $3\times3$ complex-valued matrices. The unimodular group of this
algebra is the familiar gauge group $U(1)\times SU(2)\times SU(3)$ of the
standard model. The action contains the Yang-Mills action and the Higgs
term. The input parameters are the masses of all fermions and the coupling
constants of the gauge group, while the (classical) mass of the Higgs boson is
a prediction of the model. Other predictions for masses at current energies can
also be made \cite{Marseille}. The actions of \cite{ChamsConnes,CostaRica} also
enable the introduction of gravity into the model.

\newsection{Quantum Spacetimes and the Fr\"ohlich-Gaw\c edzki Construction}

In this section we will now try to generalize the material developed in the
previous section to the generalizations of spaces suggested by noncommutative
geometry. We first discuss some generalities on noncommutative spaces, leading
up to the spacetime of string theory. We then begin the construction of the
spectral triples pertinent to string theory, as suggested by the work of
Fr\"ohlich and Gaw\c edzki \cite{FG}. In the latter half of this section we
will present a systematic construction of the string Hilbert space, following
\cite{FG} for the most part, by analyzing in detail the basic quantum fields
which will compose the quantum spacetime.

\subsubsection*{Noncommutative Spaces}

There are various examples of noncommutative spaces which are close in
character to the context which will be described in the
following. Some of them are the {\em Fuzzy Sphere} \cite{fuzzy}, {\em
noncommutative lattices} \cite{NCL}, and the possibility of having a
spacetime with noncommuting coordinates as described in
\cite{Dfield,Matrix,DFR}. Quantum groups \cite{QGroups} are
noncommutative spaces, in the sense that the algebra of functions
defined on them is noncommutative, while another example is the
quantum plane \cite{QuantumPlane}.

What is common to this variety of noncommutative spaces is that they
are described by the noncommutative algebra defined upon them. It
might in general be a misnomer to call them ``spaces'', in the sense
that the concept of point is not appropriate.  The problem in dealing
with noncommutative \cstars is that the various sets described in the
previous section, of maximal ideals, primitive ideals, irreducible
representations, characters, and pure states, and the topologies that
these concepts induce, are no longer equivalent. The identification of
the points of the space therefore becomes ambiguous, and has to be
abandoned. However, another important feature, which is common to
these examples, is that there exists some regime, usually when a scale
goes to zero, in which it is possible to recognize a topological
space. Such a ``low energy'' regime is of course necessary if one
wants to identify at some level the space in which we live and do
experiments.

In string theory a low energy regime is one in which no excited, vibrational
states of the string are present and, in the case of closed strings, no string
modes wind around the spacetime. In this regime the theory is well described by
an ordinary (point particle) quantum field theory. In this respect we look for
a noncommutative algebra which describes the ``space'' of interacting strings.
{\it Vertex operators} were originally introduced in string theory to describe
the interactions of strings. They operate on the Hilbert space of strings
as insertions on the worldsheet corresponding to the emission or absorption of
string states. The ``coordinates'' of a string are the
Fubini-Veneziano fields $X$ (which we will construct formally in the next
subsection), and the basic vertex operators are objects of the form $\e^{ipX}$.
These vertex operators act as a basis of a ``smeared'' set of vertex operators
which form a noncommutative algebra. For toroidally compactified strings
restrictions are
imposed on the momenta $p$, which have to lie on an even self-dual lattice. We
will discuss this algebra in more detail in section 5, and we have described
some of the more formal algebraic properties of vertex operator algebras in the
appendix. This is the spacetime that was proposed by Fr\"ohlich and Gaw\c edzki
in \cite{FG}. In their construction the algebra is the vertex operator algebra
of the string theory under consideration. The spacetime is thus described by
the operator algebra which describes the relations among the quantum fields of
the conformal field theory.

String theories come with a scale (the string tension), which is usually set to
be of the order of the Planck mass. Usually the dimensions of compactified
directions are taken to be of a similar scale, so that ``low energy'' in this
context means to neglect higher (vibrational) excited states of the strings, as
well as the non-local states which correspond to the strings winding around a
compactified direction. In this limit the theory becomes a theory of point
particles, and we should thus expect a commutative spacetime. In fact, a
quantum theory of point particles on a spin-manifold $M$ naturally supplies the
Hilbert space ${\cal H}=L^2({\rm spin}(M))$ of physical states and the
commutative $*$-algebra $\alg=C^\infty(M)$ of observables. Thus the
low-energy limit of the noncommutative string spacetime should be represented
by a spectral triple $(C^\infty(M),L^2({\rm
spin}(M)),ig^{\mu\nu}\gamma_\mu\nabla_\nu)$ corresponding to an ordinary
spacetime manifold $M$ at large distance scales.

Let us briefly remark that there is another approach to the noncommutative
spacetime which is inspired by D-brane field theory
\cite{fgrleshouches,Dbranes}. An effective low-energy description of open
superstrings is provided by supersymmetric $U(N)$ Yang-Mills theory in ten
dimensions. Dimensional reduction of this gauge theory down to a
$(p+1)$-dimensional manifold $M_{p+1}$ describes the low-energy dynamics of $N$
D$p$-branes, with $U(N)$ gauge symmetry. In \cite{fgrleshouches,Dbranes} the
spacetime is thus described in analogy with the Connes-Lott formulation of the
standard model by choosing as algebra
\beq
\alg_D=C^\infty(M_{p+1})\otimes M(N,\complex)
\label{Dalg}\eeq
This construction is rather different in spirit from the one which we shall
present in the following, in that it utilizes a different 
low-energy regime of the string theory.

\subsubsection*{Fubini-Veneziano Fields}

We now begin the construction of the spectral triple describing the
noncommutative spacetime of string theory. From the point of view of the
corresponding conformal field theory, it is more natural to start with the
Hilbert space of physical states rather than the algebra since the former, with
its oscillatory Fock space component, is one of the immediate characterizations
of the string theory. We shall take this to be the space of states on
which the quantum string configurations act. We consider a linear
sigma model with target space the flat $n$-torus
\beq
T^n=(S^1)^n\cong\real^n/2\pi\Gamma
\label{ntorus}\eeq
where $\Gamma$ is a lattice of rank $n$ with inner product $g_{\mu\nu}$ of
Euclidean signature. The action of the model is (in units where $l_P=1$)
\beq
S[X]=\frac1{2\pi}\int_\Sigma\left[g_{\mu\nu}dX^\mu\wedge\star
dX^\nu+2X^*(\beta)\right]
\label{sigmaaction}\eeq
where $X^*(\beta)=\frac12\beta_{\mu\nu}dX^\mu\wedge dX^\nu$ is the pull-back of
the constant two-form $\beta$ to the worldsheet $\Sigma$ by the embedding
fields $X:\Sigma\to T^n$, and $\star$ denotes the Hodge dual. The kinetic term
in (\ref{sigmaaction}) is defined by the constant Riemannian metric
$g\equiv\frac12g_{\mu\nu}dX^\mu\otimes dX^\nu$ on $T^n$, and it leads to local
propagating degrees of freedom in the field theory. The second term in
(\ref{sigmaaction}) is the topological instanton term and it depends only on
the cohomology class of $\beta$. In fact, as we shall show, at the quantum level
the two-form $\beta$ takes values in the torus $\beta\in
H^2(T^n;\real)/H^2(T^n;\zed)$, where the real cohomology represents the local
gauge transformations $\beta\to\beta+ d\lambda$ while the integer cohomology
represents the large gauge transformations $\beta\to\beta+4\pi^2C$ with $C$ a
closed two-form with integer periods. We shall find it convenient to introduce
the non-singular `background' matrices
\beq
[d_{\mu\nu}^\pm]=[g_{\mu\nu}\pm\beta_{\mu\nu}]
\label{ddef}\eeq
which, when $n$ is even, determine both the complex and K\"ahler structures of
$T^n$.

To highlight the relevant features of the construction, in this paper we shall
consider only the simplest non-trivial case where the Riemann surface $\Sigma$
is taken to be an infinite cylinder with local coordinates
$(\tau,\sigma)\in\real\times S^1$. On a cylinder the action (\ref{sigmaaction})
can be written in terms of the local worldsheet coordinates as
\beq
S[X]=\frac1{4\pi}\int d\tau~\oint d\sigma~\left[g_{\mu\nu}\left(\partial_\tau
X^\mu\partial_\tau X^\nu-\partial_\sigma X^\mu\partial_\sigma
X^\nu\right)-2\beta_{\mu\nu}\partial_\sigma X^\mu\partial_\tau X^\nu\right]
\label{sigmacyl}\eeq
and we assume periodic boundary conditions for the embedding fields
$\{X^\mu(\tau,\sigma)\}\in(S^1)^n$ (corresponding to closed strings). Canonical
quantization identifies the momentum conjugate to the field
$X^\mu(\tau,\sigma)$ as
\beq
\Pi_\mu=\mbox{$\frac1{2\pi}$}\left(g_{\mu\nu}\partial_\tau
X^\nu+\beta_{\mu\nu}\partial_\sigma X^\nu\right)
\label{canmomX}\eeq
which leads to the canonical equal-time quantum commutator
\beq
\left[X^\mu(\tau,\sigma),\Pi_\nu(\tau,\sigma')\right]=-i~\delta_\nu^\mu~\delta
(\sigma,\sigma')
\label{cancommX}\eeq

The one-forms $dX^\mu$ have winding numbers $w^\mu$ which represent the
number of times that the worldsheet circle $S^1$ wraps around the 
$\mu^{\rm th}$ circle of the torus $(S^1)^n$, i.e.
\beq
\frac1{2\pi}\oint_{S^1}dX^\mu=w^\mu~~~~~{\rm with}~~~~~\{w^\mu\}\in\Gamma
\label{windings}\eeq
These winding numbers define a homotopy invariant of the target space which, as
we will see below, label the connected components of the Hilbert space. In
addition to the periodicity condition (\ref{windings}), we also require that
\beq
\frac1{2\pi}\int_\tau dX^\mu=g^{\mu\nu}p_\nu~~~~~{\rm
with}~~~~~\{p_\mu\}\in\Gamma^*
\label{momquant}\eeq
where $\Gamma^*$ is the lattice dual to $\Gamma$ (obtained by joining the
centers of all plaquettes of $\Gamma$). Equations (\ref{momquant}) and
(\ref{canmomX}) define the momenta $p_\mu$ of the winding modes in
the target space. The periodicity condition follows from the fact that the
translation generator $\e^{ip_\mu X^\mu}$ on the target space should be
invariant under the windings of $X^\mu\in S^1$. In addition to these momentum
windings, from (\ref{canmomX}) and \eqn{windings} it follows that there are
also the translations generated by the instanton windings which are given by
$\beta_{\mu\nu}w^\nu$.

The closed one-form $dX$ can be written as
\beq
dX=dY+2\pi h
\label{dXdecomp}\eeq
where $Y:\Sigma\to\real^n$ is a single-valued function and $h\in
H^1(\Sigma;\Gamma)$ is a harmonic vector-valued one-form, $dh=d\star h=0$. In
the
case of the cylinder, we have from (\ref{windings}) that $h^\mu=w^\mu d\sigma$.
The Euler-Lagrange equations determine the local configurations $Y$ as the
solutions of the two-dimensional wave equation
\beq
\Box Y^\mu(\tau,\sigma)=0
\label{waveeq}\eeq
whose solutions are given by meromorphic expansions of the chiral spin-1
currents
\beq
-i\partial_\pm
Y_\pm^\mu(\tau\pm\sigma)=\sum_{k=-\infty}^\infty\alpha_k^{(\pm)\mu}~\e^{i(k+1)
(\tau\pm\sigma)}
\label{Ymodeexp}\eeq
with $(\alpha_k^{(\pm)\mu})^\dagger=\alpha_{-k}^{(\pm)\mu}$ and the light-cone
derivatives are $\partial_\pm=\partial_\tau\pm\partial_\sigma$.

Combining the above configurations, and splitting the fields up into chiral
sectors, we find that the configurations of the sigma-model (\ref{sigmacyl})
are given by the Fubini-Veneziano fields
\beq
X^\mu_\pm(\tau\pm\sigma)=x^\mu_\pm+g^{\mu\nu}p_\nu^\pm(\tau\pm\sigma)
+\sum_{k\neq0}\frac1{ik}~\alpha^{(\pm)\mu}_k~\e^{ik(\tau\pm\sigma)}
\label{chiralmultfields} \label{FubVen}\eeq
where, from the canonical quantum commutator (\ref{cancommX}), the zero-modes
$x^\mu_\pm$ (the center of mass coordinates of the string)
and the (center of mass) momenta $p_\mu^\pm$ are canonically conjugate
variables,
\beq
[x^\mu_\pm,p_\nu^\pm]=-i\delta_\nu^\mu
\label{cancommmomx}\eeq
with all other commutators vanishing. The left-right momenta are
\beq
p_\mu^\pm=\mbox{$\frac1{\sqrt2}$}\left(p_\mu\pm d_{\mu\nu}^\pm w^\nu\right)
\label{momlattice}\eeq
The set of momenta $\{(p_\mu^+,p_\mu^-)\}$ along with the integer-valued
quadratic form
\beq
\langle p,q\rangle_\Lambda\equiv p_\mu^+g^{\mu\nu}q_\nu^+-p^-_\mu
g^{\mu\nu}q_\nu^-=p_\mu v^\mu+q_\mu w^\mu~~~~,
\label{quadform}\eeq
where $q_\mu^\pm=\frac1{\sqrt2}(q_\mu\pm d_{\mu\nu}^\pm v^\nu)$, form an even
self-dual Lorentzian lattice
\beq
\Lambda=\Gamma^*\oplus\Gamma
\label{Lambdadef}\eeq
of rank $2n$ and signature $(n,n)$ which is called the Narain lattice
\cite{narain}. The even and self-duality properties, $\langle
p,p\rangle_\Lambda=2p_\mu w^\mu\in2\zed$ and $\Lambda^*=\Lambda$, guarantee
modular invariance of the worldsheet theory \cite{modinv}. The functions
(\ref{chiralmultfields}) define chiral multi-valued quantum fields of the
sigma-model. The oscillatory modes $\alpha_k^{(\pm)\mu}$ in
(\ref{chiralmultfields}) yield bosonic creation and annihilation operators
(acting on some vacuum states $|0\rangle_\pm$) with the non-vanishing
commutation relations (see (\ref{cancommX}))
\beq
\left[\alpha^{(\pm)\mu}_k,\alpha^{(\pm)\nu}_m\right]=k~g^{\mu\nu}~
\delta_{k+m,0}
\label{creannalg}\eeq
where $g^{\mu\lambda}g_{\lambda\nu}=\delta^\mu_\nu$.

The Hilbert space of states of this quantum field theory is thus
\beq
{\cal
H}_X=L^2((S^1)^n,\mbox{$\prod_{\mu=1}^n\frac{dx^\mu}{\sqrt{2\pi}}$})^\Gamma
\otimes{\cal F}^+\otimes{\cal F}^-
\label{sigmahilbert}\eeq
where
\beq
L^2((S^1)^n,\mbox{$\prod_{\mu=1}^n\frac{dx^\mu}{\sqrt{2\pi}}$})^\Gamma
=\bigoplus_{\{w^\mu\}\in\Gamma
}L^2((S^1)^n,\mbox{$\prod_{\mu=1}^n\frac{dx^\mu}{\sqrt{2\pi}}$})
\label{L2space}\eeq
with $L^2((S^1)^n,\mbox{$\prod_{\mu=1}^n\frac{dx^\mu}{\sqrt{2\pi}}$})$ the
space of square integrable functions on $T^n$ with its Riemannian volume form.
Periodic functions of $x^\mu=\frac1{\sqrt2}(x_+^\mu+x_-^\mu)$ act in
(\ref{L2space}) by multiplication and $p_\mu$ as the derivative operator
$i\frac\partial{\partial x^\mu}$ in each $L^2$-component labelled by the
winding numbers $w^\mu$. The Hilbert space (\ref{L2space}) is spanned by the
eigenvectors $|p^+;p^-\rangle=\e^{-ip_\mu x^\mu}$ of $i\frac\partial{\partial
x^\mu}$ in each component of the direct sum. The spaces ${\cal F}^\pm$ are two
commuting copies of the bosonic Fock space
\beq
{\cal F}^\pm=\bigoplus_{k>0}<~\alpha_{n_1}^{(\pm)\mu_1}\cdots\alpha_{n_k}
^{(\pm)\mu_k}|0\rangle_\pm~|~n_j<0~~>
\label{bosfock}\eeq
built on the vacuum states $|0\rangle_\pm$ with
$\alpha_k^{(\pm)\mu}|0\rangle_\pm=0$ for $k>0$. The unique vacuum state of
${\cal H}_X$ is
\beq
|\mbox{vac}\rangle\equiv|0;0\rangle\otimes|0\rangle_+\otimes|0\rangle_-
\label{vacstate}\eeq
with
$p_\mu^\pm|\mbox{vac}\rangle=\alpha_k^{(\pm)\mu}|\mbox{vac}\rangle=0~~\forall
k>0$.

\newsection{Spacetime Symmetries and Dirac Operator}

In this section we shall introduce a Dirac operator that will yield the
Riemannian geometry of the eventual string spacetime. The Dirac operator that
we construct is related to the two fundamental, infinite-dimensional continuous
symmetries of the conformal field theory (\ref{sigmacyl}). We shall also
briefly show how this Dirac operator is related to some previous approaches to
describing the geometrical and topological properties of a spacetime using
supersymmetric sigma-models. Superconformal field theories have been
emphasized recently as the correct field theoretical structure for the
description of stringy spacetimes in the framework of noncommutative geometry
\cite{FG}--\cite{Chamseddine},\cite{fgr}.

\subsubsection*{Gauge Symmetry and the Dirac-Ramond Operator}

The first fundamental symmetry of the string theory is the target space
reparametrization symmetry
$X^\mu_\pm(\tau\pm\sigma)\to X^\mu_\pm(\tau\pm\sigma)+\delta
X_\pm^\mu(\tau\pm\sigma)$, where $\delta X_\pm^\mu(\tau\pm\sigma)$ are
arbitrary
periodic functions. Varying the action (\ref{sigmacyl}) shows that this
symmetry is generated on the Hilbert space (\ref{sigmahilbert}) by the
$u(1)^n_+\oplus u(1)^n_-$ Kac-Moody algebra at level 2 with conserved currents
\beq
J_\pm^\mu(\tau\pm\sigma)=\partial_\pm
X_\pm^\mu(\tau\pm\sigma)=\sum_{k=-\infty}^\infty
\alpha_k^{(\pm)\mu}~\e^{ik(\tau\pm\sigma)}
\label{currentexp}\eeq
obeying the commutation relations
(\ref{creannalg}), where we have defined $\alpha_0^{(\pm)\mu}\equiv
g^{\mu\nu}p_\nu^\pm$. Then the Hilbert space (\ref{sigmahilbert}) is a direct
sum of the irreducible highest-weight representations of the current algebra
acting in $|p^+;p^-\rangle\otimes{\cal F}^+\otimes{\cal F}^-$ and labelled by
the $U(1)_\pm^n$ charges $p_\mu^\pm$.

The currents (\ref{currentexp}) can be used to define the Dirac operator which
describes the DeRham cohomology and Riemannian geometry of the effective
spacetime of the sigma-model (\ref{sigmacyl}). For this, 
we endow the toroidal spacetime $T^n$ with a spin structure. Note
that there are $2^n$ possibilities corresponding to a choice of Neveu-Schwarz
or Ramond fermionic boundary conditions around each of the $n$ circles of
$T^n$. We then introduce two anti-commuting copies of the ${\rm
spin}(n)$ Clifford algebra ${\cal C}(T^n)^\pm$ whose corresponding Dirac
generators $\gamma_\mu^\pm=(\gamma_\mu^\pm)^*$ obey the
non-vanishing anti-commutation relations
\beq
\left\{\gamma_\mu^\pm,\gamma_\nu^\pm\right\}=2g_{\mu\nu}
\label{cliffalgs}\eeq
To define the appropriate Dirac operator we must first enlarge the Hilbert
space (\ref{sigmahilbert}) to include the spin structure. Thus we replace
${\cal H}_X$ by
\beq
{\cal H}=\bigoplus_{S[{\cal C}(T^n)]}L^2({\rm spin}(T^n))^\Gamma
\otimes{\cal F}^+\otimes{\cal F}^-
\label{sigmahilbertspin}\eeq
where
\beq
L^2({\rm spin}(T^n))^\Gamma=\bigoplus_{\{w^\mu\}\in\Gamma}S_{\{w^\mu\}}[{\cal
C}(T^n)]\otimes L^2((S^1)^n,\mbox{$\prod_{\mu=1}^n\frac{dx^\mu}{\sqrt{2\pi}}$})
\label{L2spacespin}\eeq
is (a local trivialization of) the space of square integrable spinors, i.e.
$L^2$-sections of the spin bundle ${\rm spin}(T^n)$ of the $n$-torus, with
$S_{\{w^\mu\}}[{\cal C}(T^n)]$ unitary irreducible representations of the
double Clifford algebra ${\cal C}(T^n)={\cal C}(T^n)^+\oplus{\cal C}(T^n)^-$.
These modules have the form
\beq
S_{\{w^\mu\}}[{\cal C}(T^n)]=\left\{\new{\begin{array}{l}S_{\{w^\mu\}}[{\cal
C}(T^n)]^+\otimes S_{\{w^\mu\}}[{\cal C}(T^n)]^-~~~~~~\mbox{for $n$
even}\\S_{\{w^\mu\}}[{\cal C}(T^n)]^+\otimes S_{\{w^\mu\}}[{\cal
C}(T^n)]^-\otimes\complex^2~~~~~~\mbox{for $n$ odd}\end{array}}\right.
\label{spinreps}\eeq
where the bundle of spinors contains both chiralities of fermion fields in the
even-dimensional case. These various representations of the Clifford algebra
need not be the same, but, to simplify notation in the following, we shall
typically omit the explicit representation labels for the spinor parts of the
Hilbert space (\ref{sigmahilbertspin}).

We now define two anti-commuting Dirac operators acting on the Hilbert space
(\ref{sigmahilbertspin}) by
\beq
\Dirac^\pm(\tau\pm\sigma)=\sqrt2~\gamma_\mu^\pm\otimes
J^\mu_\pm(\tau\pm\sigma)=\sum_{k=-\infty}^\infty\Dirac
_k^\pm~\e^{ik(\tau\pm\sigma)}
\label{diracrampm}\eeq
where
\beq
\Dirac_k^\pm=\sqrt2~\gamma_\mu^\pm\otimes\alpha_k^{(\pm)\mu}
\label{dirramk}\eeq
That these Dirac operators are appropriate for the target space geometry
can be seen by
noting that $J^\mu_\pm\sim\frac\delta{\delta X_\pm^\mu}$ when acting on
spacetime functions of $X_\pm^\mu$. Furthermore, as we shall see, their
squares determine the appropriate Laplace-Beltrami operator for the target
space geometry which can also be found from conventional conformal field
theory.

The Dirac operator introduced above is a low-energy limit of Witten's
Dirac-Ramond operator  for the full superstring theory corresponding
to the two-dimensional $N=1$ supersymmetric sigma-model \cite{witten1,witten2}.
For the simple linear sigma-model under consideration here, this field theory is
obtained by adding to the action (\ref{sigmacyl}) a free fermion term, so that
the total action is
\beq
S[X,\psi]=S[X]+\frac i{2\pi}\int d\tau~\oint
d\sigma~g_{\mu\nu}\left(\psi_+^\mu\partial_-\psi_+^\nu+\psi_-^\mu\partial_+
\psi_-^\nu\right)
\label{susysigmacyl}\eeq
where $\psi_\pm^\mu(\tau,\sigma)$ are Majorana spinor fields. Varying
(\ref{susysigmacyl}) we find that the worldsheet supersymmetry generators are
the $N=1$ supercharges
\beq
Q^\pm=\mbox{$\frac1{\sqrt2}$}~g_{\mu\nu}~\psi^\mu_\pm\partial_\pm X^\nu
\label{wssupercharge}\eeq
and canonical quantization of the action (\ref{susysigmacyl}) leads to the
non-vanishing equal-time canonical anti-commutators
\beq
\left\{\psi_\pm^\mu(\tau,\sigma),\psi_\pm^\nu(\tau,\sigma')\right\}=
g^{\mu\nu}~\delta(\sigma,\sigma')
\label{cananti}\eeq
The equations of motion $\partial_-\psi^\mu_+=\partial_+\psi_-^\mu=0$ for the
fermion fields imply that they have the Weyl mode decompositions
\beq
\psi_\pm^{(\epsilon)\mu}(\tau\pm\sigma)=\sum_{k\in\zeds
+\epsilon}\psi_k^{(\pm)\mu}~\e^{i(k+1/2)(\tau\pm\sigma)}
\label{fermmodes}\eeq
so that the supercharges (\ref{wssupercharge}) are also Majorana-Weyl spinor
fields, i.e. $\partial_\mp Q^\pm=0$. Here
$(\psi_k^{(\pm)\mu})^\dagger=\psi_{-k}^{(\pm)\mu}$, and $\epsilon=\frac12$ for
Neveu-Schwarz boundary conditions while $\epsilon=0$ for Ramond boundary
conditions corresponding to the two possible choices of spin structure on the
worldsheet circle $S^1$. The canonical anti-commutator (\ref{cananti}) implies
that the fermionic modes obey the non-vanishing anti-commutation relations
\beq
\left\{\psi_k^{(\pm)\mu},\psi_m^{(\pm)\nu}\right\}=g^{\mu\nu}~\delta_{k+m,0}
\label{psikanticomm}\eeq

In the Ramond sector the fermionic zero modes $\psi_0^{(\pm)\mu}$ generate the
Clifford algebras (\ref{cliffalgs}) and thus coincide with the gamma-matrices
introduced above in specific irreducible representations $S_F[{\cal
C}(T^n)]^\pm$, i.e.
\beq
\gamma_\mu^\pm=\mbox{$\frac1{\sqrt2}$}~g_{\mu\nu}\,\psi_0^{(\pm)\nu}
\label{gammapsidef}\eeq
The Ramond Fock space for the fermion fields is thus built
\beq
{\cal F}^\pm_{\rm R}=\bigoplus_{k>0}<~\psi_{n_1}^{(\pm)\mu_1}\cdots\psi_{n_k}
^{(\pm)\mu_k}|\Psi_\pm\rangle~|~n_j\in\zed^-~,~\Psi_\pm\in S_F[{\cal
C}(T^n)]^\pm~~>
\label{Rfermfock}\eeq
on the Clifford vacua $|\Psi_\pm\rangle$ with
$\psi_k^{(\pm)\mu}|\Psi_\pm\rangle=0$ for $k>0$. The Neveu-Schwarz Fock space
is built
\beq
{\cal F}^\pm_{\rm NS}=\bigoplus_{k>0}<~\psi_{n_1}^{(\pm)\mu_1}\cdots\psi_{n_k}
^{(\pm)\mu_k}|0\rangle_\pm^F~|~n_j\in\zed^-+\mbox{$\frac12$}~~>
\label{NSfermfock}\eeq
on the vacuum states $|0\rangle_\pm^F$ with
$\psi_k^{(\pm)\mu}|0\rangle_\pm^F=0$ for $k>0$. The total fermionic Hilbert
space of the model is the Fock space
\beq
{\cal F}_F={\cal F}_F^+\otimes{\cal F}_F^-~~~~~~{\rm with}~~~{\cal
F}_F^\pm={\cal F}^\pm_{\rm R}\oplus{\cal F}^\pm_{\rm NS}
\label{totfermfock}\eeq
and the total Hilbert space of the supersymmetric sigma-model
(\ref{susysigmacyl}) is
\beq
{\cal H}_{X,\psi}={\cal F}_F\otimes{\cal H}_X
\label{susyhilbert}\eeq
where the bosonic Hilbert space ${\cal H}_X$ is defined in
(\ref{sigmahilbert}).

In the Ramond sector the quantized fermionic zero-modes of the worldsheet
supercharges (\ref{wssupercharge}) coincide with the generalized Dirac
operators (\ref{diracrampm}) acting in the representation sector of the Hilbert
space (\ref{sigmahilbertspin}) determined by the fermion fields. More
precisely, we define a mode expansion
$Q_\epsilon^\pm(\tau\pm\sigma)=\sum_{k\in\zeds+\epsilon}Q_{k,\epsilon}^\pm~
\e^{i(k+1/2)(\tau\pm\sigma)}$ of the worldsheet supercharges with the operators
\beq
Q_{k,\epsilon}^\pm=\sum_{m\in\zeds+\epsilon}g_{\mu\nu}~\psi_m^{(\pm)\mu}\alpha
_{k-m}^{(\pm)\nu}
\label{susymodes}\eeq
acting on the full Hilbert space (\ref{susyhilbert}) of the supersymmetric
sigma-model. The orthogonal projection of the Hilbert space ${\cal H}_{X,\psi}$,
with projector ${\cal P}_{\rm R}^{(0)}$, onto the fermionic zero-modes
is the subspace
\beq
{\cal H}_{\rm R}^{(0)}\equiv{\cal P}_{\rm R}^{(0)}{\cal H}_{X,\psi}={\cal
F}_{\rm R}^{(0)+}\otimes{\cal F}_{\rm R}^{(0)-}\otimes{\cal H}_X
\label{hsusy0}\eeq
where
\beq
{\cal F}_{\rm R}^{(0)\pm}=\left\{|\Psi_\pm\rangle~|~\Psi_\pm\in S_F[{\cal
C}(T^n)]^\pm\right\}\cong S_F[{\cal C}(T^n)]^\pm\
\label{fock0}\eeq
{}From (\ref{gammapsidef}) it follows that under this orthogonal projection
the Hilbert space (\ref{susyhilbert}) reduces to (\ref{sigmahilbertspin}) and
the supersymmetry charges (\ref{wssupercharge}) coincide with the Dirac
operators (\ref{diracrampm}) in the representation $S_F[{\cal C}(T^n)]$ of the
Clifford algebra determined by the fermion fields,
\beq
{\cal H}\bigm|_{S_F[{\cal C}(T^n)]}\cong{\cal H}_{\rm
R}^{(0)}~~~~~~,~~~~~~\Dirac^\pm_k={\cal P}_{\rm R}^{(0)}\,Q_{k,0}^\pm\,{\cal
P}_{\rm R}^{(0)}
\label{DiracQcorr}\eeq
Note that the analogous projection onto the Neveu-Schwarz sector of
(\ref{susyhilbert}),
\beq
{\cal H}_{\rm NS}^{(0)}=|0\rangle_+^F\otimes|0\rangle_-^F\otimes{\cal
H}_X\cong{\cal H}_X
\label{NS0}\eeq
projects out the original bosonic Hilbert space (\ref{sigmahilbert}). This
supersymmetric construction thus exhibits an algebraic, field-theoretical
origin for the Dirac operators introduced above.

\subsubsection*{Conformal Symmetry and the Witten Complex}

The other basic symmetry that the field theory (\ref{sigmacyl}) possesses is
worldsheet conformal invariance under transformations which act by
reparametrization of the light-cone coordinates $\tau\pm\sigma$. At the
quantum level, this symmetry is represented on the spin-extended Hilbert space
(\ref{sigmahilbertspin}) by a commuting pair of Virasoro algebras with the
conserved stress-energy tensors
\beq\new{\begin{array}{c}
T^\pm(\tau\pm\sigma)=-\mbox{$\frac12$}\,:\Dirac^\pm(\tau\pm\sigma)^2:~=-
\mbox{$\frac12$}~\id\otimes g_{\mu\nu}\,:\partial_\pm
X_\pm^\mu(\tau\pm\sigma)\partial_\pm
X_\pm^\nu(\tau\pm\sigma):\\=\sum_{k=-\infty}^
\infty\id\otimes L_k^\pm~\e^{i(k+2)(\tau\pm\sigma)}\end{array}}
\label{stressen}\eeq
where $\id$ is the identity operator and the Sugawara-Virasoro generators
\beq
L_k^\pm=\frac12\sum_{m=-\infty}^\infty
g_{\mu\nu}~:\alpha_m^{(\pm)\mu}\alpha_{k-m}^{(\pm)\nu}:
\label{sugvirgens}\eeq
act on the bosonic Hilbert space (\ref{sigmahilbert}). The Wick normal ordering
is defined by
\beq\new{\begin{array}{lll}
:\alpha_k^{(\pm)\mu}\alpha_m^{(\pm)\nu}:&=&\alpha_k^{(\pm)\mu}
\alpha_m^{(\pm)\nu}~~\mbox{for}~~k<m\\
&=&\alpha_m^{(\pm)\nu}\alpha_k^{(\pm)\mu}~~\mbox{for}~~k>m\end{array}}
\label{wickorder}\eeq
and also by putting the operators $x_\pm^\mu$ to the left of $p_\mu^\pm$. The
operators (\ref{sugvirgens}) generate the Virasoro algebra
\beq
\left[L_k^\pm,L_m^\pm\right]=(k-m)L_{k+m}^\pm+
\mbox{$\frac c{12}$}\left(k^3-k\right)\delta_{k+m,0}
\label{viralg}\eeq
with central charge $c=n$, the dimension of the toroidal spacetime.

Using these Virasoro operators we can construct the global spacetime symmetry
generators of the Poincar\'e algebra. The momentum operator generating
space translations is $P=L_0^+-L_0^-$. The Hamiltonian operator is
$H=L_0^++L_0^--\frac c{12}$, which can be written explicitly as
\beq
H=\frac12g^{\mu\nu}p_\mu^+p_\nu^++\frac12g^{\mu\nu}p_\mu^-p_\nu^-
+\sum_{k>0}g_{\mu\nu}~\alpha_{-k}^{(+)\mu}
\alpha_k^{(+)\nu}+\sum_{k>0}g_{\mu\nu}~
\alpha_{-k}^{(-)\mu}\alpha_k^{(-)\nu}-\frac n{12}
\label{ham}\eeq
The Hamiltonian (\ref{ham}) has unique vacuum eigenstate (\ref{vacstate}) (with
eigenvalue $-\frac n{12}$) at the bottom of its spectrum. It determines the
Laplace-Beltrami operator of the Riemannian geometry of the effective string
spacetime \cite{FG}, and it is the square of the Dirac operator 
(\ref{diracrampm})
in the sense of (\ref{stressen}). This geometrical property can be made
somewhat more precise by turning to
the supersymmetric sigma-model (\ref{susysigmacyl}). The fermion fields also
generate two commuting Virasoro algebras of central charge
$c=n/2$ defined by the fermionic stress-energy tensors
\beq
t^\pm_\epsilon(\tau\pm\sigma)=-\frac12~g_{\mu\nu}\,:\psi^{(\epsilon)\mu}_\pm
(\tau\pm\sigma)\partial_\mp\psi^{(\epsilon)\nu}_\pm(\tau\pm\sigma):~=
\sum_{k\in\zeds+\epsilon}\ell_{k,\epsilon}^\pm~\e^{i(k+2)(\tau\pm\sigma)}
\label{fermstressen}\eeq
where the Virasoro operators are
\beq
\ell_{k,\epsilon}^\pm=-\frac12\sum_{m\in\zeds+\epsilon}m~g_{\mu\nu}\,
:\psi_m^{(\pm)\mu}\psi_{k-m}^{(\pm)\nu}:\,+\frac{3n}{48}~\delta_{\epsilon,0}
\delta_{k,0}
\label{fermvirs}\eeq
The supersymmetric modes (\ref{susymodes}) together with the Virasoro
generators
\beq
{\cal L}_{k,\epsilon}^\pm=L_k^\pm+\ell_{k,\epsilon}^\pm
\label{virtot}\eeq
generate the $N=1$ supersymmetric extension of the $c=3n/2$ Virasoro algebra
(\ref{viralg}),
\beq\new{\begin{array}{rrl}
\left[{\cal L}_{k,\epsilon}^\pm,Q_{m,\epsilon}^\pm\right]&=&\left(\mbox{$\frac
k2$}-m\right)Q_{k+m,\epsilon}^\pm\\
\left\{Q_{k,\epsilon}^\pm,Q_{m,\epsilon}^\pm\right\}&=&2{\cal
L}^\pm_{k+m,\epsilon}+\mbox{$\frac
c3$}\left(k^2-\mbox{$\frac14$}\right)\delta_{k+m,0}\\\left[Q_{k,\epsilon}^\pm,
{\cal L}_{m,\epsilon}^\mp\right]&=&\left\{Q_{k,\epsilon}^\pm,Q_{m,\epsilon}^\mp
\right\}=0\end{array}}
\label{susyvir}\eeq

In particular, in the Ramond sector we have $(Q_{0,0}^\pm)^2={\cal
L}_{0,0}^\pm-\frac c{24}$ and $\{Q_{0,0}^+,Q_{0,0}^-\}=0$ which defines the
global supersymmetry algebra associated with the spacetime Poincar\'e group.
Thus the spacetime Poincar\'e generators of the bosonic sigma-model
(\ref{sigmacyl}) are given by the projections
\beq
\id\otimes(P\pm H)=\mbox{$\frac12$}{\cal P}_{\rm
R}^{(0)}\,(Q^\pm_{0,0})^2\,{\cal P}_{\rm R}^{(0)}
\label{stQpm}\eeq
onto the spin-extended Hilbert space (\ref{sigmahilbertspin}). The operators
$(Q^\pm_{0,0})^2$ also annihilate all states of the form
$|\Psi_\pm\rangle\otimes|{\rm vac}\rangle$, so that
\beq
\ker Q_{0,0}^\pm\cong S_F[{\cal C}(T^n)]^\pm
\label{kernelQ}\eeq
Thus the global supersymmetry is unbroken and there exists a whole family of
supersymmetric ground states of the field theory (\ref{susysigmacyl}). When
restricted to Hilbert space states of spacetime momentum $P=0$, the fields of
\eqn{susysigmacyl} generate
the DeRham complex of the target space $T^n$ \cite{witten1}. The $P=0$
projected supercharges ${\cal P}_{P=0}Q_0{\cal P}_{P=0}$ and ${\cal
P}_{P=0}\bar Q_0{\cal P}_{P=0}$, with
\beq
Q_0=\mbox{$\frac1{\sqrt2}$}\left(Q_{0,0}^++Q_{0,0}^-\right)
{}~~~~~~,~~~~~~\bar Q_0=\mbox{$\frac1{\sqrt2}$}\left(Q_{0,0}^+-Q_{0,0}^-\right)
\label{ddstarQ}\eeq
realize the exterior derivative d and the co-derivative $\star{\rm d}\star$,
respectively, when acting on the Ramond sector of the Hilbert space of the
supersymmetric sigma-model. Moreover, the projected fermionic zero-modes ${\cal
P}_{P=0}\psi^\mu{\cal P}_{P=0}$ and ${\cal P}_{P=0}\bar\psi^\mu{\cal P}_{P=0}$,
with
\beq
\psi^\mu=\psi_{0,0}^{(+)\mu}+\psi_{0,0}^{(-)\mu}~~~~~~,
{}~~~~~~\bar\psi^\mu=\psi_{0,0}^{(+)\mu}-\psi_{0,0}^{(-)\mu}
\label{psiforms}\eeq
correspond, respectively, to basis differential one-forms and basis
vector fields, and Poincar\'e-Hodge duality is realized by (Hermitian) 
conjugation in the sense of mappings $\psi^\pm\to\pm\psi^\pm$ of left and 
right chiral sectors in the light-cone parametrizations above.

\newsection{Vertex Operator Algebra}

We finally introduce an appropriate operator algebra acting on the Hilbert
space (\ref{sigmahilbert}) which will give the necessary topology and
differentiable structure to the string spacetime. We want to use an algebra
that acts on (\ref{sigmahilbert}) densely, i.e. it maps a dense subspace of
${\cal H}_X$ into itself, so as to capture the full structure
of the string spacetime as determined by the Fubini-Veneziano fields
(\ref{FubVen}). For this, we now introduce the basic (single-valued) quantum
fields of the sigma-model. We define the holomorphic coordinates
$z_\pm=\e^{-i(\tau\pm\sigma)}$, which after a Wick rotation of the worldsheet
temporal coordinate maps the cylinder onto the complex plane. We then define
the mutually local holomorphic and anti-holomorphic vertex operators
\beq
V_{q^\pm}(z_\pm)=~:\e^{-iq_\mu^\pm X_\pm^\mu(\tau\pm\sigma)}:
\label{Vops}\eeq
where single-valuedness restricts their momenta to
\beq
q^\pm_\mu=\mbox{$\frac1{\sqrt2}$}\left(q_\mu\pm d_{\mu\nu}^\pm
v^\nu\right)~~~~{\rm with}~~\{q_\mu\}\in\Gamma^*,\{v^\nu\}\in\Gamma
\label{vmomenta}\eeq
so that $(q^+,q^-)\in\Lambda$. The complete, left-right symmetric local vertex
operators of the conformal field theory are
\beq
V_{q^+q^-}(z_+,z_-)=c_{q^+q^-}(p^+,p^-)V_{q^+}(z_+)V_{q^-}(z_-)=
c_{q^+q^-}(p^+,p^-):\e^{-iq^+_\mu X_+^\mu(\tau+\sigma)-iq^-_\mu
X_-^\mu(\tau-\sigma)}:
\label{vertexops}\eeq
The operator-valued phases
\beq
c_{q^+q^-}(p^+,p^-)=(-1)^{[(d^+)^{\lambda\rho}q_\rho^++(d^-)^{\lambda\rho}
q_\rho^-]d_{\lambda\mu}g^{\mu\nu}[p_\nu^+-p_\nu^-]}=(-1)^{q_\mu w^\mu}
\label{cocycles}\eeq
where $d=[(d^+)^{-1}+(d^-)^{-1}]^{-1}$, are 2-cocycles of the lattice algebra
generated by the complexification $\Lambda^c=\Lambda\otimes_\zeds\,\bb C$, and
they are inserted to correct the algebraic transformation properties (both
gauge and conformal) of the vertex operators. They also enable the vertex
operator construction of affine Kac-Moody algebras \cite{go}.

The local vertex operators generate an important algebraic structure which we
now describe in some detail. We consider the $n$-fold Heisenberg-Weyl 
operator algebra spanned
by the oscillator modes,
\beq
\hat
h_\pm=\left\{q_\mu^\pm\alpha_m^{(\pm)\mu}~\Bigm|~(q^+,q^-)\in\Lambda~,~m\in{\bb
Z}\right\}
\label{heisenalg}\eeq
and the algebra of polynomials on the bosonic creation operators which is the
symmetric vector space
\beq
S(\hat
h_\pm^{(-)})=\bigoplus_{k>0}\left\{\prod_{i=1}^kq_\mu^{(i)\pm}\alpha
_{-m_i}^{(\pm)\mu}~\Bigm|~(q^{(i)+},q^{(i)-})\in\Lambda~,~m_i>0\right\}
\label{symmalg}\eeq
Next we consider the group algebra ${\bb
C}[\Lambda]=\complex[\Lambda]^+\times\complex[\Lambda]^-$ of the
complexified lattice $\Lambda^c$, which is the abelian group generated by the
translation operators $\e^{iq_\mu^\pm x_\pm^\mu}$, $(q^+,q^-)\in\Lambda$.
Because of the inclusion of the lattice cocycles in the
definition of the complete vertex operators (\ref{vertexops}), we
``twist" these group algebra generators by defining the operators
\beq
\varepsilon_{q^+q^-}
\equiv\e^{iq_\mu^+x_+^\mu+iq_\mu^-x_-^\mu}~c_{q^+q^-}(p^+,p^-)
\label{twistops}\eeq
Using the Baker-Campbell-Hausdorff formula it is straightforward to show that
\beq
\varepsilon_{q^+q^-}~\varepsilon_{r^+r^-}=c_{q^+q^-}(r^+,r^-)~\varepsilon_
{(q^++r^+)(q^-+r^-)}
\label{epprodsum}\eeq
and hence that the operators (\ref{twistops}) generate the clock algebra
\beq
\varepsilon_{q^+q^-}~\varepsilon_{r^+r^-}=(-1)^{\langle
q,r\rangle_\Lambda}~\varepsilon_{r^+r^-}~\varepsilon_{q^+q^-}
\label{epclockalg}\eeq
with the 2-cocycles $(-1)^{\langle q,r\rangle_\Lambda}$ in the lattice algebra
generated by $\Lambda^c$.

Notice that the 2-cocycles (\ref{cocycles}) are maps
$c:\Lambda^c\oplus\Lambda^c\to{\bb Z}_2$. This means that the twisted operators
(\ref{twistops}) generate the group algebra of the double cover
$\hat\Lambda^c={\bb Z}_2\times\Lambda^c$ of the complexified lattice
$\Lambda^c$. The multiplication in the lattice algebra of $\hat\Lambda^c$ is
$(\rho;q^+,q^-)\cdot(\sigma;r^+,r^-)=(c_{q^+q^-}(r^+,r^-)\rho\sigma
;q^++r^+,q^-+r^-)$ for $\rho,\sigma\in{\bb Z}_2$ and
$(q^+,q^-),(r^+,r^-)\in\Lambda^c$. Thus taking the twisted group algebra
${\bb C}\{\Lambda\}$ generated by the operators $\varepsilon_{q^+q^-}$, the
vertex operators (\ref{vertexops}) are formally endomorphisms of the ${\bb
Z}$-graded Fock space of operators
\beq
\widehat{\cal H}_X(\Lambda)={\bb C}\{\Lambda\}\otimes S(\hat h_+^{(-)})\otimes
S(\hat
h_-^{(-)})
\label{vertexopalg}\eeq
which defines the space of twisted endomorphisms of the Hilbert space
(\ref{sigmahilbert}). The oscillator modes $\alpha_k^{(\pm)\mu}$ for $k<0$ act
in the latter tensor products of (\ref{vertexopalg}) by multiplication, and for
$k>0$ their (adjoint) action is defined by (\ref{creannalg}). 
The zero mode operators $\alpha_0^{(\pm)\mu}$ act on $\complex\{\Lambda\}$ by
\beq
\alpha_0^{(\pm)\mu}~\varepsilon_{q^+q^-}=g^{\mu\nu}q_\nu^\pm~
\varepsilon_{q^+q^-}
\label{0modeaction}\eeq
while the action of $\varepsilon_{q^+q^-}$ on the twisted group algebra is
given by (\ref{epprodsum}).

Explicitly, we can expand the exponentials in (\ref{vertexops}) in terms of the
Schur polynomials $P_m[t_1,\dots,t_m]$ defined by
\beq
\exp\left(\sum_{k>0}\frac{t_k}k~z^k\right)=\sum_{m=0}^\infty
P_m[t_1,\dots,t_m]~z^m
\label{schur}\eeq
to get
\beq\new{\begin{array}{lll}
V_{q^+q^-}(z_+,z_-)&=&\varepsilon_{q^+q^-}\sum_{k,m=0}^\infty
P_k[q_{\mu_1}^+\alpha_{-1}^{(+)\mu_1},\dots,q_{\mu_k}^+\alpha_{-k}^{(+)\mu_k}]
\\& &~~~~~~~~~~~~~~~~\times
P_m[q_{\nu_1}^-\alpha_{-1}^{(-)\nu_1},\dots,q_{\nu_m}^-\alpha_{-m}^{(-)\nu_m}]
{}~z_+^kz_-^m~~\in\widehat{\cal H}_X(\Lambda)[z_+,z_-]\end{array}}
\label{vertexopexp}\eeq
More generally, to a typical homogeneous element
\beq
\Psi=\varepsilon_{q^+q^-}\otimes\prod_jr_\mu^{(j)+}\alpha_{-n_j}^{(+)\mu}
\otimes\prod_kr_\nu^{(k)-}\alpha_{-m_k}^{(-)\nu}
\label{helt}\eeq
of $\widehat{\cal H}_X(\Lambda)$ with
$(q^+,q^-),(r^{(i)+},r^{(i)-})\in\Lambda$,
we associate the higher-spin vertex operator
\beq\new{\begin{array}{lll}
{\cal V}(\Psi;z_+,z_-)&\equiv&{\cal V}_{q^+q^-}^\Omega(z_+,z_-)\\
&=&:~i\,V_{q^+q^-}(z_+,z_-)~\prod_j\frac{r_\mu^{(+)j}}{(n_j-1)!}~\partial_{z_+}
^{n_j}X_+^\mu~\prod_k\frac{r_\nu^{(-)k}}{(m_k-1)!}~
\partial_{z_-}^{m_k}X_-^\nu~:\end{array}}
\label{spinvertex}\eeq
where $\Omega=\{(r^{(i)+},r^{(i)-});n_j,m_k\}$ labels the fields and
$\partial_{z_\pm}
X_\pm^\mu(z_\pm)=\sum_{k\in\zeds}i\alpha_k^{(\pm)\mu}z_\pm^{-k-1}$. Extending
this by linearity we obtain a well-defined one-to-one mapping from
$\widehat{\cal H}_X(\Lambda)$ into the algebra of endomorphism-valued Laurent
power series in the variables $z_\pm$,
\beq
\widehat{\cal H}_X(\Lambda)\to({\rm End}~\widehat{\cal
H}_X(\Lambda))[z_+^{\pm1},z_-^{\pm1}]
\label{1-1map}\eeq
which defines the space of quantum conformal fields of the sigma-model. For
example, the stress-energy tensors (\ref{stressen}) are given by
\beq
T^\pm(z_\pm)=\id\otimes{\cal V}(\omega^\pm;z_\pm,1)~~~~~{\rm
with}~~\omega^\pm=\mbox{$\frac12$}\,\id\otimes g_{\mu\nu}(\vec
e^{\,\mu})_\lambda(\vec
e^{\,\nu})_\rho~\alpha_{-1}^{(\pm)\lambda}\alpha_{-1}^{(\pm)\rho}
\label{stressenvec}\eeq
where $\{\vec e^{\,\mu}\}$ is an arbitrary basis of the lattice $\Gamma$.

The fundamental (tachyonic) vertex operators (\ref{vertexops}) have some
noteworthy algebraic properties which define the algebra of the primary
fields. Since the sigma-model under consideration is essentially a free field
theory, these features follow from the normal-ordering property
\beq\new{\begin{array}{l}
V_{q^+q^-}(z_+,z_-)V_{r^+r^-}(w_+,w_-)\\
=~:V_{q^+q^-}(z_+,z_-)V_{r^+r^-}(w_+,w_-):~\exp\left\{\langle{\rm
vac}|q_\mu^+X^\mu_+(z_+)+r_\mu^+X_+^\mu(w_+)|{\rm
vac}\rangle\right\}\\~~~~~~~~~~~~\times\exp\left\{\langle{\rm
vac}|q_\mu^-X_-^\mu(z_-)+r_\mu^-X_-^\mu(w_-)|{\rm
vac}\rangle\right\}\end{array}}
\label{vertexopid1}\eeq
Evaluating the vacuum expectation values explicitly yields the operator product
algebra
\beq\new{\begin{array}{lll}
V_{q^+q^-}(z_+,z_-)V_{r^+r^-}(w_+,w_-)&=&(z_+-w_+)^{g^{\mu\nu}q_\mu^+r_\nu^+}
(z_--w_-)^{g^{\mu\nu}q_\mu^-r_\nu^-}\\& &~~~~\times~
:V_{q^+q^-}(z_+,z_-)V_{r^+r^-}(w_+,w_-):\end{array}}
\label{vertexopalg1}\eeq
which represents the product of two vertex operators in terms of another one
(in general with singular local coefficient). Interchanging the order of the
two vertex operators in (\ref{vertexopalg1}) and using (\ref{epclockalg}) we
then find the local cocycle relation
\beq\new{\begin{array}{l}
V_{q^+q^-}(z_+,z_-)V_{r^+r^-}(w_+,w_-)\\~~~~~=
V_{r^+r^-}(w_+,w_-)V_{q^+q^-}(z_+,z_-)
{}~\exp\left\{-i\pi g^{\mu\nu}q_\mu^+r_\nu^+~{\rm sgn}({\rm arg}~z_+-{\rm
arg}~w_+)\right\}\\~~~~~~~~~~~~~~~~\times\exp\left\{-i\pi
g^{\mu\nu}q_\mu^-r_\nu^-~{\rm sgn}({\rm arg}~z_--{\rm
arg}~w_-)\right\}\end{array}}
\label{vertexopalg2}\eeq
where we have chosen the branch of $\log(-1)$ on the worldsheet for which the
imaginary part of the logarithm lies in the interval $[-i\pi,+i\pi]$.

The algebraic relations (\ref{vertexopalg1}) and (\ref{vertexopalg2})
determine, by differentiation of the appropriate vertex operators,
corresponding relations
among the general higher-spin vertex operators (\ref{spinvertex}). For
instance, from the identities
\beq\new{\begin{array}{rrl}
\partial_{z_\pm}V_{q^+q^-}(z_+,z_-)&=&\sum_{k=-\infty}^\infty:~q_\mu^\pm
\alpha_k^{(\pm)\mu}z_\pm^{-k-1}~V_{q^+q^-}(z_+,z_-)~:\\\partial_{z_\pm}{\cal
V}(\alpha_{-m}^{(\pm)\mu};z_+,z_-)&=&{\cal
V}(m\alpha_{-m-1}^{(\pm)\mu};z_+,z_-)
\end{array}}
\label{derivids}\eeq
we have
\beq
\partial_{z_\pm}{\cal V}(\Psi;z_+,z_-)={\cal
V}(L_{-1}^\pm\Psi;z_+,z_-)~~~~~~\forall\Psi\in\widehat{\cal H}_X(\Lambda)
\label{translprop}\eeq
In fact, the coefficients of the monomials in the variables $z_\pm$, that arise
in an expansion of products of the basis vertex operators
\beq
{\cal V}_{[q^+q^-]}^{(R)}[z_+,z_-]=~:\prod_{i=1}^RV_{q^{(i)+}q^{(i)-}}
(z_+^{(i)},z_-^{(i)})~:
\label{vertexopprods}\eeq
in terms of Schur polynomials, span the linear space (\ref{vertexopalg}) as $R$
and the momenta $(q^{(i)+},q^{(i)-})\in\Lambda$ are varied. The complete set of
relations of the conformal field algebra is therefore determined by the
operator product expansion formula for the operators (\ref{vertexopprods}),
which is the local cocycle relation
\beq\new{\begin{array}{l}
{\cal V}_{[q^+q^-]}^{(R)}[z_+,z_-]{\cal V}_{[r^+r^-]}^{(S)}[w_+,w_-]={\cal
V}_{[r^+r^-]}^{(S)}[w_+,w_-]{\cal
V}_{[q^+q^-]}^{(R)}[z_+,z_-]\\~~~~~~\times\prod_{1\leq(i,j)\leq(R,S)}
\exp\left\{-i\pi g^{\mu\nu}q_\mu^{(i)+}r_\nu^{(j)+}~{\rm sgn}({\rm
arg}~z_+^{(i)}-{\rm arg}~w_+^{(j)})\right\}\\~~~~~~~~~~~~~~~\times\exp\left\{
-i\pi g^{\mu\nu}q_\mu^{(i)-}r_\nu^{(j)-}~{\rm sgn}({\rm arg}~z_-^{(i)}-{\rm
arg}~w_-^{(j)})\right\}\end{array}}
\label{opprodsexp}\eeq

The fundamental vertex operators (\ref{vertexops}) are primary
fields of both the Kac-Moody and Virasoro algebras of the sigma-model.
Specifically, $V_{q^+q^-}(z_+,z_-)$ transform as local fields of $U(1)^n_\pm$
charges $q_\mu^\pm$ under the local gauge transformations generated by the
Kac-Moody currents,
\beq
\left[\alpha_k^{(\pm)\mu},V_{q^+q^-}(z_+,z_-)\right]=
g^{\mu\nu}q_\nu^\pm~z_\pm^k~V_{q^+q^-}(z_+,z_-)
\label{u1transfv}\eeq
Moreover,
\beq
\left[L_k^\pm,V_{q^+q^-}(z_+,z_-)\right]=\left(z_\pm^{k+1}\partial_{z_\pm}
+(k+1)\Delta_{q^\pm}z_\pm^k\right)V_{q^+q^-}(z_+,z_-)
\label{conftransfv}\eeq
so that the local vertex operators (\ref{vertexops}) also transform under
conformal transformations as tensors of weight
\beq
\Delta_{q^\pm}=\mbox{$\frac12$}g^{\mu\nu}q_\mu^\pm q_\nu^\pm
\label{primweights}\eeq
However, (\ref{u1transfv}) holds only at $k=0$ and (\ref{conftransfv}) only for
$k=0$ and $k=-1$ in general for the higher-spin vertex operators ${\cal
V}_{q^+q^-}^\Omega(z_+,z_-)$. Thus the general operators (\ref{spinvertex})
also have $U(1)_\pm^n$ charges $q_\mu^\pm$ and scaling dimensions
\beq
\Delta_{q^+}^\Omega=\mbox{$\frac12$}g^{\mu\nu}q_\mu^+q_\nu^++\mbox{$\sum_jn_j$}
{}~~~~~~,~~~~~~\Delta_{q^-}^\Omega=\mbox{$\frac12$}g^{\mu\nu}q_\mu^-q_\nu^-
+\mbox{$\sum_km_k$}
\label{spinweights}\eeq
The general transformation property (\ref{conftransfv}) holds for the vertex
operators (\ref{spinvertex}) which are primary fields of the sigma-model (of
weights $\Delta^\Omega_{q^\pm}$), i.e. the endomorphisms acting on the
subspaces
\beq
\widehat{\cal P}_{\Delta^\Omega_q}(\Lambda)\equiv\left\{\Psi\in\widehat{\cal
H}_X(\Lambda)~|~L_0^\pm\Psi=\Delta_{q^\pm}^\Omega\Psi~~,~~L_k^\pm\Psi
=0~~\forall
k>0\right\}
\label{primaryops}\eeq
of highest-weight operators in $\widehat{\cal H}_X(\Lambda)$. The space
(\ref{vertexopalg}) is then a direct sum of the irreducible highest-weight
representations of the Virasoro algebra acting in the subspaces
(\ref{primaryops}) and labelled by the conformal weights
$\Delta_{q^\pm}^\Omega$. Similarly, one can decompose (\ref{vertexopalg}) into
irreducible highest-weight representations of the current algebra acting in
subspaces of definite $U(1)_\pm^n$ charges $q_\mu^\pm$.

In fact, in addition to defining the grading of the space $\widehat{\cal
H}_X(\Lambda)$, the conformal dimension and spacetime momentum eigenvalues
grade the Hilbert space (\ref{sigmahilbert}). This follows from the
operator-state correspondence which relates the states
\beq
|\varphi_{q^+q^-}^\Omega\rangle=|q^+;q^-\rangle\otimes\prod_j
r_\mu^{(j)+}\alpha_{-n_j}^{(+)\mu}
|0\rangle_+\otimes\prod_kr_\nu^{(k)-}\alpha_{-m_k}^{(-)\nu}|0\rangle_-\in{\cal
H}_X
\label{phistate}\eeq
to the higher-spin vertex operators (\ref{spinvertex}), where $\Omega$ labels
the quantum numbers of the states. For example, the spin-0 tachyon state
corresponds to the basis vertex operators themselves. In this case, the basis
primary fields correspond to the Hilbert space states
\beq
|q^+;q^-\rangle\otimes|0\rangle_+\otimes|0\rangle_-=\lim_{z_\pm\to0}
V_{q^+q^-}(z_+,z_-)|{\rm vac}\rangle\in
L^2((S^1)^n,\mbox{$\prod_{\mu=1}^n\frac{dx^\mu}{\sqrt{2\pi}}$})
\label{primarystates}\eeq
The (primary) tachyon vectors (\ref{primarystates}) are eigenstates of
$\alpha_0^{(\pm)\mu}$ with $U(1)_\pm^n$ charge eigenvalues $q_\mu^\pm$ and of
$L_0^\pm$ with conformal weight eigenvalues $\Delta_{q^\pm}$.

Likewise, the general states (\ref{phistate}) have spacetime momentum
eigenvalues $q_\mu^\pm$ and conformal dimension eigenvalues
(\ref{spinweights}).  An important state is the graviton state which
corresponds to the minimal spin-2 operator
\beq
{\cal
V}_{q^+q^-}^{\mu\nu}(z_+,z_-)=~:~i\,V_{q^+q^-}(z_+,z_-)~\partial_{z_+}X_+^\mu~
\partial_{z_-}X_-^\nu~:
\label{gravop}\eeq
which represents the basic conformal structure embedded into the content of
the algebra under construction here (see \eqn{stressenvec}). 
The operator (\ref{gravop}) creates a graviton of polarization $(\mu\nu)$ and
represents the Fourier modes of the background matrices
$d_{\mu\nu}^\pm$, i.e. of
the metric $g_{\mu\nu}$ and instanton form $\beta_{\mu\nu}$. It corresponds to
the string state
$|q^+;q^-\rangle\otimes\alpha^{(+)\mu}_{-1}|0\rangle_+\otimes
\alpha^{(-)\nu}_{-1}|0\rangle_-\in{\cal H}_X$, which is the lowest-lying vector
with non-trivial string oscillations and will play an important 
role in section 7
(see also the appendix). As another
example, the massless dilaton state corresponds to the minimal spin-0 operator
$:iV_{q^+q^-}(z_+,z_-)\,\frac12g_{\mu\nu}\partial_{z_+}X_+^\mu\partial_{z_-}
X_-^\nu:$ which is orthogonal to the tachyon field. In the general case, the
operator-state correspondence is achieved by the action of the smeared vertex
operators
\beq
V_\Omega(q^+,q^-)=\int\frac{dz_+~dz_-}{4\pi z_+z_-}~{\cal
V}^\Omega_{q^+q^-}(z_+,z_-)f_S(z_+,z_-)\in\widehat{\cal H}_X(\Lambda)
\label{smearedops}\eeq
on the vacuum state of the Hilbert space, so that
\beq
|\varphi^\Omega_{q^+q^-}\rangle=V_\Omega(q^+,q^-)|{\rm vac}\rangle
\label{opstates}\eeq
Here $f_S$ is an appropriate Schwartz space test function which smears out the
operator-valued distributions ${\cal V}^\Omega_{q^+q^-}(z_+,z_-)$
\footnote{For ease of notation in the following, we suppress the explicit
dependence on the test function $f_S$ in \eqn{smearedops}.}. The operator
(\ref{smearedops}) is said to create a string state of type
$\Omega$ and momentum $(q^+,q^-)\in\Lambda$. The smeared vertex operators
represent spacetime Fourier transformations of the string state operators
inserted on the worldsheet. They span
the space of primary fields of the conformal sigma model and generate the set
of all functionals on the vector space (\ref{vertexopalg}), i.e. they span the
twisted dual of the Hilbert space (\ref{sigmahilbert}). The quantities
(\ref{smearedops}) are naturally (non-local) elements of the operator algebra
(\ref{vertexopalg}) and are well-defined operators on (\ref{sigmahilbert}). In
fact, they map a dense subspace of ${\cal H}_X$ into itself, as
follows from (\ref{opstates}) and the completeness of the primary states
(\ref{phistate}) in ${\cal H}_X$. The operator-state correspondence is thus
the Fock space mapping between the Hilbert space
(\ref{sigmahilbert}) and its twisted endomorphisms (\ref{vertexopalg})
\beq
|\varphi^\Omega_{q^+q^-}\rangle\leftrightarrow V_\Omega(q^+,q^-):{\cal
H}_X\leftrightarrow\widehat{\cal H}_X(\Lambda)
\label{opstateiso}\eeq
Note that this mapping is not one-to-one as there are many smeared vertex
operators which can be associated to a given conformal state of the Hilbert
space (\ref{sigmahilbert}).

The algebraic make-up of the vertex operators described above has the
formal mathematical structure of a vertex operator algebra \cite{flm} (see
also \cite{huang,gebert} for concise introductions). A vertex operator algebra
is the formal mathematical definition of a chiral algebra in conformal field
theory \cite{ms1}, which is defined to be the operator product algebra of
(primary) holomorphic fields in the conformal field theory (see
(\ref{vertexopalg1}) in the present case). In this context the Klein
transformations (\ref{cocycles}), which in the above were introduced to correct
signs in the Kac-Moody and Virasoro commutation relations with the vertex
operators, are needed to adjust some signs in the so-called `Jacobi identities'
of the vertex operator algebra. In conformal field theory, these
Jacobi identities are the precise statement of the Ward identities associated
with the conformal invariance on the 3-punctured Riemann sphere. The general
properties of vertex operator algebras as they pertain to this
paper are briefly discussed in the appendix.

The smeared vertex operators $V_\Omega(q^+,q^-)$ generate a noncommutative
unital $*$-algebra ${\cal A}_X$ which contains two Virasoro and two
Kac-Moody subalgebras. The
operator-state correspondence (\ref{opstateiso}) then implies that
\beq
\alg_X{\cal H}_X={\cal H}_X
\label{Adensemap}\eeq
The noncommutativity of $\alg_X$ is expressed in terms of the cocycle relation
(\ref{opprodsexp}) which determines the complete set of relations between the
smeared vertex operators. As a vertex operator algebra, the identity
(\ref{opprodsexp}) in fact leads immediately to the Jacobi identity, which in
turn encodes many other non-trivial algebraic relations among the elements of
$\alg_X$, including the noncommutativity, associativity, locality of matrix
elements, and the duality (or crossing symmetry) of 4-point (and $m$-point)
correlation functions on the Riemann sphere. The complicated nature of the
3-term Jacobi relation among the elements of the vertex operator algebra
$\alg_X$ is what distinguishes the string spacetime from not only a classical
(commutative) spacetime, but also from the conventional examples of
noncommutative spaces. The full set of relations of the vertex operator algebra
$\alg_X$ are presented in a more general context in the appendix, where it is
also discussed how the algebraic structure of $\alg_X$ leads to a construction
of the corresponding string spacetime along the lines described in section 2. In
particular, they illustrate the complicated nature of the string spacetime
determined by the algebra $\alg_X$, and how more general string spacetimes
constructed from other conformal field theories arise from the general
structure of vertex operator algebras.

Note that the chiral and anti-chiral vertex operators (\ref{Vops}) generate
local chiral algebras ${\cal E}^\pm$ whose products combine into the full
algebra ${\cal E}={\cal E}^+\otimes{\cal E}^-$ corresponding to the usual
decomposition of the operator product algebra of primary holomorphic fields in
conformal field theory. As mentioned before, the twisting of this
chirally-symmetric algebraic structure by the cocyle factors \eqn{cocycles} is
required to compensate signs in the Jacobi identity for $\alg_X$. But we shall
see that the non-trivial duality structure of the effective string spacetime is
represented essentially as a left-right chirality symmetry between the Dirac
operators introduced in section 4. The twistings then yield more complicated
duality transformations, and, as we will discuss in section 7, at certain
special points in the quantum moduli space of the sigma-model the chiral
symmetry is restored and the algebra $\cal E$ controls the structure of the
quantum spacetime.

\newsection{Quantum Geometry of Toroidal Spacetimes}

We shall now begin applying the formalism developed thus far to a
systematic analysis of the geometrical properties of the string spacetime.
With the above constructions, we obtain the spectral triple
\beq
{\cal T}\equiv({\cal A}~,~{\cal H}~,~D)
\label{sigmatriple}\eeq
associated to the sigma-model with target space the $n$-torus $(S^1)^n$ with
metric $g_{\mu\nu}$ and torsion form $\beta_{\mu\nu}$. The triple
(\ref{sigmatriple}) encodes the effective target space geometry of the linear
sigma-model, i.e. the moduli space of this class of conformal field theories.
Here $D$ is an appropriately defined Dirac operator on the spin-extended
Hilbert space ${\cal H}$ which encodes the effective geometry and topology of
the string spacetime described by (\ref{sigmatriple}). It will be constructed
below from the two Dirac operators presented in section 4. The algebra
$\alg\equiv\id\otimes\alg_X$ acts trivially on the spinor part of $\cal H$
\footnote{An algebraic, field-theoretical construction of (\ref{sigmatriple})
can be naturally given as a low-energy limit of the canonical spectral triple
associated with the $N=1$ superconformal field theory \eqn{susysigmacyl}. There
we take the Hilbert space (\ref{susyhilbert}), the algebra generated by the
superconformal primary fields corresponding to states of ${\cal H}_{X,\psi}$,
and a Dirac operator $Q$ constructed from the worldsheet supersymmetry
generators (\ref{wssupercharge}). The resulting unital $*$-algebra is
described as a vertex operator superalgebra (see the appendix). The projection
onto fermionic zero-modes gives ${\cal H}_{\rm R}^{(0)}$, the Dirac operator
${\cal P}_{\rm R}^{(0)}Q{\cal P}_{\rm R}^{(0)}$, and the algebra $\alg{\cal
P}_{\rm R}^{(0)}$ where $\alg=\id\otimes\alg_X$ is generated by the primary
conformal fields corresponding to states of ${\cal H}_{\rm NS}^{(0)}$. This is
the construction that has been elucidated on in
\cite{FG,ChamFro,fgrleshouches,fgr}.}. As
we have pointed out in section 2, different selections of $D$ lead to different
geometries for the string spacetime. Below we shall examine the properties of
the string spacetime with appropriate selections of this operator.

The topology and differentiable structure of the spacetime is encoded via the
complicated, noncommutative vertex operator algebra ${\cal A}_X$ that we
described in the previous section. The effective spacetime of the strings is in
this sense horribly complicated. 
The smearing of the vertex operators effectively
achieves the non-locality property of noncommuting spacetime coordinate fields.
We can get some insight into the various symmetries of the
stringy geometries by viewing the emergence of ordinary spacetime (i.e. one
with a commutative geometry) as a low-energy limit of the quantum spacetime of
the strings. The low-energy limit of the sigma-model is the limit wherein the
oscillatory modes $\alpha_k^{(\pm)\mu}$ of the string vanish, leaving only the
particle-like, center of mass degrees of freedom $x_\pm^\mu,p_\mu^\pm$. Thus in
the low-energy limit of the string theory, the spectral triple
(\ref{sigmatriple}) should contain a subspace
\beq
{\cal T}_0=\left(C^\infty(T^n)~,~L^2({\rm
spin}(T^n))~,~ig^{\mu\nu}\gamma_\mu\partial_\nu\right)
\label{spectral0}\eeq
which represents the ordinary
spacetime manifold $T^n$ at large distance scales.
We shall see that these commutative spacetimes are indeed small subspaces of
the
larger string spacetime given by the spectral triple $\cal T$. The various
string theoretic symmetries that determine the quantum 
moduli space of the toroidal
sigma-models will then appear naturally via the possibility of introducing more
than one Dirac operator $D$ corresponding to isometries of $\cal T$.

The moduli space of linear sigma-models is determined by the symmetries of the
lattice $\Gamma$ and its dual $\Gamma^*$ which define the toroidal spacetime.
The symmetry group of the lattice $\Lambda$ with inner product (\ref{quadform})
is the non-compact orthogonal group $O(n,n)$ in $(n\,+\,n)$-dimensions.
However,
the Hamiltonian (\ref{ham}) and hence the spectrum of the toroidal sigma-model
change under an $O(n,n)$ rotation. Since the chiral momenta
$p_\mu^\pm$ transform as vectors under
$O(n,n)$, the quantum sigma-model is invariant under rotations by the maximal
compact subgroup $O(n)\times O(n)\subset O(n,n)$ representing the rotations of
the lattice $\Gamma$ defining the torus $T^n$ itself (i.e. the set of allowed
winding modes) and the rotations of the dual lattice $\Gamma^*$ (i.e. the set
of allowed momentum modes). Thus the ``classical" moduli space is the right
coset manifold
\beq
{\cal M}_{\rm cl}=O(n,n)/(O(n)\times O(n))
\label{classmod}\eeq
The homogeneous space (\ref{classmod}) is isomorphic to the Grassmannian ${\rm
Gr}(n,n)$ of maximal positive subspaces of $\real^{n,n}$ which is parametrized
by the constant $n\times n$ matrix $d_{\mu\nu}^+$ (equivalently
$d_{\mu\nu}^-$). In the following we shall see how the appropriate quantum
string modification of the moduli space 
(\ref{classmod}) appears naturally in the
framework of the noncommutative geometry of the sigma-model. We will see that
the discrete group of automorphisms of the quantum sigma-model, the so-called
``duality group", can be readily identified by viewing the string spacetime as
the spectral triple (\ref{sigmatriple}). By construction, this point of view
automatically establishes the equivalence between the original quantum field
theory and its dual (i.e. that the correlation functions of the models
are also equivalent to
each other), this being encoded in the quantum field algebra $\alg$.

The spacetime duality maps are, by definition, those which lead to isomorphisms
between inequivalent low-energy spectral triples (\ref{spectral0}). As we will
show, they emerge from the possibility of defining the two independent
(smeared) Dirac operators
\beq\new{\begin{array}{l}
\Dirac=\int\frac{dz_+~dz_-}{4\pi z_+z_-}~\mbox{$\frac1{\sqrt2}$}
\left(\Dirac^+(z_+)+\Dirac^-(z_-)\right)f_S(z_+,z_-)\\
\bar\Dirac=\int\frac{dz_+~dz_-}{4\pi z_+z_-}~\mbox{$\frac1{\sqrt2}$}
\left(\Dirac^+(z_+)-\Dirac^-(z_-)\right)f_S(z_+,z_-)\end{array}}
\label{DDbar}\eeq
in the spectral triple (\ref{sigmatriple}). The main point is that there exists
several unitary transformations $T:{\cal H}\to{\cal H}$ with
\beq
\bar\Dirac=T\,\Dirac\,T^{-1}
\label{TDmap}\eeq
which define automorphisms of the vertex operator algebra, i.e. $T\alg
T^{-1}=\alg$. This then immediately leads to the isomorphism of spectral
triples
\beq
{\cal T}_\Dirac\equiv(\alg~,~{\cal H}~,~\Dirac)\cong(\alg~,~{\cal
H}~,~\bar\Dirac)\equiv{\cal T}_{\bar\Dirac}
\label{Tdualiso}\eeq
The isomorphism (\ref{Tdualiso}) is a special case of the spectral action
principle \cite{ChamsConnes} which describes Riemannian manifolds which are
isospectral (i.e. their Dirac K-cycles have the same spectrum) but not
necessarily isometric. It states that the noncommutative string spacetime
determined by the spectral triple $\cal T$ with $D=\Dirac$ is identical to that
defined with $D=\bar\Dirac$. As such a change of Dirac operator in
noncommutative geometry simply represents a change of metric structure on the
spacetime, the isomorphism (\ref{Tdualiso}) is simply the statement of general
covariance of the noncommutative string spacetime represented as an isometry of
$\cal T$.

{}From this point of view, target space duality can be represented symbolically
by the commutative diagram
\beq\new{\begin{array}{rrl}
{\cal T}_\Dirac&{\buildrel T\over\longrightarrow}&{\cal
T}_{\bar\Dirac}\cong{\cal T}_\Dirac\\{\scriptstyle{\cal P}_0}\downarrow&
&\downarrow{\scriptstyle\bar{\cal P}_0}\\{\cal T}_0&{\buildrel
T_0\over\longrightarrow}&\bar{\cal T}_0\end{array}}
\label{dualitydiag}\eeq
The operators ${\cal P}_0$ and $\bar{\cal P}_0$ project the full spectral
triples onto their respective low-energy subspaces ${\cal T}_0$ and $\bar{\cal
T}_0$ (these projections will be defined formally below). The triple ${\cal
T}_0$ is the commutative spacetime \eqn{spectral0} while $\bar{\cal T}_0$
represents a duality transformed commutative spacetime. From the point of view
of classical general relativity, these two spacetimes are inequivalent.
However, the isomorphism $T$ in the top line of (\ref{dualitydiag}) makes the
diagram commutative, so that $T_0{\cal P}_0=\bar{\cal P}_0T$ and $T_0$ is an
isomorphism of subspaces of the noncommutative spacetime. Thus as subspaces of
the full quantum spacetime (\ref{sigmatriple}), the commutative spacetimes
${\cal T}_0$ and $\bar{\cal T}_0$ are equivalent. This is the essence of
duality and the stringy modification of classical general relativity.

\subsubsection*{T-duality, Low-energy Projections and Spin Structures}

The first symmetry of the string spacetime that we explore is T-duality which
relates large and small radius tori on equal footing and is responsible for the
fundamental length scale in string theory. In terms of the isomorphism
(\ref{Tdualiso}), it is the unitary mapping $T\equiv T_S\otimes T_X:{\cal
H}\to{\cal H}$ in (\ref{TDmap}) which is defined as follows\footnote{In the
next section we shall present explicit operator expressions for the duality
transformations $T$.}. $T_X$ acts trivially on the spinor part of $\cal H$ and
on the bosonic Hilbert space ${\cal H}_X$ itself it is defined by
\beq\new{\begin{array}{rrl}
T_X\,|p^+;p^-\rangle\otimes|0\rangle_+\otimes|0\rangle_-&=&c_{p^+p^-}(p^+,p^-)
|(d^+)^{-1}p^+;-(d^-)^{-1}p^-\rangle\otimes|0\rangle_+\otimes|0\rangle_-\\
T_X\,\alpha_k^{(\pm)\mu}\,T_X^{-1}&=&\pm
g_{\nu\lambda}(d^\mp)^{\mu\nu}\alpha_k^{(\pm)\lambda}\end{array}}
\label{Topbos}\eeq
$T_S$ acts trivially on ${\cal H}_X$ and on the generators of the spin bundle
as
\beq
T_S\,\gamma_\mu^\pm\,
T_S^{-1}=g^{\nu\lambda}d_{\mu\nu}^\mp\gamma_\lambda^\pm
\label{TSgamma}\eeq
Thus the operator $T$ simply redefines the bosonic basis of the Hilbert
space, and it also changes the choice of spin structure on the target space by
defining a different representation of the double Clifford algebra
(\ref{cliffalgs}). This latter property implies that the target space metric
$g_{\mu\nu}$ changes to its dual
\beq
\tilde g^{\mu\nu}=(d^+)^{\mu\lambda}g_{\lambda\rho}(d^-)^{\rho\nu}
\label{dualmetric}\eeq
which defines an inner product on the dual lattice $\Gamma^*$. Note that in the
case where the spacetime has torsion ($\beta\neq0$), the duality (and other
properties) are determined by the mutually inverse background matrices
$(d^\pm)^{-1}$, rather than just $g^{\mu\nu}$ itself which controls the dual
torsion-free theory. The quadratic form (\ref{quadform}) of the lattice
$\Lambda$ is invariant under these transformations.

The Hilbert space (\ref{sigmahilbertspin}) is thus invariant under this mapping
between the two Dirac operators (\ref{DDbar}), as is the Hamiltonian
(\ref{ham}) (and hence the spectrum of the theory). Furthermore, the action of
$T$ on the smeared vertex operator algebra is given by
\beq
T_X\,V_\Omega(q^+,q^-)\,T_X^{-1}=c_{q^+q^-}(p^+,p^-)
V_\Omega(q^+(d^+)^{-1},-q^-(d^-)^{-1})
\label{TonVOm}\eeq
which amounts to simply producing the same type of vertex operator with a
redefined twisted momentum in $\hat\Lambda^c$. The algebra ${\cal A}=T{\cal
A}T^{-1}$ is thus invariant under the unitary mapping $T$. The
transformation $T$ defined by (\ref{Topbos})--(\ref{TonVOm}) which yields an
explicit realization of the isomorphism (\ref{Tdualiso}) is the noncommutative
geometry version of the celebrated `T-duality' transformation of string theory
which exchanges the torus $T^n$ with its dual $(T^n)^*$, and at the same time
interchanges momenta and winding numbers in the spectrum of the compactified
string theory. It corresponds to an inversion of the background
matrices $d^\pm\to(d^\pm)^{-1}$ and is the $n$-dimensional analog of the
$R\to1/R$ circle duality \cite{Tdualrev}. 
{}From the point of view of the noncommutative geometry
formalism, T-duality thus appears quite naturally as a very simple geometric
invariance property of the noncommutative spacetime. As we will see, the 
choice and independence of spin structure in the string theory is also
intimately related to the T-duality symmetry.

We now describe the low-energy projections in detail and explore the
consequences of the above duality map in this sector of the string spacetime.
For this, we consider the subspace
\beq
\bar{\cal H}_0\equiv{\rm
ker}~\Dirac\cong\left\{|\psi;p^+,p^-\rangle\in\bigoplus_{S[{\cal
C}(T^n)]}L^2({\rm
spin}(T^n))^\Gamma~\biggm|~\Dirac_0|\psi;p^+,p^-\rangle=0\right\}
\subset{\cal H}
\label{smallH}\eeq
where
\beq
\Dirac_0=\mbox{$\frac1{\sqrt2}$}~g^{\mu\nu}\left[\left(\gamma_\mu^++
\gamma_\mu^-\right)\otimes p_\nu+\left(d_{\nu\lambda}^+\gamma_\mu^+-
d_{\nu\lambda}^-\gamma^-_\mu\right)\otimes w^\lambda\right]
\label{Dirac0}\eeq
The equivalence in (\ref{smallH}) shows that all oscillator modes are
suppressed, leaving only the center of mass zero modes of the strings. The
subspace $\bar{\cal H}_0$ can be decomposed into a direct sum of $2^n$ smaller
subspaces via
\beq
\bar{\cal H}_0=\bigotimes_{\mu=1}^n\left(\bar{\cal H}_0^{(+)\mu}\oplus\bar{\cal
H}_0^{(-)\mu}\right)
\label{spinspaces}\eeq
where
\beq\new{\begin{array}{lll}
\bar{\cal H}_0^{(+)\mu}&=&\left\{|\psi\rangle\otimes|p^+;p^-\rangle\in\bar{\cal
H}_0~|~g^{\nu\lambda}d_{\lambda\mu}^+\gamma_\nu^+|\psi\rangle=
g^{\nu\lambda}d_{\lambda\mu}^-\gamma_\nu^-|\psi\rangle~~,~~p_\mu=0\right\}
\\\bar{\cal
H}_0^{(-)\mu}&=&\left\{|\psi\rangle\otimes|p^+;p^-\rangle\in\bar{\cal
H}_0~|~\gamma_\mu^+|\psi\rangle=-\gamma_\mu^-|\psi\rangle~~,~~w^\mu=0
\right\}\end{array}}
\label{chiralsubsps}\eeq
Each subspace (\ref{chiralsubsps}) represents a particular choice of chiral or
anti-chiral representation of the double Clifford algebra ${\cal C}(T^n)={\cal
C}(T^n)^+\oplus{\cal C}(T^n)^-$, respectively (see (\ref{spinreps})). Each such
representation in turn encodes a particular choice of one of the $2^n$ possible
spin structures of the torus.

The Hilbert space $\bar{\cal H}_0$ contains the subspace of highest-weight
vectors which belong to complex-conjugate pairs of left-right representations
of the $u(1)_+^n\oplus u(1)_-^n$ current algebra. The $2^n$ subspaces
in
(\ref{spinspaces}) are all naturally isomorphic to the canonical anti-chiral
subspace
\beq
\bar{\cal H}_0^{(-)}=\bar{\cal H}_0^{(-)1}\otimes\bar{\cal
H}_0^{(-)2}\otimes\cdots\otimes\bar{\cal H}_0^{(-)n}
\label{antichiral0}\eeq
The explicit isomorphism maps $\bar{\cal H}_0^{(+)\mu}\leftrightarrow\bar{\cal
H}_0^{(-)\mu}$ by $g^{\nu\lambda}d_{\lambda\mu}^\pm
\gamma_\nu^\pm\leftrightarrow\pm\gamma_\mu^\pm$
and $g^{\mu\nu}p_\nu\leftrightarrow w^\mu$ (or equivalently
$g^{\mu\nu}p_\nu^\pm\leftrightarrow\pm(d^\pm)^{\mu\nu}p_\nu^\pm$). These
isomorphisms themselves determine a sort of partial T-duality
transformation that exchanges $m\leq n$ momenta with winding
numbers. We will see below that they are related to another type of
duality called `factorized duality'. Thus the various chiral and
anti-chiral subspaces (\ref{chiralsubsps}) are all naturally
isomorphic to one another under such ``partial" T-duality
transformations. Here we shall make the canonical
choice of anti-chiral low-energy subspace (\ref{antichiral0})
corresponding to the representation of the double Clifford algebra for
which $\gamma_\mu^+=-\gamma_\mu^-\equiv\gamma_\mu~~\forall\mu$. The
isomorphisms above demonstrate explicitly the independence of quantities on the
choice of spin structure for the spacetime. It is intriguing that a change of
spin structure is manifested as a T-duality symmetry of the string theory.

In the subspace $\bar{\cal H}_0^{(-)}$, we have $w^\mu=0~~\forall\mu$, as
required since in the low-energy sector there should be no global winding modes
of the string around the compactified directions of the target space. It
follows that $\sqrt2p_\mu^+=\sqrt2p_\mu^-=p_\mu\in\Gamma^*$, and thus the
subspace (\ref{antichiral0}) is naturally isomorphic to the Hilbert space
\beq
\bar{\cal H}_0^{(-)}\cong\varrho^-[{\cal C}(T^n)]\otimes
L^2((S^1)^n,\mbox{$\prod_{\mu=1}^n\frac{dx^\mu}{\sqrt{2\pi}}$})
\label{lowenhilbert}\eeq
which is (a local trivialization of) the bundle of anti-chiral
square-integrable spinors on the torus. Here $\varrho^-$ denotes the
anti-chiral spinor representation and the explicit $L^2$-isomorphism in
(\ref{lowenhilbert}) maps $|p;p\rangle\leftrightarrow\e^{-ip_\mu x^\mu}$ in the
restriction to the $L^2$-component of winding number 0. The anti-holomorphic
Dirac operator $\bar\Dirac$ acting in (\ref{lowenhilbert}) is
\beq
\bar\Dirac\,\bar{\cal
P}_0^{(-)}=ig^{\mu\nu}\gamma_\mu\otimes\mbox{$\frac\partial{\partial x^\nu}$}
\label{diraco-}\eeq
where $\bar{\cal P}_0^{(-)}$ is the operator that projects ${\cal H}$
orthogonally onto $\bar{\cal H}_0^{(-)}$, and we have defined
$x^\mu=\frac1{\sqrt2}(x_+^\mu+x_-^\mu)$. In particular, $\bar{\cal H}_0^{(-)}$
is an invariant subspace of the Dirac operator, $\bar\Dirac\bar{\cal
H}_0^{(-)}=\bar{\cal H}_0^{(-)}$, and the full low-energy Hilbert space
(\ref{spinspaces}) is a maximal subspace of ${\cal H}$ with this invariance
property.

Next we need a corresponding low-energy projection $\bar{\cal A}_0$ of the
vertex operator algebra $\cal A$. We define $\bar{\cal A}_0$ to be the commutant
of the Dirac operator restricted to the Hilbert space $\bar{\cal H}_0$,
\beq
\bar{\cal A}_0=\bar{\cal P}_0\,({\rm comm}~\Dirac)\,\bar{\cal
P}_0\equiv\{V\in{\cal A}~|~[\Dirac,V]\,\bar{\cal P}_0=0\}
\label{A0comm}\eeq
It is the largest subalgebra of $\cal A$ with the property
\beq
\bar{\cal A}_0\bar{\cal H}_0=\bar{\cal H}_0
\label{A0def}\eeq
To describe the restriction of $\bar{\cal A}_0$ to $\bar{\cal H}_0^{(-)}$,
consider a typical homogenous smeared vertex operator
$V_\Omega(q^+,q^-)\in{\cal A}_X$ of type $\Omega$ and momentum
$(q^+,q^-)\in\Lambda$. Since
\beq
\bar{\cal P}_0^{(-)}\,\left[\Dirac,\id\otimes
V_\Omega(q^+,q^-)\right]\,\bar{\cal
P}_0^{(-)}=g^{\mu\nu}\gamma_\mu\otimes
\left(q_\nu^+-q_\nu^-\right)\bar{\cal P}_0^{(-)}\,V_\Omega(q^+,q^-)\,\bar{\cal
P}_0^{(-)}
\label{DVOmcomm}\eeq
it follows that the subalgebra $\bar{\cal P}_0^{(-)}\bar{\cal A}_0\bar{\cal
P}_0^{(-)}$ consists of those vertex operators which create string states of
identical left and right chiral momentum, i.e.
$\sqrt2q_\mu^+=\sqrt2q_\mu^-=q_\mu\in\Gamma^*$, which again agrees 
heuristically
with the zero winding number restriction of the low-energy sector. For the
basis (tachyon) vertex operators of $\cal A$ we have in general that
\beq
\bar{\cal P}_0\,(\id\otimes
V_{q^+q^-}(1,1))|\psi;p^+,p^-\rangle=|\psi;q^++p^+,q^-+p^-\rangle
\label{V11proj}\eeq
It follows that $V(q,q)$ generate $\bar{\cal P}_0^{(-)}\bar{\cal A}_0\bar{\cal
P}_0^{(-)}$ and, in particular, we have
\beq
(\id\otimes V_{qq}(z_+,z_-))\,\bar{\cal P}_0^{(-)}=\e^{-iq_\mu
x^\mu}\left(z_+z_-\right)^{q_\mu g^{\mu\nu}p_\nu/2}
\label{Vqqproj}\eeq
Thus the smeared tachyon generators of $\bar{\cal P}_0^{(-)}\bar{\cal
A}_0\bar{\cal P}_0^{(-)}$ coincide with the spacetime functions $\e^{-iq_\mu
x^\mu}$ which constitute a basis for the algebra $C^\infty(T^n)$ of smooth
(single-valued) functions on the toroidal target space. Thus the low-energy
algebra $\bar{\cal A}_0$ yields a natural isomorphism with the abelian algebra
\beq
\bar{\cal P}_0^{(-)}\,\bar{\cal A}_0\,\bar{\cal P}_0^{(-)}\cong C^\infty(T^n)
\label{A0C*}\eeq

To summarize then, we have proven that there is a natural isomorphism between
the spectral triples:
\beq
\bar{\cal P}_0^{(-)}{\cal T}_{\bar\Dirac}\equiv\left(\bar{\cal
P}_0^{(-)}\,\bar{\cal A}_0\,\bar{\cal P}_0^{(-)}~,~\bar{\cal
H}_0^{(-)}~,~\bar\Dirac\,\bar{\cal
P}_0^{(-)}\right)\cong\left(C^\infty(T^n)~,~L^2({\rm
spin}^-(T^n))~,~ig^{\mu\nu}\gamma_\mu\partial_\nu\right)
\label{tripiso}\eeq
This says that the low-energy projection of the spectral triple
(\ref{sigmatriple}) determined by the kernel of the Dirac operator $\Dirac$
coincides with the spectral triple that describes the ordinary (commutative)
spacetime geometry of the $n$-torus $T^n\cong\real^n/2\pi\Gamma$ with metric
$g_{\mu\nu}$. Thus the full noncommutative spacetime (\ref{sigmatriple}) of the
string theory can be projected onto an ordinary, commutative spacetime via an
explicit choice of Dirac operator. A key feature of these low-energy
projections is that the corresponding algebras $\bar\alg_0$ consist only of the
zero-mode components of the tachyon vertex operators. The full
noncommutative spacetime is then built from the highest-weight states of the
current algebra which are eigenstates of the Dirac operator $\Dirac$,
\beq
\left[\Dirac,\id\otimes
V_\Omega(q^+,q^-)\right]=g^{\mu\nu}\left(\gamma_\mu^+\otimes
q_\nu^++\gamma_\mu^-\otimes q_\nu^-\right)V_\Omega(q^+,q^-)
\label{Diracspec}\eeq
which is just another feature of the spectral action principle 
\cite{ChamsConnes}.

Now let us treat the second Dirac operator $\bar\Dirac$ in (\ref{DDbar}) in an
analogous way. It yields another low-energy subspace of the Hilbert space
${\cal H}$,
\beq
{\cal H}_0\equiv{\rm
ker}~\bar\Dirac\cong\bigotimes_{\mu=1}^n\left({\cal
H}_0^{(+)\mu}\oplus{\cal H}_0^{(-)\mu}\right)
\label{spinspacesbar}\eeq
where
\beq\new{\begin{array}{lll}
{\cal H}_0^{(+)\mu}&=&\left\{|\psi\rangle\otimes|p^+;p^-\rangle\in{\cal
H}_0~|~\gamma_\mu^+|\psi\rangle=\gamma_\mu^-|\psi\rangle~~,~~w^\mu=0\right\}
\\{\cal
H}_0^{(-)\mu}&=&\left\{|\psi\rangle\otimes|p^+;p^-\rangle\in{\cal
H}_0~|~g^{\nu\lambda}d_{\lambda\mu}^+\gamma_\nu^+|\psi\rangle=-
g^{\nu\lambda}d_{\lambda\mu}^-\gamma_\nu^-|\psi\rangle~~,~~p_\mu=0
\right\}\end{array}}
\label{chiralsubspsbar}\eeq
define the chiral and anti-chiral subspaces of the kernel of $\bar\Dirac_0$
which is obtained from (\ref{Dirac0}) by the replacements
$\gamma_\mu^\pm\leftrightarrow\pm\gamma_\mu^\pm$. Again the $2^n$ spin
structure subspaces in (\ref{spinspacesbar}) are all naturally isomorphic under
partial T-duality, and we therefore take the canonical anti-chiral subspace
${\cal H}_0^{(-)}={\cal H}_0^{(-)1}\otimes{\cal
H}_0^{(-)2}\otimes\cdots\otimes{\cal H}_0^{(-)n}$ with the representation
of the double Clifford algebra for which
$g^{\nu\lambda}d_{\lambda\mu}^+\gamma_\nu^+=-g^{\nu\lambda}d_{\lambda\mu}^-
\gamma_\nu^-\equiv\tilde\gamma_\mu~~\forall\mu$ (which is dual to the
anti-chirality condition of the subspace (\ref{antichiral0})). In this subspace
we have $p_\mu=0~~\forall\mu$ so that
$\sqrt2(d^+)^{\mu\nu}p_\nu^+=-\sqrt2(d^-)^{\mu\nu}p_\nu^-=w^\mu\in\Gamma$.
${\cal H}_0^{(-)}$ is also naturally isomorphic to the low-energy particle
Hilbert space (\ref{lowenhilbert}), with the dual anti-chiral spinor
representation $(\varrho^-)^*$, under the identification
$|d^+w;-d^-w\rangle\leftrightarrow\e^{-id^-_{\mu\lambda}g^{\lambda\rho}
d_{\rho\nu}^+w^\nu x^\mu}$, and the holomorphic Dirac operator action is given
by
\beq
\Dirac\,{\cal P}_0^{(-)}=i\tilde
g^{\mu\nu}\tilde\gamma_\mu\otimes\mbox{$\frac\partial{\partial x^\nu}$}
\label{dirac0-bar}\eeq
where the dual metric $\tilde g_{\mu\nu}$ is defined by (\ref{dualmetric}).
Again the full low-energy Hilbert space ${\cal H}_0$ is a maximal invariant
subspace for the Dirac operator $\Dirac$.

Finally, we define the maximal subalgebra
\beq
{\cal A}_0={\cal P}_0\,({\rm comm}~\bar\Dirac)\,{\cal P}_0
\label{A0bardef}\eeq
of the smeared vertex operator algebra $\cal A$ with the property
${\cal A}_0{\cal H}_0={\cal H}_0$. Since
\beq
{\cal P}_0^{(-)}\,\left[\bar\Dirac,\id\otimes V_\Omega(q^+,q^-)\right]\,{\cal
P}_0^{(-)}=\tilde\gamma_\mu\otimes
\left((d^+)^{\nu\mu}q_\nu^++(d^-)^{\nu\mu}q_\nu^-\right){\cal
P}_0^{(-)}\,V_\Omega(q^+,q^-)\,{\cal P}_0^{(-)}
\label{DVOmcommbar}\eeq
it follows that the subalgebra ${\cal P}_0^{(-)}{\cal A}_0{\cal P}_0^{(-)}$
consists of smeared vertex operators $V_\Omega(q^+,q^-)$ with
$\sqrt2(d^+)^{\nu\mu}q_\nu^+=-\sqrt2(d^-)^{\nu\mu}q_\nu^-=v^\mu\in\Gamma$.
${\cal P}_0^{(-)}{\cal A}_0{\cal P}_0^{(-)}$ is generated by the smeared
tachyon
vertex operators $V(d^+v,-d^-v)$ which coincide with the basis spacetime
functions $\e^{-i\tilde g_{\mu\nu}v^\nu x^\mu}$ of $C^\infty(T^n)$. Thus there
is also the natural isomorphism of spectral triples,
\beq
{\cal P}_0^{(-)}{\cal T}_\Dirac\equiv\left({\cal P}_0^{(-)}\,{\cal A}_0\,{\cal
P}_0^{(-)}~,~{\cal H}_0^{(-)}~,~\Dirac\,{\cal
P}_0^{(-)}\right)\cong\left(C^\infty(T^n)^*~,~L^2({\rm
spin}^-(T^n)^*)~,~i\tilde g^{\mu\nu}\tilde\gamma_\mu\partial_\nu\right)
\label{tripisobar}\eeq
which identifies the low-energy projection of $\cal T$ determined by the kernel
of the Dirac operator $\bar\Dirac$ with the commutative spacetime geometry of
the dual $n$-torus $(T^n)^*\cong\real^n/2\pi\Gamma^*$ with metric $\tilde
g^{\mu\nu}$.

According to the T-duality symmetry (\ref{Tdualiso}) of the effective string
spacetime, as subspaces of the quantum spacetime we have the isomorphism
\beq
\left(C^\infty(T^n)~,~L^2({\rm
spin}^-(T^n))~,~ig^{\mu\nu}\gamma_\mu\partial_\nu\right)
\cong\left(C^\infty(T^n)^*~,~L^2({\rm spin}^-(T^n)^*)~,~i\tilde
g^{\mu\nu}\tilde\gamma_\mu\partial_\nu\right)
\label{Tdualitycomm}\eeq
which is the usual statement of the T-duality $T^n\leftrightarrow(T^n)^*$ of
string theory compactified on an $n$-torus. Notice how the statement
that this duality symmetry corresponds to the target space symmetry
$g_{\mu\nu}\leftrightarrow\tilde g^{\mu\nu}$ and the symmetry under interchange
of momentum and winding numbers in the compactified string spectrum arise very
naturally from the point of view of noncommutative geometry. It appears as a
discrete $\zed_2$-symmetry of the noncommutative geometry.

\subsubsection*{Worldsheet Parity}

T-duality was shown above to be the linear isomorphism of the string spacetime
which relates the anti-chiral low-energy subspaces defined by the pair of Dirac
operators (\ref{DDbar}). There are several other symmetries of the string
spacetime which also arise from the above construction, represented by the
isomorphisms between other pairs of subspaces in (\ref{spinspaces}) and
(\ref{spinspacesbar}). For example, suppose we consider the chiral subspace of
(\ref{spinspacesbar}),
\beq
{\cal H}_0^{(+)}={\cal H}_0^{(+)1}\otimes{\cal
H}_0^{(+)2}\otimes\cdots\otimes{\cal H}_0^{(+)n}
\label{chiralspinspace}\eeq
In this subspace we have $\gamma_\mu^+=\gamma_\mu^-\equiv\gamma_\mu$ and
$w^\mu=0$ for all $\mu$. The Hilbert space ${\cal H}_0^{(+)}$ is also
isomorphic to (\ref{lowenhilbert}), with the chiral spinor representation
$\varrho^+$, and the holomorphic Dirac operator action is now
\beq
\Dirac\,{\cal
P}_0^{(+)}=-ig^{\mu\nu}\gamma_\mu\otimes\mbox{$\frac\partial{\partial
x^\nu}$}=-\bar\Dirac\,\bar{\cal P}_0^{(-)}
\label{parityop}\eeq
Moreover, one finds that again ${\cal P}_0^{(+)}\alg_0{\cal P}_0^{(+)}$ is
generated by the smeared tachyon vertex operators $V(q,q)\sim\e^{-iq_\mu
x^\mu}$. According to the equivalences discussed above, we then have the
isomorphism of low-energy spectral triples
\beq
\left(C^\infty(T^n)~,~L^2({\rm
spin}^-(T^n))~,~ig^{\mu\nu}\gamma_\mu\partial_\nu\right)
\cong\left(C^\infty(T^n)~,~L^2({\rm
spin}^+(T^n))~,~-ig^{\mu\nu}\gamma_\mu\partial_\nu\right)
\label{parityisolow}\eeq

This quantum symmetry of the string spacetime is the worldsheet parity symmetry
of string theory. It is the left-right chirality symmetry which reflects
the worldsheet spatial coordinate $\sigma\to-\sigma$ and thus acts on the
string background as $\beta\to-\beta$, i.e. $d_{\mu\nu}^\pm\to d_{\mu\nu}^\mp$.
It acts on the Hilbert space ${\cal H}$ as
$\alpha_k^{(\pm)\mu}\to\alpha_k^{(\mp)\mu}$, $p_\mu^\pm\to p_\mu^\mp$ and
$\gamma_\mu^\pm\to\pm\gamma^\mp_\mu$. Because it interchanges the left and
right chirality sectors of the string theory, it flips the sign of the
Lorentzian quadratic form (\ref{quadform}) and is thus not an automorphism of
the lattice $\Lambda$. In terms of the full spectral triple describing the
quantum spacetime, it is the $\zed_2$-transformation $T\equiv W_S\otimes
W_X:{\cal H}\to{\cal H}$ that achieves (\ref{TDmap}) and (\ref{Tdualiso}) by
mapping
\beq\new{\begin{array}{rrl}
W_X\,\alpha_k^{(\pm)\mu}\,W_X^{-1}&=&\alpha_k^{(\mp)\mu}
\\W_S\,\gamma_\mu^\pm\,W_S^{-1}&=&\pm\gamma_\mu^\mp\\W_X\,V_\Omega(q^+,q^-)\,
W_X^{-1}&=&V_\Omega(q^-,q^+)\end{array}}
\label{diracparity}\eeq
which amounts to the interchange $\pm\leftrightarrow\mp$ on both ${\cal H}$ and
$\cal A$. This worldsheet quantum symmetry of the sigma-model thus also arises
as a change of Dirac operator for the noncommutative geometry, and so the
isometries of the spectral triple (\ref{sigmatriple}) account for both target
space {\em and} discrete worldsheet symmetries of the quantum geometry.

\subsubsection*{Factorized Duality and Spacetime Topology Change}

The next generalization of the T-duality isomorphism of low-energy sectors
is to compare the anti-chiral subspace (\ref{antichiral0}) with the subspaces
\beq
{\cal H}_0^{[+;\mu]}={\cal H}_0^{(+)1}\otimes\cdots\otimes{\cal
H}_0^{(+)\mu-1}\otimes{\cal H}_0^{(-)\mu}\otimes{\cal
H}_0^{(+)\mu+1}\otimes\cdots\otimes{\cal H}_0^{(+)n}
\label{chiral+mu}\eeq
which are defined for each $\mu=1,\dots,n$. In ${\cal H}_0^{[+;\mu]}$ we
have $\gamma_\nu^+=\gamma_\nu^-\equiv\gamma_\nu$ and $w^\nu=0$ for all
$\nu\neq\mu$, while
$g^{\lambda\rho}d_{\lambda\mu}^+\gamma_\rho^+=-g^{\lambda\rho}
d_{\lambda\mu}^-\gamma_\rho^-\equiv\tilde\gamma_\mu$ and $p_\mu=0$. This
Hilbert space is isomorphic to (\ref{lowenhilbert}) with the mixed chirality
spinor representation $\varrho^{(\mu)}$ determined by the spinor conditions of
(\ref{chiral+mu}), where the explicit $L^2$-isomorphism maps the bosonic states
of (\ref{chiral+mu}) to the functions
\beq
f^{(\mu)}(x)=\exp\left(-i\,\mbox{$\sum_{\nu\neq\mu}$}\,p_\nu x^\nu-i\,
\mbox{$\sum_\lambda$}\,\tilde
g_{\mu\lambda}w^\mu x^\lambda\right)
\label{mixedbasisfns}\eeq
The restriction of the holomorphic Dirac operator to (\ref{chiral+mu}) is
\beq
\Dirac\,{\cal P}_0^{[+;\mu]}=-i\,\mbox{$\sum_{\nu\neq\mu}$}
\mbox{$\sum_\lambda$}
\,g^{\nu\lambda}\gamma_\nu\otimes\mbox{$\frac\partial{\partial
x^\lambda}$}+i\,\mbox{$\sum_\lambda$}\,\tilde
g^{\mu\lambda}\tilde\gamma_\mu\otimes\mbox{$\frac\partial{\partial x^\lambda}$}
\label{Diracmixed}\eeq
The algebra ${\cal P}_0^{[+;\mu]}\alg_0{\cal P}_0^{[+;\mu]}$ is generated by
the
smeared tachyon vertex operators which are given by the basis spacetime
functions (\ref{mixedbasisfns}) of $C^\infty(T^n)$, and we arrive at the
spectral triple isomorphism
\beq\new{\begin{array}{l}
\left(C^\infty(T^n)~,~L^2({\rm
spin}^-(T^n))~,~ig^{\mu\nu}\gamma_\mu\partial_\nu\right)
\\~~~~~~~~~~~~\cong\left(C^\infty(T^n)~,~L^2({\rm
spin}^{(\mu)}(T^n))~,~-i\,\mbox{$\sum_{\nu\neq\mu}\,\sum_\lambda$}
\,g^{\nu\lambda}\gamma_\nu\partial_\lambda+i\,\mbox{$\sum_\lambda$}\,\tilde
g^{\mu\lambda}\tilde\gamma_\mu\partial_\lambda\right)
\end{array}}
\label{partialiso}\eeq

The symmetry (\ref{partialiso}) of the string spacetime is called `factorized
duality'. For each $\mu=1,\dots,n$ it is a generalization of the $R\to1/R$
circle duality in the $X^\mu$ direction of $T^n$. This becomes more transparent
if we choose a particular basis of the lattice $\Gamma$ that splits the
$n$-torus into a product of a circle $S^1$ of radius $R_\mu$ and an
$(n-1)$-dimensional background $T^{n-1}$. The factorized duality map then takes
$R_\mu\to1/R_\mu$ leaving $T^{n-1}$ unchanged, and at the same time
interchanges the $\mu^{\rm th}$ momentum and winding mode
$g^{\mu\nu}p_\nu\leftrightarrow w^\mu$ leaving all others invariant. Acting on
the full spectral triple of the noncommutative spacetime it is described
formally as follows. Let
$(E_\mu)^{\nu\lambda}=\delta^\nu_\mu\delta^\lambda_\mu$ be the $n$-dimensional
step operators, and consider the unitary transformation $T\equiv{\cal
D}^{(\mu)}_S\otimes{\cal D}^{(\mu)}_X:{\cal H}\to{\cal H}$ in \eqn{TDmap}
defined by
\beq\new{\begin{array}{rrl}
{\cal
D}_X^{(\mu)}\,|p^+,p^-\rangle\otimes|0\rangle_+\otimes|0\rangle_-&=&(-1)^{p_\mu
w^\mu}|\tilde p^+;\tilde p^-\rangle\otimes|0\rangle_+\otimes|0\rangle_-\\{\cal
D}_X^{(\mu)}\,\alpha_k^{(\pm)\nu}\,{\cal
D}_X^{(\mu)-1}&=&\left[\left(\delta^\nu_\lambda-g_{\lambda\rho}
(E_\mu)^{\nu\rho}\right)\pm(E_\mu)^{\nu\rho}d_{\rho\lambda}^\mp\right]
\alpha_k^{(\pm)\lambda}\\{\cal D}_S^{(\mu)}\,\gamma_\nu^\pm\,{\cal
D}_S^{(\mu)-1}&=&\left[\left(\delta_\nu^\lambda-g_{\nu\rho}
(E_\mu)^{\rho\lambda}\right)+g_{\nu\sigma}g^{\alpha\lambda}
(E_\mu)^{\sigma\rho}d_{\rho\alpha}^\mp\right]\gamma_\lambda^\pm\\{\cal
D}^{(\mu)}_X\,V_\Omega(q^+,q^-)\,{\cal D}^{(\mu)-1}_X&=&(-1)^{q_\mu
w^\mu}V_\Omega(\tilde q^+,\tilde q^-)\end{array}}
\label{factdualtransfs}\eeq
where $\tilde p_\nu^\pm=p_\nu^\pm~~\forall\nu\neq\mu$ and $\tilde p_\mu^\pm=\pm
g_{\mu\lambda}(d^\pm)^{\lambda\rho}p_\rho^\pm$ (or equivalently
$g^{\mu\nu}p_\nu\leftrightarrow w^\mu$). The mapping (\ref{factdualtransfs})
acts on the background matrices and the metric tensor as
\beq\new{\begin{array}{l}
d_{\nu\rho}^\pm\to\left[\left(\delta_\nu^\lambda-g_{\nu\alpha}
(E_\mu)^{\alpha\lambda}\right)d_{\lambda\gamma}^\pm+g_{\nu\alpha}g_{\gamma
\beta}(E_\mu)^{\alpha\beta}\right]\left([E_\mu\cdot
d^\pm+(I_n-g\cdot E_\mu)]^{-1}\right)^\gamma_\rho\\g_{\nu\rho}\to\left[\left(
\delta^\lambda_\nu-g_{\nu\alpha}(E_\mu)^{\alpha\lambda}\right)+g_{\nu\alpha}
g^{\lambda\beta}(E_\mu)^{\alpha\gamma}d_{\gamma\beta}^-\right]g_{\lambda\sigma}
\left[\left(\delta^\sigma_\rho-g_{\rho\alpha}(E_\mu)^{\alpha\sigma}\right)+
g_{\rho\alpha}g^{\beta\sigma}(E_\mu)^{\alpha\gamma}d_{\gamma\beta}^+\right]
\end{array}}
\label{metricmutransfs}\eeq
where $I_n$ is the $n\times n$ identity matrix. As before, the transformation
(\ref{factdualtransfs}) leads to the isomorphism (\ref{Tdualiso}) of the full
noncommutative geometry.

When $n$ is even, the factorized duality map yields the famous `mirror
symmetry' of string theory which expresses the equivalence of string
spacetimes under the interchange of the complex and K\"ahler structures of
$T^n$. It is equivalent to the interchange of the Dolbeault cohomology groups
$H^{p,q}(T^n)$ and $H^{\frac n2-p,q}(T^n)$ which gives a mirror reflection
along the diagonals of the corresponding Hodge diamonds. The change of Dirac
operator in \eqn{TDmap} therefore also yields the stringy phenomenon of
{\it spacetime topology change}. It arises in the
present point of view from the non-trivial chirality structure which is present
in the spin bundle of $T^n$ when $n$ is even (see \eqn{spinreps})
\footnote{More complicated duality symmetries, such as mirror symmetry
between distinct, curved Calabi-Yau manifolds \cite{Tdualrev}, can be obtained
by introducing a larger set of Dirac operators which are related to, for instance,
an $N=2$ supersymmetric sigma-model. The resulting spectral triple contains a
larger symmetry than just the chiral-antichiral one used in this paper. Some of
these points are addressed in \cite{FG,fgrleshouches}.}. Furthermore,
the above analysis shows that generally a factorized duality transformation in
the $\mu^{\rm th}$ direction must be accompanied by a worldsheet parity
transformation in all of the other $n-1$ directions. This is somewhat
anticipated from our earlier remarks about the relationship between the spin
structures of $T^n$ and T-duality, and it agrees with some basic statements
concerning mirror symmetry when $n$ is even. Thus the noncommutative geometry
formulation of the quantum geometry shows that worldsheet parity is in fact a
crucial part of the duality symmetries. The remaining isomorphisms between
pairs of subspaces in (\ref{spinspaces}) and (\ref{spinspacesbar}) are then
combinations of the discrete duality transforms exhibited above.

\subsubsection*{Lattice Isomorphism}

The duality symmetries which were essentially deduced above from the various
isomorphisms that exist between the low-energy spectral triples exhaust the
transformations (\ref{TDmap}) which lead to non-trivial equivalences between
spacetimes of distinct geometry and topology. There are, however, two other
discrete ``internal" spacetime symmetries that leave each of the Dirac
operators in (\ref{DDbar}) invariant and trivially leave the corresponding
spectral triples unaffected. These transformations do not affect the classical
spacetimes, but they do lead to non-trivial effects in the quantum field theory
and are therefore associated with symmetries of the quantum geometry.

The first one is a change of basis of the compactification lattice $\Gamma$,
which is described by an invertible, integer-valued matrix
$[A^\mu_\nu]\in GL(n,\zed)$ that acts on the spacetime metric as
\beq
g_{\mu\nu}\to(A^{-1})^\lambda_\mu~g_{\lambda\rho}~(A^{-1})^\rho_\nu
\label{basischange}\eeq
In general, all covariant tensors from which the spectral triples are built
transform under $A^{-1}$ while all contravariant tensors transform under $A$.
This leads to a simple reparametrization
of all quantities composing the spectral triples, thus leaving them unaltered.
This change of basis is therefore also trivially a symmetry of the low-energy
commutative string spacetime \eqn{spectral0}. For instance, some of these
$GL(n,\zed)$ transformations simply permute the spacetime dimensions, while
others
reflect the configurations $X^\mu\to-X^\mu$. T-duality can be shown to be the
composition of a succession of factorized dualities and dimension permutations
in all of the directions of $T^n$. This is naturally evident from the
isomorphisms between (\ref{spinspaces}) and (\ref{spinspacesbar}), where
appropriate permutations of the spin structures map factorized duality onto
T-duality.

\subsubsection*{Torsion Cohomology}

The final quantum symmetry is the shift
\beq
\beta_{\mu\nu}\to\beta_{\mu\nu}+C_{\mu\nu}
\label{betashift}\eeq
of the spacetime torsion form by an antisymmetric, integer-valued matrix
$C_{\mu\nu}$. This shift corresponds to a change of integer cohomology class of
the instanton form, and thus only affects the winding numbers in the target
space $T^n$. It can be absorbed by a shift $p_\mu\to p_\mu-C_{\mu\nu}w^\nu$
which simply yields a reparametrization of the momenta
$\{p_\mu\}\in\Gamma^*$. All other quantities are left invariant by this shift,
and the action of the Dirac operators on $(\alg,{\cal H})$ is unaffected by
this discrete transformation. Again we trivially have the equivalence between
the corresponding string spacetimes.

\subsubsection*{Quantum Moduli Space}

It can be shown that the discrete transformations exhibited in this section
generate the duality group of the string theory, which is the semi-direct
product
\beq
G_d=O(n,n;\zed)\otimes_{\rm S}\zed_2
\label{dualitygp}\eeq
of the lattice automorphism group $O(n,n;\zed)$ (i.e. the group of
transformations of $\Lambda$
that preserve the quadratic form \eqn{quadform}) by the action
of the reflection group $\zed_2$ corresponding to worldsheet parity. The group
$O(n,n;\zed)$ is the arithmetic subgroup of $O(n,n)$ and it acts on the
background matrices $d_{\mu\nu}^\pm$ by linear fractional transformations. Note
that inside the duality group $G_d$ lies the discrete geometrical subgroup
$SL(n,\zed)\subset O(n,n;\zed)$ which represents the group of large
diffeomorphisms of $T^n$. The {\it quantum} modification of \eqn{classmod} is
therefore given by the Narain moduli space \cite{narain}
\beq
{\cal M}_{\rm qu}=O(n,n;\zed)\setminus O(n,n)/(O(n)\times O(n))\otimes_{\rm
S}\zed_2
\label{qumoduli}\eeq
where the quotient by the infinite discrete group $O(n,n;\zed)$ acts on
$O(n,n)$ from the left. As we have discussed, the duality group
(\ref{dualitygp}) is a discrete subgroup of the group ${\rm Aut}(\alg)$ of
automorphisms of the vertex operator algebra $\alg$. In the next section we
shall discuss the structure of ${\rm Aut}(\alg)$ in a somewhat more general
setting.

\newsection{Symmetries of the Noncommutative String Spacetime}

In the previous section we examined the symmetries of the quantum spacetime by
determining the discrete automorphisms of the spectral triples which led to
isomorphisms between their low-energy projection subspaces. The full duality
group (\ref{dualitygp}) was thus determined as the set of all isomorphisms
between subspaces of (\ref{spinspaces}) and (\ref{spinspacesbar}). Along the
way we showed that this way of viewing the target space duality in
noncommutative geometry led to new insights into the relationships between
choices of spin structure, worldsheet parity, and T-duality. However, there are
many more possible automorphisms of the vertex operator algebra $\alg$ than
just those which preserve the commutative subspaces. In fact, the elements of
the duality group $G_d$ arise from the equivalences between the zero-mode
eigenspaces of the Dirac operators (\ref{DDbar}), while the general isospectral
automorphisms of the spectral triple (\ref{sigmatriple}) are determined by
unitary transformations (such as \eqn{TDmap}) between different Dirac operators
that have the same spectrum \cite{ChamsConnes}.  Indeed, the structure of the
full string spacetime is determined by the spectrum of the Dirac K-cycle $({\cal
H},\Dirac)$ or $({\cal H},\bar\Dirac)$ (see \eqn{Diracspec})), which in turn
incorporates the non-zero oscillatory modes of the strings. 
In this final section we shall
examine the geometrical symmetries of the string spacetime in a more general
setting by viewing them as automorphisms of the vertex operator algebra. This
point of view will, among other things, naturally establish the framework for
viewing duality as a gauge symmetry \cite{Tdualrev}. We shall also comment on
some non-metric aspects of the noncommutative geometry.

\subsubsection*{Automorphisms of the Vertex Operator Algebra}

The basic symmetry group of the noncommutative string spacetime
(\ref{sigmatriple}) is ${\rm Aut}(\alg)$. An {\it automorphism} of the vertex
operator algebra is a unitary transformation $g:{\cal H}\to{\cal H}$ which
preserves both the vacuum state and conformal vectors $\omega^\pm$ defined in
(\ref{stressenvec}), i.e. $g|{\rm vac}\rangle=|{\rm vac}\rangle$ and
$g\omega^\pm=\omega^\pm$, such that the actions
of $g$ and ${\cal V}(\Psi;z_+,z_-)$ on $\widehat{\cal H}_X(\Lambda)$ are
compatible in the sense that
\beq
g\,{\cal V}(\Psi;z_+,z_-)\,g^{-1}={\cal
V}(g\Psi;z_+,z_-)~~~~,~~~~\forall\Psi\in\widehat{\cal H}_X(\Lambda)
\label{autodef}\eeq
Thus $g$ preserves the stress-energy tensors, and hence the
representations of the Virasoro subalgebras, and the mapping $\cal V$ on
\eqn{1-1map} is equivariant with respect to the natural adjoint actions of 
$g$. The
automorphism $g$ also preserves the grading of $\widehat{\cal H}_X(\Lambda)$ 
(i.e. the
subspaces \eqn{primaryops}) which can be decomposed into a direct sum of the
eigenspaces of $g$,
\beq
\widehat{\cal H}_X(\Lambda)=\bigoplus_{j\in\zeds_r}\widehat{\cal
H}_X^{(j)}(\Lambda)
\label{eigendecompH}\eeq
where $r>0$ is the order of $g$ and $\widehat{\cal
H}_X^{(j)}(\Lambda)=\{\Psi\in\widehat{\cal H}_X(\Lambda)~|~g\Psi=\eta^j\Psi\}$
with $\eta$ the generator of $\zed_r$. Note that, from (\ref{stressen}), the
invariance of the stress-energy tensors automatically implies invariances among
the Dirac operators $\Dirac^\pm$. Two immediate examples with $r=\infty$ are
provided by the Kac-Moody and Virasoro transformation groups of the
sigma-model. The former group decomposes $\alg$ in terms of the spectrum of
the Dirac operators. Generally, given a subgroup $G\subset{\rm Aut}(\alg)$, the
Fock space $\widehat{\cal H}_X(\Lambda)$ decomposes into a direct sum of
irreducible $G$-modules $\widehat{\cal H}_X^{[R(G)]}(\Lambda)$, $\widehat{\cal
H}_X(\Lambda)=\bigoplus_{R(G)}\widehat{\cal H}_X^{[R(G)]}(\Lambda)$.

In the ordinary commutative case of a manifold $M$, the group ${\rm Diff}(M)$
of diffeomorphisms of $M$ is naturally isomorphic to the group of automorphisms
of the abelian algebra $\alg=C^\infty(M)$. To each $\varphi\in{\rm Diff}(M)$
one associates the algebra-preserving map $g_\varphi:\alg\to\alg$ by
$g_\varphi(f)=f\circ\varphi^{-1}~~\forall f\in\alg$. In the general
noncommutative case, the group ${\rm Aut}(\alg)$ has a natural normal subgroup
${\rm Inn}(\alg)\subset{\rm Aut}(\alg)$ consisting of {\it inner} automorphisms
of $\alg$, i.e. the algebra-preserving maps $g_u:\alg\to\alg$ that act on the
algebra as conjugation by elements of the group \eqn{unitarygp} of unitary
operators in $\alg$,
\beq
g_u(a)=uau^\dagger~~~,~~~\forall a\in\alg
\label{innerautodef}\eeq
The exact sequence of groups
\beq
\id\to{\rm Inn}(\alg)\to{\rm Aut}(\alg)\to{\rm Out}(\alg)\to\id
\label{exactauto}\eeq
defines the remaining {\it outer} automorphisms in ${\rm Aut}(\alg)$ such that
the automorphism group is the semi-direct product
\beq
{\rm Aut}(\alg)={\rm Inn}(\alg)\otimes_{\rm S}{\rm Out}(\alg)
\label{autosemi}\eeq
of ${\rm Inn}(\alg)$ by the natural action of ${\rm Out}(\alg)$.

Note that for an {\em abelian} algebra $\alg$ the group of inner automorphisms
${\rm Inn}(\alg)=\{\id\}$ is trivial, so that in the case of a commutative
space $M$ the diffeomorphisms of the manifold correspond to outer
automorphisms. Recall from section 2 that the group
(\ref{unitarygp}) of unitary elements of an algebra 
$\alg$ defines a natural gauge
group of the space. In this context inner automorphisms then correspond to
gauge transformations. For example, the automorphism group (\ref{autosemi}) of
the noncommutative algebra (\ref{standardalg}) of the standard model is the
semi-direct product
\beq
{\rm Aut}(\alg_{SM})=C^\infty(M,U(1)\times SU(2)\times U(3))\otimes_{\rm S}{\rm
Diff}(M)
\label{autostandard}\eeq
of the group of local gauge transformations by the natural action of the
diffeomorphism group of $M$. The inner automorphisms in this case are therefore
associated with the local internal gauge invariance of the model while the
outer automorphisms represent the spacetime general covariance dictated by
general relativity. In fact, (\ref{autostandard}) is the canonical invariance
group of the standard model coupled to Einstein gravity, modulo an overall
$U(1)$ phase group which can be eliminated by restricting to the unimodular
group of $\alg_{SM}$.
In the general case then, one can identify the outer automorphisms of the
noncommutative string spacetime as general coordinate transformations and the
inner automorphisms as internal gauge symmetry transformations
\cite{connesauto}, i.e. internal fluctuations of the noncommutative geometry
corresponding to the rotations (\ref{innerautodef}) of the elements of $\alg$.
We shall see that these symmetry structures of the string spacetime have some
remarkable features.

\subsubsection*{Duality Transformations as Inner Automorphisms and Gauge
Symmetries}

We shall now begin to explore the structure of the symmetry group ${\rm
Aut}(\alg)$. For illustration, we start by giving the explicit duality maps $T$
that were exhibited in the previous section and show that they correspond to
inner automorphisms of the vertex operator algebra. A general formalism for
viewing symmetries of string theory as inner automorphisms of the vertex
operator algebra has been given in \cite{strsyms} and applied to duality
transformations in \cite{strsymsdual}. The basic idea behind this approach is
that such an inner automorphism represents a deformation of the conformal field
theory by a marginal operator, and as such it represents the same point in the
corresponding moduli space. Here we wish to stress the fact that such
automorphisms arise quite naturally from the point of view of the
noncommutative geometry formalism and lead immediately to the well-known
interpretation of duality as a gauge symmetry \cite{Tdualrev}.

Recall that the Dirac operators were constructed from the generators
(\ref{currentexp}) of the fundamental $U(1)_+^n\times U(1)_-^n$ gauge symmetry
of the theory. It turns out that this symmetry group is augmented at the fixed
point of the T-duality transformation of the string spacetime. T-duality is
tantamount to the inversion $d^\pm\to(d^\pm)^{-1}$ of the background matrices.
This transformation has a unique fixed point $(d^\pm)^2=I_n$ given by
$g_{\mu\nu}=\delta_{\mu\nu}$ and $\beta_{\mu\nu}=0$. At this single fixed point
the generic $U(1)_+^n\times U(1)_-^n$ gauge symmetry is `enhanced' to a level 1
representation of the affine Lie group
$\widehat{SU(2)}_+^n\times\widehat{SU(2)}_-^n$ \cite{ginsparg}. Thus the fixed
point $\Lambda_0\in{\cal M}_{\rm qu}$ in the Narain moduli space of toroidal
compactification coincides with the occurrence of `enhanced gauge symmetries'.
It is due to the appearance of extra dimension $(\Delta^+,\Delta^-)=(1,0)$ and
$(0,1)$ operators in the theory. To describe this structure, let $k_\mu^{(i)}$,
$i=1,\dots,n$, be a suitable basis of (constant) Killing forms on $T^n$ which
are the generators of isometries of the spacetime metric $g_{\mu\nu}$. Then the
operators
\beq
J_\pm^{\alpha(i)}(z_\alpha)=~:~\e^{\pm ik_\mu^{(i)}X_\alpha^\mu(z_\alpha)}~:
{}~~~~~~,~~~~~~J_3^{\alpha(i)}(z_\alpha)=ik_\mu^{(i)}J_\alpha^\mu(z_\alpha)
\label{SU2currents}\eeq
where $\alpha=\pm$, generate a level 1 $su(2)_+^n\oplus su(2)_-^n$ Kac-Moody
algebra,
\beq
\left[J_{3,k}^{\alpha(i)},J_{\pm,m}^{\alpha(i)}\right]=\pm
J_{\pm,k+m}^{\alpha(i)}~~~~,~~~~\left[J_{+,k}^{\alpha(i)},J_{-,m}^{\alpha(i)}
\right]=2J_{3,k+m}^{\alpha(i)}+2k\delta_{k+m,0}
\label{su2kmalg}\eeq
with all other commutators vanishing, and where we have defined the
mode expansions $J_a^{\alpha(i)}(z_\alpha)
= \sum_{k\in\zeds}J_{a,k}^{\alpha(i)}
z^{-k-1}_\alpha$.

Associated with the $SU(2)_+^n\times SU(2)_-^n$ gauge symmetry of the theory is
the corresponding gauge group element
\beq
g=\e^{i{\cal G}_\chi}
\label{gaugeauto}\eeq
where the generator ${\cal G}_\chi$ is defined as the smeared operator
\beq
{\cal
G}_\chi=\int\frac{dz_+~dz_-}{4\pi
z_+z_-}~\left(\chi^a_{+,\mu}[X]\,J^{+(\mu)}_a(z_+)
+\chi^a_{-,\mu}[X]\,J^{-(\mu)}_a(z_-)\right)f_S(z_+,z_-)
\label{gaugegen}\eeq
and the gauge parameter functions $\chi^a_{\pm,\mu}[X]$, $a=1,2,3$,
$\mu=1,\dots,n$, are sections of the spin bundle of $T^n$. Here and in the
following we define $X=\frac1{\sqrt2}(X_++X_-)$. The unitary operators
(\ref{gaugeauto}) locally decompose as $g=g_S\otimes g_X$, as in the previous
section, where the operators $g_X$ act as inner automorphisms of $
\alg_X$. The automorphisms $g_S$ are defined by their
corresponding actions on the generators $\gamma_\mu^\pm$ and lead to 
reparametrization of the spin structure of the target space.

The operators (\ref{gaugeauto}) that implement spacetime duality
transformations of the string theory have been constructed in
\cite{strsymsdual}. The $\mu^{\rm th}$ factorized duality map corresponds to
the action of the inner automorphism (\ref{gaugeauto}) with ${\cal
G}_\chi={\cal G}^{(\mu)}$ where
\beq
{\cal G}^{(\mu)}=\frac\pi{2i}\int\frac{dz_+~dz_-}{4\pi
z_+z_-}~\left(J_+^{+(\mu)}J_+^{-(\mu)}-J_-^{+(\mu)}J_-^{-(\mu)}\right)f_S
\label{factauto}\eeq
Another duality map which comes from the enhanced gauge symmetry follows from
choosing $k_\mu^{(i)}=\delta_\mu^i$ and ${\cal G}_\chi={\cal G}_+^{(\mu)}+{\cal
G}_-^{(\mu)}$ where
\beq
{\cal G}_\pm^{(\mu)}=\frac\pi{2i}\int\frac{dz_+~dz_-}{4\pi
z_+z_-}~\left(J_+^{\pm(\mu)}-J_-^{\pm(\mu)}\right)f_S
\label{reflauto}\eeq
The inner automorphism generated by (\ref{reflauto}) corresponds to a
reflection $X^\mu\to-X^\mu$ of the coordinates of $T^n$ and is part of the
lattice isomorphism group $GL(n,\zed)$ . Thus factorized dualities and
spacetime reflections are enhanced gauge symmetries of the noncommutative
geometry, and as such they are intrinsic properties of the string spacetime.

The remaining $O(n,n;\zed)$ transformations are abelian gauge symmetries. By
the definition of the currents (\ref{currentexp}), a general spacetime
coordinate transformation $X\to\xi(X)$, with $\xi(X)$ a local section of ${\rm
spin}(T^n)$, is generated by ${\cal G}_\chi={\cal G}_\xi$ with
\beq
{\cal
G}_\xi=\int\frac{dz_+~dz_-}{4\pi
z_+z_-}~\xi_\mu(X)\left(J_+^\mu(z_+)+J_-^\mu(z_-)\right)f_S(z_+,z_-)
\label{gencoordauto}\eeq
Taking the large diffeomorphism $\xi_\mu(X)=\xi_\mu^{(\pi)}(X)=\frac\pi2{\rm
sgn}(\pi)g_{\pi(\mu),\nu}X^\nu$ then yields a permutation $\pi\in S_n$ of the
coordinates of $T^n$ (corresponding to another lattice isomorphism). Combining
this with the factorized duality transformations above yields T-duality in the
form of an inner automorphism. As such, T-duality corresponds to the global
gauge transformation in the Weyl subgroup $\zed_2$ of $SU(2)$. Next, the local
gauge transformations $\beta\to\beta+d\lambda$ of the torsion two-form are
generated by ${\cal G}_\chi={\cal G}_\lambda$ with
\beq
{\cal G}_\lambda=\int\frac{dz_+~dz_-}{4\pi
z_+z_-}~\lambda_\mu(X)\left(J_+^\mu(z_+)-J_-^\mu(z_-)\right)f_S(z_+,z_-)
\label{torsionauto}\eeq
Taking the gauge transformation $\lambda_\mu(X)=C_{\mu\nu}X^\nu$, with
$C_{\mu\nu}$ a constant antisymmetric matrix, gives effectively the torsion
shift (\ref{betashift}). Singlevaluedness of the corresponding group element
(\ref{gaugeauto}) then forces $C_{\mu\nu}\in\zed~~\forall\mu,\nu$ yielding a
large gauge transformation. The final $O(n,n;\zed)$ transformations correspond
to large diffeomorphisms $\xi_\mu(X)=T_{\mu\nu}X^\nu$ of the $n$-torus. Again
singlevaluedness of the corresponding gauge group element puts $[T_{\mu\nu}]\in
SL(n,\zed)$.

As anticipated, the $\zed_2$ part of the duality group $G_d$ representing
worldsheet parity corresponds to an {\it outer} automorphism of the vertex
operator algebra. This is because it corresponds to the automorphism
$W_X\in{\rm Aut}(\alg)$ that interchanges the left and right chiral algebras
${\cal E}={\cal E}^+\otimes{\cal E}^-~{\buildrel W_X\over\longrightarrow}~{\cal
E}^-\otimes{\cal E}^+$. Clearly no inner automorphism of $\alg$ can achieve
this transformation, and indeed worldsheet parity is the outer automorphism of
$\alg$ represented by the $\zed_2$ generator
\beq
W_X=\pmatrix{0&I_n\cr I_n&0\cr}
\label{WXouter}\eeq
acting in the two-dimensional space labelled by the chiral components ${\cal
E}={\cal E}^+\otimes{\cal E}^-$. Thus worldsheet parity cannot be
interpreted in terms of any gauge symmetry and represents a large
diffeomorphism of the noncommutative string spacetime. This $\zed_2$-symmetry
is actually a discrete
subgroup of the $U(1)$ worldsheet symmetry group that acts by rotating the
chiral sectors among each other. Associated to the spin structure of the string
worldsheet there is a representation of ${\rm spin}(2)\cong\real$ on the
Hilbert space ${\cal H}_X$ that implements the group $SO(2)\cong U(1)$ with
generator
\beq
W_\theta=\pmatrix{\cos\theta&\sin\theta\cr-\sin\theta&\cos\theta}~~~~
,~~~~\theta\in[0,2\pi)
\label{thetagen}\eeq

The discrete vertex operator automorphisms, corresponding to large gauge
transformations of the internal string spacetime, also represent inner
automorphisms of the finite-dimensional
Lie group $O(n,n)$. In this context the target space
duality group $O(n,n;\zed)$ is generated by the groups $G_{\rm Weyl}(n)$ and
$\zed_2$, where the Weyl group $G_{\rm Weyl}(n)$ of order $n$ is generated by
the spacetime coordinate permutations $\xi_\mu^{(\pi)}$, the $SL(n,\zed)$
generators $T_{\mu\nu}$, and the torsion cohomology shifts (\ref{betashift}),
while the reflection group $\zed_2$ is the inner automorphism of $O(n,n)$ that
permutes two coordinates at the point $\Lambda_0$ in the Narain moduli space
with identity background matrices. This is the usual description of the duality
group $O(n,n;\zed)$ as a spontaneously broken gauge symmetry of the toroidal
sigma-model. On the other hand, worldsheet parity $\zed_2$ cannot be
interpreted in terms of any gauge symmetry and represents a large
diffeomorphism of the noncommutative string spacetime.

Thus, as a start, we can identify the infinite-dimensional symmetry algebras
which contain the target space duality group $O(n,n;\zed)$ as inner
automorphisms at the fixed point $\Lambda_0\in{\cal M}_{\rm qu}$,
\beq
{\rm Inn}(\alg_{\Lambda_0})\supset\left(\widehat{SU(2)}_+^n\times\widehat
{SU(2)}_-^n\right)\otimes_{\rm S}\left({\rm Vir}^+\times{\rm Vir}^-\right)
\supset\left(\widehat{U(1)}_+^n\times\widehat{U(1)}_-^n\right)\otimes_{\rm S}
\left({\rm Vir}^+\times{\rm Vir}^-\right)
\label{innduality}\eeq
where ${\rm Vir}^\pm$ denotes the chiral Virasoro groups.
The abelian gauge symmetry group in \eqn{innduality} (the maximal torus of the
nonabelian one) is present at a generic point $\Lambda\neq\Lambda_0$. 
Furthermore, worldsheet parity is a finite subgroup of the outer automorphism
group of $\alg$ which contains a finite-dimensional worldsheet rotation group,
\beq
{\rm Out}(\alg)\supset O(2)
\label{outduality}\eeq

It is interesting to note what happens to these automorphisms when projected
onto the low-energy sector $\bar{\cal P}_0^{(-)}\bar{\cal A}_0\bar{\cal
P}_0^{(-)}$ representing the ordinary spacetime $T^n$. Only the inner
automorphisms (\ref{gencoordauto}) act non-trivially on this subalgebra of
$\alg$ and represent the generators of ${\rm Diff}(T^n)$ in terms of the
canonically conjugate center of mass variables $x^\mu,p_\mu$. The other
transformations when restricted to this subalgebra act as the identity $\id$,
i.e. as inner automorphisms. Thus, the subgroup (\ref{innduality}) of the {\em
inner} automorphism group of $\alg$, representing internal gauge symmetries of
the string spacetime, is projected onto the full group of {\em outer}
automorphisms of the low-energy target space $T^n$, corresponding to
diffeomorphisms of the manifold. This approach therefore naturally identifies
the usual invariance principles of general relativity as a {\it gauge symmetry}
of the stringy modification. The diffeomorphisms of the full string spacetime
are completely unobservable in the low-energy sector (for instance in the
anti-chiral projection onto $\bar{\cal H}_0^{(-)}$ the operator (\ref{WXouter})
acts as $I_n$), as are the gauge symmetries corresponding to the duality
transformations.

This is indeed another essence of the target space duality in string theory. It
is only observable as a symmetry of the huge noncommutative spacetime
represented by the full spectral triple (\ref{sigmatriple}) and acts trivially
on the corresponding low-energy projections representing the conventional
spacetimes. In fact, the duality automorphisms naturally partition the full
vertex operator algebra into sectors, each of which project onto the
various low-energy
spacetimes we described in the last section. Each such low-energy sector is
distinct at the classical level but related to the other ones by the duality
maps. At the level of the full spectral triple, there are two sectors
corresponding to the two eigenspaces of the duality maps as dictated by the
decomposition \eqn{eigendecompH}. For example, in the case of worldsheet
parity, the two eigenspaces consist of holomorphic and anti-holomorphic
combinations, respectively, of the chirality sectors of the vertex operator
algebra. In a low-energy projection, where the notion of chirality is absent,
the effects of duality are unobservable. Similar decompositions can also be
made for the larger symmetry groups in (\ref{innduality}) and
(\ref{outduality}) in terms of their irreducible representations.
Note that, given a subgroup $G$ of automorphisms, the
$G$-invariant subspace $\widehat{\cal H}_X^{(0)}(\Lambda)$ in
(\ref{eigendecompH}) (corresponding to the one-dimensional trivial
representation of $G$) defines a vertex operator subalgebra and hence leads to a
subspace of the string spacetime which is invariant under the
$G$-transformations. In particular, the corresponding low-energy subspace from
the decomposition with respect to a duality map is then completely unaffected
by the duality transformation. Thus, the above presentation of duality (and
other symmetries of the string spacetime) naturally leads to a systematic
construction of the low-energy projective subspaces that we presented in the
previous section.

\subsubsection*{Universal Gauge Groups and Monster Symmetry}

At present, the general structure of the unitary group of the vertex operator
algebra $\alg$ is not known, nor are its general automorphisms. The group
${\cal U}(\alg)$ represents the complete internal (gauge) symmetry group of the
noncommutative spacetime and appears to be quite non-trivial. Even at the
commutative level where $\alg=C^\infty(M)$, the unitary group is the
complicated, infinite-dimensional loop group $C^\infty(M,S^1)$ of the manifold
$M$. The inner automorphism group of the noncommutative string spacetime
includes spacetime diffeomorphisms, two copies of the Virasoro group, and the
Kac-Moody symmetry groups in (\ref{innduality}) which contain
the spacetime duality
symmetries. There are a number of additional infinite-dimensional subalgebras
of $\alg$ that have been identified as subspaces of the inner automorphism
algebra ${\rm inn}(\alg)$, such as the algebras of area-preserving ($W_\infty$)
and volume-preserving diffeomorphisms in $n=2$ dimensions \cite{wz} and also
the weighted tensor algebras described in \cite{wtalg}. In all of these
instances the inner automorphisms define appropriate mixings among the chiral
Dirac operators $\Dirac^\pm$ which preserve the conformal invariance of the
theory. Indeed, the chiral and conformal properties of the worldsheet theory
are, as we have extensively shown in this paper, crucial aspects of the string
spacetime.

A classification of ${\cal U}(\alg)$ would ultimately lead to a `universal
symmetry group' of string theory that would contain all unbroken gauge groups
and represent the true stringy symmetries of the quantum spacetime.
The problem with such a classification scheme though is that the ``size" of
${\cal U}(\alg)$ appears to be very sensitive to the lattice $\Lambda\in{\cal
M}_{\rm qu}$ from which the vertex operator algebra is built. A natural
Lie group $G_\Lambda$ of automorphisms arises from exponentiating the Lie
algebra ${\cal L}_\Lambda$ associated with the lattice $\Lambda$ \cite{flm}
(see the appendix). Then $G_\Lambda$ acts continuously and faithfully on the
vertex operator algebra $\alg$. The construction of ${\cal U}(\alg)$ has
been discussed by Moore in \cite{moore} who considered the appearance of
enhanced symmetry points in the Narain moduli space, i.e. points
$\Lambda\in{\cal M}_{\rm qu}$ at which extra dimension (1,0) and (0,1)
operators appear and generate new symmetries of the conformal field theory. For
this, we analytically continue in the spacetime momenta and extend the
lattice $\Lambda$ to a module over the Gaussian integers,
\beq
\Lambda^{\rm (G)}=\Lambda\otimes_\zeds\zed[i]\subset\Lambda^c
\label{Gaussints}\eeq
We then form the corresponding operator Fock space \eqn{vertexopalg} based on
$\Lambda^{\rm (G)}$ by
\beq
\widehat{\cal H}_X(\Lambda^{\rm(G)})={\bb C}\{\Lambda^{\rm(G)}\}\otimes S(\hat
h_+^{(-)})\otimes S(\hat h_-^{(-)})
\label{vertexopalgG}\eeq
so that the corresponding Lie algebra of dimension 1 primary fields
(\ref{primaryops}) is (see the appendix)
\beq
{\cal L}_U\equiv\widehat{\cal
P}_1(\Lambda^{\rm(G)})/~\mbox{$\bigcup_{k\geq1}$}\,\widehat{\cal
P}_1(\Lambda^{\rm(G)})\cap\left(L^+_{-k}\otimes L_{-k}^-\right)
\widehat{\cal H}_X(\Lambda^{\rm(G)})
\label{LUdef}\eeq
Moore proved that, since the action of
$O(n,n;\zed)$ on ${\cal M}_{\rm qu}$ is transitive, the Lie algebra
(\ref{LUdef}) generates a {\it universal} symmetry group of the string theory,
i.e. if ${\cal L}_\Lambda$ is the (affine) Lie algebra that appears at an
enhanced symmetry point $\Lambda$, then there is a natural Lie subalgebra
embedding ${\cal L}_\Lambda\hookrightarrow{\cal L}_U$. We refer to \cite{moore}
for the details.

We would like to stress that, from the point of view of the noncommutative
geometry formalism that we have discussed, not only is the interpretation of
duality symmetries as being part of some mysterious gauge group \cite{Tdualrev}
now clarified, but the Lie group generated by (\ref{LUdef}) now has a natural
geometrical description in terms of the theory of vertex operator algebras and
the noncommutative geometry of $\alg$. The Lie algebra ${\cal L}_U$ naturally
overlies all symmetries of the string spacetime obtained from marginal
deformations of the conformal field theory, and geometrically it contains many
of the internal rotational symmetries of the noncommutative geometry. This by
no means exhausts all of the inner automorphisms of $\alg$, but it provides a
geometric, universal way of identifying gauge symmetries. Note that, in
contrast to the low-energy subspaces which were determined by the tachyon
sector of the vertex operator algebra $\alg$, the universal gauge symmetries
are determined by the graviton sector of $\alg$.

To get an idea of how large the gauge group ${\cal U}(\alg)$ can be, it is
instructive to consider a specific example. We consider an $n=(25\,+\,1)$
dimensional toroidal spacetime defined by Wick rotating the target space
coordinate $X^{26}$. The change from Euclidean to hyperbolic
compactification lattices $\Gamma$ is well-known to have dramatic effects on
the structures of the corresponding vertex operator algebra \cite{flm,gebert}
and on the Narain moduli space \cite{moore}. Consider the unique 26-dimensional
even unimodular Lorentzian
lattice $\Gamma={\mit\Pi}_{25,1}$. It can be shown \cite{moore}
that $\Lambda_*={\mit\Pi}_{25,1}\oplus{\mit\Pi}_{25,1}\in{\cal M}_{\rm qu}$ is
the unique point in the Narain moduli space at which the vertex operator
algebra $\alg$ completely factorizes between its left and right chiral sectors,
\beq
\widehat{\cal H}_X(\Lambda_*)={\cal C}^+\otimes{\cal C}^-
\label{HXfactor}\eeq
where
\beq
{\cal C}^\pm=\complex\{{\mit\Pi}_{25,1}\}\otimes S(\hat h_\pm^{*(-)})
\label{chiralfact}\eeq
and $\hat h_\pm^*$ is the Heisenberg-Weyl algebra (\ref{heisenalg}) built on
$\Lambda_*$. The distinguished point $\Lambda_*\in{\cal M}_{\rm qu}$ is an
enhanced symmetry point and the corresponding Lie algebra ${\cal L}_
{\Lambda_*}$ is a
{\it maximal} symmetry algebra, in the sense that it contains all unbroken
gauge symmetry algebras. ${\cal L}_{\Lambda_*}$
is not, however, universal since the
gauge symmetries are not necessarily embedded into it as Lie subalgebras. Again
the framework of noncommutative geometry naturally constructs 
${\cal L}_{\Lambda_*}$ as a
symmetry algebra of the string theory.

The Lie algebra ${\cal L}_{\Lambda_*}={\cal B}\oplus{\cal B}$ is an example of a
mathematical entity known as a {\it Borcherds} or {\it generalized Kac-Moody}
algebra \cite{borcherds}, where
\beq
{\cal B}=\widehat{\cal P}_1({\mit\Pi}_{25,1})/\ker\langle\cdot,\cdot\rangle
\label{monster}\eeq
and $\langle\cdot,\cdot\rangle$ is the bilinear form on the Lie algebra of
primary fields of weight one defined in the appendix. The root lattice of $\cal
B$ is ${\mit\Pi}_{25,1}$ along with the set of positive integer multiples of
the Weyl vector $\vec\rho=(1,2,\dots,25;70)\in{\mit\Pi}_{25,1}$, each of
multiplicity 24. It is generated by $\varepsilon_{q^\pm}$,
$\varepsilon_{-q^\pm}$, $q_\mu^\pm\alpha_{-1}^{(\pm)\mu}$ (in each chiral
sector) and $\e^{m\vec\rho}$ where $q^\pm\in{\mit\Pi}_{25,1}$ and $m\in\zed$.
The first three of these generators span an infinite-dimensional Kac-Moody
algebra of infinite rank. The simple roots of $\cal B$ are the simple roots of
this Kac-Moody algebra, and the positive-norm simple roots of the lattice
${\mit\Pi}_{25,1}$ lie in the Leech lattice $\Gamma_{\rm Leech}$, which is the
unique 24-dimensional even unimodular Euclidean lattice with no vectors of
square length two. The symmetries of its Dynkin diagram can be classified
according to the automorphism group of the Leech lattice. The Lie algebra $\cal
B$ is called the {\it fake Monster Lie algebra} \cite{flm,gebert,borcherds}.

Thus the fake Monster Lie algebra (\ref{monster}) is
a maximal symmetry algebra of the string theory, so that Borcherds algebras,
when interpreted as generalized symmetry algebras of the noncommutative
geometry, seem to be relevant for the construction of a universal symmetry of
string theory. These algebras, being a natural generalization of affine Lie
algebras, may emerge as new symmetry algebras for string spacetimes within the
unified framework of vertex operator algebras and noncommutative geometry. The
fake Monster Lie algebra can also be used to construct the Monster Lie algebra
\cite{flm}, (see also \cite{harvey}). 
Mathematically, the most interesting aspect of this construction
is that a subgroup of the automorphism group of the Monster vertex operator
algebra is the celebrated Monster group, which is the full automorphism group
of the 196884-dimensional Griess algebra that is constructed from the Monster
Lie algebra and the moonshine module along the lines described above and in the
appendix. The Monster group is the largest finitely-generated simple sporadic
group.

The appearance of this Monster symmetry as a gauge symmetry of the
noncommutative spacetime emphasizes the point that these exotic mathematical
structures, such as those contained in the content of Borcherds algebras, might
play a role as a sort of dynamical Lie algebra which changes the Dirac
operators and the Fock space gradings. But the underlying noncommutative
geometrical structure of the string spacetime remains unchanged. We know of no
complete classification of such vertex operator algebra automorphisms, and, in
the context of this paper, this remains an important problem to be carried out
in order to understand the full set of geometrical symmetries that underlie the
stringy modification of classical general relativity.

\subsubsection*{Differential Topology of the Quantum Spacetime}

As a final example of the formalism developed in this paper, we look briefly at
the problem of computing the cohomology groups of the noncommutative string
spacetime and compare them with the known (DeRham) cohomology groups of the
ordinary $n$-torus
\beq
H^k(T^n;\real)=\left\{\new{\begin{array}{l}\real^{n\choose k}~~~~{\rm
for}~~0\leq k\leq n\\\{0\}~~~~~{\rm otherwise}\end{array}}\right.
\label{toruscoh}\eeq
The rigorous way to describe the non-metric aspects of noncommutative geometry
is through noncommutative K-theory \cite{Book}, but we shall not enter into
this formalism here. Here we shall simply compute the cohomology groups in
analogy with the example of a manifold that was presented in section 2. This 
approach is based on a natural generalization of the Witten complex of section 4
which describes the cohomology \eqn{toruscoh}.

We assume, for simplicity, that $n$ is even and that a basis for the
compactification lattice $\Gamma$ has been chosen so that
$g_{\mu\nu}=\delta_{\mu\nu}$. In light of our analysis above, no loss of
generality occurs with this choice of point in the Narain moduli space. In that
case the ${\rm spin}(n)$ Clifford algebras ${\cal C}(T^n)^\pm$ each possess a
chirality matrix
\beq
\gamma_c^\pm=\gamma_1^\pm\gamma_2^\pm\cdots\gamma_n^\pm
\label{chirality}\eeq
whose actions on the generators of the spin bundle are
\beq
\left\{\gamma_c^\pm,\gamma_\mu^\pm\right\}=0~~~~~~,~~~~~~\left[\gamma_c^\pm,
\gamma_\mu^\mp\right]=0
\label{ggccomms}\eeq
The chirality matrices are of order 2, $(\gamma_c^\pm)^2=\id$.

Our first observation is that the two Dirac operators in 
(\ref{DDbar}) are related
by
\beq
\bar\Dirac=\gamma_c^-\,\Dirac\,\gamma_c^-
\label{DDbargamma}\eeq
As we shall see below, the chirality operators (\ref{chirality}) define a Klein
operator $\tilde\gamma$, which provides a natural $\zed_2$-grading, 
and a Hodge duality operator $\star$
by\footnote{For a definition of the Hodge duality operator in a more
general setting which can be applied to the cases where $n$ is odd, see
\cite{fgr}.}
\beq
\tilde\gamma=\gamma_c^+\gamma_c^-~~~~~~,~~~~~~\star=\gamma_c^-
\label{Z2stardefs}\eeq
Recalling the description of section 2 and the discussion in section 3
concerning the Witten complex, we can identify the holomorphic Dirac operator
$\Dirac$ as an exterior derivative operator d. According to (\ref{DDbargamma})
and (\ref{Z2stardefs}), the anti-holomorphic Dirac operator $\bar\Dirac$ can
then be identified with the co-derivative ${\rm d}^\dagger=\star{\rm d}\star$.
The duality isomorphisms of section 6 then state that the string spacetime is
invariant under the exchange between the exterior derivative and its dual ${\rm
d}\leftrightarrow{\rm d}^\dagger=\star{\rm d}\star$, which is another
well-known characterization of target space duality in string theory.

We can now proceed to construct the complex of differential forms of the
noncommutative string spacetime as described in section 2. As always our
starting point is
\beq
\Omega_\Dirac^0\alg=\alg
\label{Om0}\eeq
Next, we can compute, for $V=\id\otimes V_\Omega\in\alg$, the exact one-form
\beq
\pi_\Dirac(dV)=[\Dirac,V]
\label{DrepdV}\eeq
Since the vertex operators of definite spacetime momentum $(q^+,q^-)\in\Lambda$
span the vertex operator algebra $\alg_X$, it follows that all commutators of
the form $[J_\pm^\mu,V_\Omega]$ sweep out the space $\alg_X$ as $V_\Omega$
is varied,
i.e. the Dirac operator $\Dirac$ acts densely on $\cal H$, just as
in the commutative case of section 2. Using the explicit form of the Dirac
operator we can thus identify the linear space of differential one-forms as
\beq
\Omega_\Dirac^1\alg=\alg\otimes_\reals\left({\cal C}(T^n)^+\oplus{\cal
C}(T^n)^-\right)~~~~~~,~~~~~~\dim_\alg\Omega_\Dirac^1\alg=2n
\label{Om1}\eeq
with basis $\{\gamma_\mu^+,\gamma_\mu^-\}_{\mu=1}^n$. Similarly, one can
proceed to calculate the higher-degree spaces $\Omega_\Dirac^k\alg$ just as in
the example of a spin-manifold described in section 2. As occurred there, we
will encounter junk forms for $k\geq2$, which can be eliminated by
antisymmetrizations of the gamma-matrices in each chiral sector. Since the left
and right chiral sector gamma-matrices already anticommute, this need not be
done for mixed chirality products of the $\gamma$'s. We therefore arrive at
\beq
\Omega_\Dirac^k\alg=\alg\otimes_\reals\left(\bigoplus_{i=0}^k{\cal
C}(T^n)^{+[i]}\otimes{\cal
C}(T^n)^{-[k-i]}\right)~~~~,~~~~\dim_\alg\Omega_\Dirac^k\alg=\sum_{i=0}^k
{n\choose i}{n\choose k-i}
\label{Omk}\eeq
where ${\cal C}(T^n)^{\pm[j]}$ is the linear space spanned by the
antisymmetrized products
$\gamma_{[\mu_1}^\pm\cdots\gamma_{\mu_j]}^\pm=\frac1{j!}\sum_{\pi\in S_j}{\rm
sgn}~\pi\prod_{l=1}^j\gamma_{\mu_{\pi(l)}}^\pm$.

The linear space (\ref{Omk}) is defined for all $0\leq k\leq n$. What is
interesting about the noncommutative differential complex is that, unlike that
of the torus $T^n$, forms of degree higher than $n$ exist. To construct
$\Omega_\Dirac^l\alg$ for $l>n$, we exploit the interpretation of the chirality
matrices above as Hodge duality operators. However, on the differential
complex, we use a slightly different representation than that given in
(\ref{Z2stardefs}) to avoid forms of negative degree. This is simply a matter
of convenience, and the entire differential topology can be instead given using
the Hodge dual in (\ref{Z2stardefs}). Thus we take
\beq
\pi_\Dirac(\star)=m_{\cal C}\circ\tilde\gamma
\label{Omstardef}\eeq
where $m_{\cal C}$ is the multiplication operator on the double Clifford
algebra ${\cal C}(T^n)$. If one now proceeds to construct differential
$l$-forms by antisymmetrizations of products of $l$ gamma-matrices
$\gamma_\mu^\pm$ for $l>n$, it is straightforward to see that
\beq
\Omega_\Dirac^l\alg\cong\tilde\gamma\cdot\Omega_\Dirac^{2n-l}\alg\cong\Omega_
\Dirac^{2n-l}\alg~~~~~~{\rm for}~~l>n
\label{Oml}\eeq
This process will terminate at $l=2n$, so that the algebra of differential
forms of the noncommutative spacetime is
\beq
\Omega_\Dirac^*\alg=\bigoplus_{k=0}^{2n}\Omega_\Dirac^k\alg
\label{diffcomplex}\eeq
The action of the Dirac operator
$\Dirac$ as defined in (\ref{DrepdV}) gives a nilpotent
linear map $d:\Omega_\Dirac^k\alg\to\Omega_\Dirac^{k+1}\alg$ with
$d(\Omega_\Dirac^{2n}\alg)=\{0\}$, while that of $\bar\Dirac$ defined by
\eqn{DrepdV} and \eqn{Omstardef} gives the adjoint nilpotent map
$d^\dagger=\star d\star:\Omega^k_\Dirac\alg\to\Omega_\Dirac^{k-1}\alg$ with
$d^\dagger(\Omega_\Dirac^0\alg)=\{0\}$.

Thus the chirality structure of the worldsheet theory leads to a ``doubling" in
the differential complex of the string spacetime. Since the $\alg$-module
$\Omega_\Dirac^1\alg$ is finitely-generated and projective, it can be viewed as
a cotangent bundle, and one can proceed to equip it with connections, although
at the level of a toroidal target space there is really no modification from
the classical (commutative) case. Since $\Omega^1_\Dirac\alg$ is free with
basis given by $\{\gamma_\mu^\pm\}_{\mu=1}^n$, we can define a connection
$\nabla_\Dirac:\Omega_\Dirac^1\alg\to\Omega_\Dirac^1\alg\otimes_\alg
\Omega_\Dirac^1\alg$ by \cite{FG,gravityNCG}
\beq
\nabla_\Dirac\,\omega=[\Dirac,\omega]
\label{connection}\eeq
Thus $\{\gamma_\mu^\pm\}_{\mu=1}^n$ constitutes a parallel basis for
$\Omega_\Dirac^1\alg$, i.e. $\nabla_\Dirac(\gamma_\mu^\pm\otimes\id)=0$, and
$\nabla_\Dirac$ has vanishing curvature. Thus the string spacetime is a
noncommutative space with flat connections of zero torsion, and the curvature
properties of the toroidal general relativity are unchanged by stringy effects.

To compute the cohomology ring of the noncommutative spacetime, we define the
cohomology group $H_\Dirac^k(\alg)$ to be the linear space spanned by the
harmonic differential $k$-forms, i.e. the $k$-forms annihilated by both
$\Dirac$ and $\bar\Dirac$ in the representation $\pi_\Dirac$ defined above. For
instance, the harmonic zero-forms are the vertex operators $V\in\alg$ with
\beq
[\Dirac,V]=0
\label{harm0}\eeq
so that
\beq
H^0_\Dirac(\alg)\cong{\rm comm}~\Dirac
\label{H0}\eeq
The situation for the higher degree cohomology groups is similar, except that
now one obtains higher-dimensional spaces corresponding to the global string
oscillations around the cycles of $T^n$. After some calculation, we find
\beq
H_\Dirac^k(\alg)\cong\left\{\new{\begin{array}{ll}{\rm comm}~\Dirac&~~~~{\rm
for}~~k=0\\\alg_{\Dirac,\bar\Dirac}\otimes_\reals\real^{\dim_\alg
\Omega_\Dirac^k\alg}&~~~~{\rm for}~~0<k<2n\\{\rm comm}~\bar\Dirac&~~~~{\rm
for}~~k=2n\\\{0\}&~~~~{\rm otherwise}\end{array}}\right.
\label{Hk}\eeq
where
\beq
\alg_{\Dirac,\bar\Dirac}\equiv{\rm comm}~\Dirac\cap{\rm comm}~\bar\Dirac
\label{ADDbar}\eeq
and the dimension of $\Omega_\Dirac^k\alg$ is given in (\ref{Omk}) and by
(\ref{Oml}).

The intermediate cohomology groups for $0<k<2n$ are characterized by vertex
operators $V\in\alg$ with
\beq
[\Dirac,V]=[\bar\Dirac,V]=0
\label{harmk}\eeq
These harmonic $k$-forms are the vertex operators which constitute `isometries'
of the string spacetime. From section 6, it follows that
$\alg_{\Dirac,\bar\Dirac}$ contains smeared tachyon vertex operators
$\id\otimes V(q^+,q^-)$ with $q^+=q^-=0$. In the low-energy projection onto
$\bar{\cal H}_0^{(-)}$, one finds that $\bar{\cal
P}_0^{(-)}\alg_{\Dirac,\bar\Dirac}\bar{\cal P}_0^{(-)}\cong\complex$. But there
are still higher-spin vertex operators of zero charge that survive in
(\ref{ADDbar}) in the general case. This is wherein most of the stringy
modification of the topology of the classical spacetime lies, in that
higher-spin oscillatory modes of the strings ``excite" the cohomology groups
(\ref{toruscoh}) leading to generalized, infinitely-many connected components
in the string spacetime. The spaces (\ref{Hk}) essentially represent the vertex
operators which are invariant under the global $U(1)_+^n\times U(1)_-^n$
Kac-Moody gauge symmetry of the string theory, and as such they represent the
globally diffeomorphism-invariant spacetime observables of the noncommutative
geometry. Explicit calculations can eliminate potential vertex operators from
belonging to (\ref{ADDbar}), for instance the graviton field (\ref{gravop}). 
This space consists of those states which belong to the simultaneous
zero-mode
eigenspaces of the two Dirac operators $\Dirac$ and $\bar\Dirac$. At
this stage though we have not found any elegant way of characterizing the
cohomology (\ref{Hk}) and it would be interesting to explore these spaces
further.

The cohomology for $k=0$ is determined by the vertex operators of zero winding
number but non-zero spacetime momentum, and vice-versa for $k=2n$. The number
of independent ``$k$-cycles" is larger in general than $n\choose k$ because in
the generic high-energy sector the string spacetime distinguishes between
chirality combinations and accounts for string oscillations and windings
about the circles of
$T^n$.  Using the $\zed_2$-grading $\tilde\gamma$ (Klein operator) and the
Hodge dual $\star$ defined in \eqn{Z2stardefs}, it is also possible to compute
topological invariants, such as the Euler characteristic and the Hirzebruch
signature, of the noncommutative geometry, in analogy to the Witten complex
\cite{fgrleshouches,fgr,witten1}.

Notice that in a low-energy projection the cohomology groups (\ref{Hk}) do not
coincide with (\ref{toruscoh}). One needs to first project the complex
\eqn{diffcomplex} and Dirac operators and {\it then} compute the cohomology
groups along the lines described in section 2. The key feature to this is that
then the chirality sectors of the spaces \eqn{Omk} become equivalent, and for
$l>n$ the spaces \eqn{Oml} ``fold" back onto the $n$ linear spaces in \eqn{Omk}.
In the general case the cohomology (\ref{Hk}) leads immediately to the
mirror symmetry of the string spacetime. We can naturally define ``Dolbeault"
cohomology groups $H_\Dirac^{k,l}(\alg)$ by using the chiral and anti-chiral
Clifford algebra decompositions in (\ref{Omk}) to split the spaces (\ref{Hk})
into holomorphic and anti-holomorphic combinations. Using the chirality matrix
we then have $\gamma_c^-\cdot H^{k,l}_\Dirac(\alg)\cong
H_\Dirac^{k,n-l}(\alg)$. Comparing the respective projections onto $\bar{\cal
H}_0^{(-)}$ and ${\cal H}_0^{[+;\mu]}$ as described in section 6 leads
immediately to the usual statement of mirror symmetry between the ``Dolbeault"
cohomology groups arising from ``foldings" in \eqn{Hk}.

Of course this analysis is only meant to be somewhat heuristic since,
as mentioned before, a complete analysis of the topological properties
of a noncommutative spacetime entails more sophisticated
techniques. But the above results show how the algebraic structures
inherent in the vertex operator algebra modify the geometry and
topology, as well as the general symmetry principles of general
relativity. More non-trivial structures could be displayed by the
Wess-Zumino-Witten (WZW) models studied in \cite{FG,fgrleshouches},
and even in the generalizations of the conformal field theory
\eqn{sigmaaction} to arbitrary compact Riemann surfaces $\Sigma$ or to
embedding fields $X^\mu$ which live in toroidal orbifold target
spaces. The methods emphasized in this paper can be more or less
straightforwardly extended to the analysis of such string theories.

\bigskip

\noindent
{\bf Acknowledgements:} We thank I. Giannakis for pointing out previous work on
vertex operator algebra automorphisms in string theory to us, and G. Mason for
comments on the manuscript. {\sc f.l.} gratefully thanks the Theoretical
Physics Department of Oxford University for hospitality during the course of
this work. The work of {\sc r.j.s.} was supported in part by the Natural
Sciences and Engineering Research Council of Canada.

\setcounter{section}{0}
\setcounter{subsection}{0}
\setcounter{equation}{0}
\renewcommand{\thesection}{Appendix}
\renewcommand{\theequation}{\Alph{section}.\arabic{equation}}

\newsection{Properties of Vertex Operator Algebras and Construction of
Quantum Spacetimes}

In string theory, vertex operators generate the algebra of observables of the
underlying two-dimensional conformal quantum field theory whose action on the
vacuum state $|{\rm vac}\rangle$ forms a dense subspace of vectors of the
corresponding Hilbert space $\cal H$ of physical states. The (anti-)chiral
algebra ${\cal E}^+$ (${\cal E}^-$) is defined to be the operator product
algebra of the (anti-)holomorphic fields in the conformal field theory. The
chiral algebras contain, in particular, copies of the Virasoro algebras
characterizing the conformal invariance of the string theory. A
rational conformal field theory is completely characterized by its chiral
algebra, and thus the classification problem of rational conformal field
theories can be reduced to that of vertex operator algebras.

Generally, in a two-dimensional conformal field theory we can always choose a
basis of primary operators $\phi_i$ of fixed conformal weights $\Delta_i^\pm$
and normalize their 2-point functions as
\beq
\langle{\rm vac}|\phi_i(z_+,z_-)\phi_j(w_+,w_-)|{\rm
vac}\rangle=\delta_{ij}~(z_+-w_+)^{-2\Delta_i^+}(z_--w_-)^{-2\Delta_i^-}
\label{2ptfnnorm}\eeq
The principles of conformal invariance suggest that the usual Wilson operator
product expansion of primary fields should converge, rather than just
representing a formal asymptotic expansion. For operators of fixed scaling
dimensions, one can define the (constant) operator product expansion
coefficients $C_{ijk}$ by
\beq
\phi_i(z_+,z_-)\phi_j(w_+,w_-)=\sum_kC_{ijk}~(z_+-w_+)^{\Delta_k^+-\Delta_i^+
-\Delta_j^+}(z_--w_-)^{\Delta_k^--\Delta_i^--\Delta_j^-}\phi_k(w_+,w_-)
\label{opephi}\eeq
for $z_\pm\to w_\pm$, where the sum runs over a complete set of primary fields.
The coefficients $C_{ijk}$ are then symmetric in $i,j,k$.

However, it is well-known that the operator product expansion is a consequence
of some more elementary relations among the vertex operators. A primary field
can be decomposed
\beq
\phi_i(z_+,z_-)\equiv\sum_{k,l}D_i^{kl}~\varphi_k^+(z_+)\otimes\varphi_l^-(z_-)
\label{sewing}\eeq
in terms of chiral and anti-chiral vertex operators $\varphi_k^\pm(z_\pm)$
which span ${\cal E}^\pm$. If ${\bf h}^\pm$ denotes the Hilbert spaces on which
${\cal E}^\pm$ act densely, then the local fields $\phi_i(z_+,z_-)$ act as
operator-valued distributions from the Hilbert space ${\cal
H}=\complex^{\{D\}}\otimes{\bf h}^+\otimes{\bf h}^-$ onto itself, where
$\complex^{\{D\}}$ is the multiplicity space which labels the different
left-right sewing determined by the complex-valued sewing coefficients
$D_i^{kl}$. After smearing they become well-defined and densely-defined
operators on ${\cal H}$. The chiral and anti-chiral vertex operators obey the
{\it braiding relations}
\beq
\varphi_i^\pm(z_\pm)\varphi_j^\pm(w_\pm)=\sum_{k,l}(R^\pm)_{ij}^{kl}
{}~\varphi_k^\pm(w_\pm)\varphi_l^\pm(z_\pm)
\label{braidrel}\eeq
when the points $z_\pm$ and $w_\pm$ are interchanged along some paths on
the worldsheet $\Sigma$, where
$(R^\pm)_{ij}^{kl}$ are braiding matrices. An example of the braiding
commutation relations are the local cocycle relations between the vertex
operators discussed in section 5. Furthermore, the vertex operators obey the
{\it fusion equations}
\beq
\varphi_i^\pm(z_\pm)\varphi_j^\pm(w_\pm)=\sum_{k,l}
(F^\pm)_{ij}^{kl}~\varphi_k^\pm(w_\pm)\circ\varphi_l^\pm(z_\pm-w_\pm)
\label{fusioneq}\eeq
where the composition of operators on the right-hand side of (\ref{fusioneq})
refers to their action on the Hilbert spaces ${\bf h}^\pm$, and
$(F^\pm)_{ij}^{kl}$ are fusion matrices. These relations then immediately lead
to the operator product expansion (\ref{opephi}) in which the coefficients
$C_{ijk}$ can be determined in terms of the sewing, braiding and fusion
coefficients introduced above.

The above relations can be further shown to lead to the property of {\it
locality}, i.e. that the quantum fields $\phi_i(\tau,\sigma)$ and
$\phi_j(\tau',\sigma')$ commute whenever their arguments are space-like
separated, and also the property of {\it duality}, i.e. crossing-symmetry of
the 4-point functions,
\beq
\sum_pC_{ijp}C_{pkl}=\sum_pC_{ipl}C_{jkp}
\label{duality4pt}\eeq
This identity immediately implies the associativity of the operator product
algebra (\ref{opephi}). The identities that we have presented here can be
expressed in terms of a single, compact relation by turning to the formal
notion of a Vertex Operator Algebra. This single identity is known as the
`Jacobi identity' and it encodes the full non-triviality of the structure of
the Vertex Operator Algebra (as the above relations do for the local conformal
field theory). The standard discussion above can be cast into such formal
form that is useful in the more algebraic applications of conformal field
theory, such as that required in the noncommutative geometry of string
spacetimes. In the remainder of
this appendix we will define formally a Vertex Operator Algebra and describe
some of its algebraic properties in the context of the construction of quantum
spacetimes. We will be very sketchy, and for more details the
reader is invited to consult, for example, the introductory reviews
\cite{huang,gebert}, the book \cite{flm}, and references therein.

\subsubsection*{Axiomatics}

A {\it vertex operator algebra} consists of a $\zed$-graded complex vector
space
\beq
{\cal F}=\bigoplus_{n\in\zeds}{\cal F}_n
\eeq
and a linear map ${\cal V}$ which associates to each element $\Psi\in{\cal
F}$ an endomorphism of ${\cal F}$ that can be expressed as a formal sum in a
variable $z$:
\beq
{\cal V}(\Psi,z)=\sum_{n\in\zeds}\Psi_n~z^{-n-1}
\label{VendPsi}\eeq
with $\Psi_n\in{\cal F}_n$. The element $\Psi\in{\cal F}$ is called a {\em
state} and the endomorphism ${\cal V}(\Psi,z)$ is called a {\em vertex
operator}. The vertex operators ${\cal V}(\Psi,z)$ are required to satisfy the
following axioms:

\begin{enumerate}

\item Given any $\Phi\in{\cal F}$ we have:
\beq
\Psi_n\Phi=0 \ ~~~~\mbox{for $n$ sufficiently large.}
\eeq

\item There is a preferred vector {\bf 1} called the {\em vacuum} such that
\beq
{\bf 1}_n=\delta_{n+1,0}
\eeq
and therefore ${\cal V}({\bf 1},z)\Phi=\Phi,\ \forall\Phi\in{\cal F}$.

\item $\Psi_n=0\ \forall n\in\zed \Longleftrightarrow \Psi=0$.

\item There exists a {\em conformal vector} 
whose component operators
$T_{n+1}=L_n$
satisfy the Virasoro Algebra \eqn{viralg} for some central charge
$c\in\complex$. This vector provides as well a translation generator:
\beq
{\cal V}(L_{-1}\Psi,z)={d\over dz}{\cal V}(\Psi,z)
\eeq
and the grading of ${\cal F}$:
\beq
L_0\Psi_n=n\Psi_n\ ~~~~\forall\Psi_n\in{\cal F}_n\ .
\label{L0grade}\eeq

\item The spectrum of $L_0$ is bounded from below.

\item The eigenspaces ${\cal F}_n$ of $L_0$ are finite-dimensional.

\item The vertex operators must also satisfy a {\em Jacobi identity}:
\beq
\sum_{i\geq 0}(-1)^i{l \choose i}\left(\Psi_{l+m-i}(\Phi_{n+i}\Xi)-
(-1)^l\Phi_{l+n-i}(\Psi_{m+i}\Xi)\right)=\sum_{i\geq 0}{m \choose i}(\Psi_{l+i}
\Phi)_{m+n-i}\Xi \label{Jacobi}
\eeq
for all $\Psi,\Phi,\Xi\in{\cal F},\ l,m,n\in\zed$.

\end{enumerate}

The axioms 1.--7. define a Vertex Operator Algebra, which contains the Virasoro
algebra (Axiom 4.). The Jacobi identity \eqn{Jacobi} contains the most
information about the algebra. Three special cases of it are
particularly interesting. They represent associativity:
\beq
(\Psi_m\Phi)_n=\sum_{i\geq 0}(-1)^i{m \choose i}\left(\Psi_{m-i}\Phi_{n+i}-
                (-1)^m\Phi_{m+n-i}\Psi_i\right)~~~, \label{associativity}
\eeq
the commutator formula:
\beq
[\Psi_m,\Phi_n]=\sum_{i\geq 0}{m \choose i}(\Psi_i\Phi)_{m+n-i}~~~,
\eeq
and skew-symmetry:
\beq
\Psi_n\Phi=(-1)^{n+1}\Phi_n\Psi+\sum_{i\geq1}\frac1{i!}~(-1)^{i+n+1}
(L_{-1})^i(\Phi_{n+i}\Psi)
\label{skewsymm}\eeq
for all $\Psi,\Phi\in{\cal F},\ m,n\in\zed$. The Jacobi identity therefore
encodes the complete noncommutativity (as well as
other nontrivial properties) of the vertex operator algebra.

\subsubsection*{Adjoint}

The adjoint of a vertex operator is defined as:
\beq
{\cal V}^\dagger(\Psi,z)=\sum_{n\in\zeds}\Psi_n^\dagger~z^{-n-1}\equiv
{\cal V}(\e^{zL_1}(-z^2)^{L_0}\Psi,z^{-1})
\eeq
It can be shown \cite{gebert} 
that with this definition $L_n^\dagger=L_{-n}$. This
yields the $*$-conjugation on the vertex operator algebra.

\subsubsection*{Translation and Scaling}

The Virasoro generators $L_{-1}$ and $L_0$ generate translations and
scale transformations, respectively:
\beq\new{\begin{array}{rrl}
\e^{wL_{-1}}\,{\cal V}(\Psi,z)~\e^{-wL_{-1}}&=&{\cal V}(\Psi,w+z)\\
\e^{wL_0}\,{\cal V}(\Psi,z)~\e^{-wL_0}&=&\e^{w\Delta_\Psi}\,{\cal
V}(\Psi,\e^wz)\end{array}}
\eeq
where $\Delta_\Psi$ is the conformal weight of the vector $\Psi$ (defined by
\eqn{L0grade}).

\subsubsection*{Conformal Highest Weight Vectors}

Vertex operators generate states when applied to the vacuum 
(this is the original motivation for their name):
\beq
{\cal V}(\Psi,z){\bf 1}=\e^{zL_{-1}}\Psi
\eeq
In general, the conformal highest weight vectors (or primary fields)
are states which satisfy:
\beq
L_0\Psi=\Delta_\Psi\Psi~~~~~~,~~~~~~L_n\Psi=0~~\forall n>0
\eeq
and in particular the vacuum is a primary state of weight zero.

\subsubsection*{Algebra of Primary Fields of Weight One}

The primary fields of weight one form a Lie algebra
\beq
{\cal L}\equiv{\cal F}_1/({\cal F}_1\cap L_{-1}{\cal F})={\cal F}_1/L_{-1}{\cal
F}_0
\label{Liealg1}\eeq
with antisymmetric bracket:
\beq
[\Psi,\Phi]\equiv\Psi_0\Phi
\label{wt1lie}\eeq
and $\cal L$-invariant bilinear form:
\beq
\langle\Psi,\Phi\rangle\equiv\Psi_1\Phi
\label{wt1bilin}\eeq
provided that the spectrum of $L_0$ is non-negative and the weight zero
subspace ${\cal F}_0$ is one-dimensional. The former constraint is
usually imposed out of physical considerations, as often $L_0$ is identified
with the Hamiltonian of the system. The quotient of ${\cal F}_1$ in
(\ref{Liealg1}) is by the set of {\it spurious} states. 
The classical Jacobi identity
for the Lie bracket (\ref{wt1lie}) follows from the Jacobi identity for the
vertex operator algebra, while the symmetry of the inner product
(\ref{wt1bilin}) follows directly from the skew-symmetry property.

Setting $\Psi=\Phi=\bf1$ in \eqn{Jacobi}, using the vacuum axiom 2. and the
definition \eqn{VendPsi}, we obtain the usual Cauchy theorem of classical
complex analysis. Thus the Jacobi identity for vertex operator algebras is a
combination of the classical Jacobi identity for Lie algebras and the Cauchy
residue formula for meromorphic functions. 
It is also possible to define, in certain
instances, more exotic subalgebras using the Jacobi identity, such as a
commutative non-associative algebraic structure
$\Psi\times\Phi\equiv\Psi_1\Phi$ on the space of fields of weight two, and a
commutative associative product $\Psi\cdot\Phi\equiv\Psi_{-1}\Phi$ 
as well as the Lie bracket \eqn{wt1lie} on the
quotient space ${\cal F}/{\cal F}_{-2}{\cal F}$.

\subsubsection*{Vertex Operators and Even Lattices}

One of the most important results in the theory of vertex operator
algebras (and the one of great relevance to the present work) is the theorem
which states that \cite{flm,dong1}:

\noindent
{\em Associated with any even positive-definite lattice $\Gamma$
there is a vertex operator algebra.}

\noindent
The proof of the theorem is a constructive procedure. The
construction is in fact the one which associates a bosonic string
theory compactified on a torus, as we did in section 5. In
this paper we had a Fock space as our starting point. In general this is
not necessary, and the Fock space can actually be constructed starting
from the lattice $\Gamma$, so that the lattice is the only ingredient
necessary for the construction. The proof that the algebra one
constructs is indeed a vertex operator algebra, as well as the details of the
formal construction, can be found in \cite{flm,gebert}.

In the case of the vertex operator algebra of section 5, the algebra one
constructs is actually the chiral algebra ${\cal E}^\pm$, the endomorphisms on
$\complex\{\Gamma\}^\pm\otimes S(\hat h^{(-)})$ (respectively on $\Gamma^*$).
The full vertex operator
algebra is then constructed using the sewing transformation (\ref{vertexops}).
For the chiral algebras, the Jacobi identity follows from the analogous
braiding relations (\ref{opprodsexp}) which lead to the fusion relations
\cite{flm}
\beq\new{\begin{array}{lll}
{\cal V}\left({\cal V}({\cal V}_{[q^\pm]}^{(R)}[z_\pm],z_\pm){\cal
V}_{[r^\pm]}^{(S)}[w_\pm],w_\pm\right)&=&\prod_{1\leq(i,j)\leq(R,S)}
\left(z_\pm+(z_\pm^{(i)}-w_\pm^{(j)})\right)^{g^{\mu\nu}q_\mu^{(i)\pm}
r_\nu^{(j)\pm}}\\& &~~~\times~:~{\cal V}({\cal
V}_{[q^\pm]}^{(R)}[z_\pm],z_\pm+w_\pm){\cal V}({\cal
V}_{[r^\pm]}^{(S)}[w_\pm],w_\pm)~:\end{array}}
\label{chiralopexp}\eeq
By the completeness of the chiral vertex operators ${\cal
V}_{[q^\pm]}^{(R)}[z_\pm]$ on the respective Hilbert spaces, the relation
(\ref{chiralopexp}) leads immediately to the Jacobi identity for the chiral
vertex operator algebras ${\cal E}^\pm$.

Note that self-duality $\Gamma=\Gamma^*$ is not a requirement. In the general
case, we have the coset decomposition 
\beq
\Gamma^*=\bigcup_{x\in\Gamma^*/\Gamma}\left(\Gamma+\lambda_x\right)
\label{cosetdecomp}\eeq
of the dual lattice with $\lambda_0=0$. It can be shown \cite{flm,dong1} that
$\{R_x~|~x\in\Gamma^*/\Gamma\}$ is the complete set of inequivalent irreducible
representations of the vertex operator algebra, where
\beq
R_x=\complex\{\Gamma+\lambda_x\}\otimes S(\hat h^{(-)})
\label{Gammareps}\eeq
Then $\widehat{\cal H}_X(\Gamma^*)=\bigoplus_{x\in\Gamma^*/\Gamma}R_x$,
and the representations (\ref{Gammareps}) can be identified with the 
``points" of
a noncommutative spacetime. Furthermore, the corresponding linear space of
characters $\phi_{R_x}\equiv{\rm tr}_{R_x}q^{L_0-c/24}$, 
where $q=\e^{\pi i\tau}$
with ${\rm Im}~\tau>0$, is modular invariant with respect to $SL(2,\zed)$
\cite{zhu}. If the lattice
$\Gamma$ is self-dual then the vertex operator algebra is {\it holomorphic},
i.e. it is its only irreducible representation. This shows how the symmetries of
$\Gamma$ control the structure of the associated spacetime.

In the cases studied throughout this paper, we can now see how the structure of
the quantum spacetime is intimately tied to the properties of the 
compactification
lattice. Those lattices associated to large symmetries of the spacetime, such as
that associated with the Monster group, give smaller quantum spacetimes that
those associated to non-symmetrical lattices, i.e. large gauge 
symmetries essentially exhaust the full structure of the spacetime. 
This increase 
in symmetry of the spacetime from a decrease in the number of its ``points"  is 
similar to the effect of increasing the number of elements of an algebra to gain
an decrease in the number of points of a topological space (see section 2).
Furthermore, given a compact automorphism group $G$ of the vertex operator
algebra as in section 7, each $G$-module $\widehat{\cal
H}_X^{[R(G)]}(\Lambda)$ is an irreducible representation of the $G$-invariant
subalgebra $\widehat{\cal H}_X^{(0)}(\Lambda)$ \cite{dong2}. 
This exemplifies, in
particular, how the construction of the commutative low-energy projective
subspaces carries through from the structure of the vertex operator algebra.
Moreover, the above results show that
more general theories than the ones we have presented in this paper
will also lead to vertex operator algebras, and thus
similar noncommutative spacetimes. For example, the allowed momenta and winding
modes of heterotic string theory live 
on an $(n+16,n)$-dimensional even self-dual
Lorentzian lattice \cite{narain}, and the construction of section 6 can be used
to show that the target space duality group in this case is isomorphic to
$O(n+16,n;\zed)$ \cite{Tdualrev}.

In this general class of vertex operator algebras, ${\cal F}_1$ is a Lie algebra
with generators $\varepsilon_q$ and $q_\mu\alpha^\mu_{-1}$, where
$q\in\Gamma$. Its root lattice is precisely the
lattice $\Gamma$, and the affinization of ${\cal F}_1$ then yields the usual
Frenkel-Kac construction of affine Lie algebras \cite{go}.  Note that, as a
subspace of the noncommutative spacetime, the subalgebra ${\cal F}_1$
contains the lowest non-trivial oscillatory modes of the strings, so that the
universal gauge symmetry of the string spacetime coincides with smallest
excitations of the commutative subspaces. It would be interesting to further
give a spacetime interpretation to the commutative structure of the quotient
space ${\cal F}/{\cal F}_{-2}{\cal F}$. Thus, more general
spacetime gauge symmetric structures can also be inputed into the constructions
presented in this paper (leading to analogs of the results of section 7), so
that our results extend naturally to a larger class of models than just the
linear sigma-models described here. The key underlying feature is always a
vertex operator algebra which leads to a horribly complicated quantum
spacetime.

\subsubsection*{Vertex Operators in other Conformal Field Theories}

At present the only other known class of examples of vertex operator algebras
arise from a restricted subalgebra of observables of WZW
conformal field theories (equivalently conformal field theory in a general
group manifold target space). The proof is due to Frenkel and Zhu
\cite{frenkzhu} who constructed vertex operator algebras corresponding to the
highest weight representations of Kac-Moody and Virasoro algebras where the
highest weights are multiples of the highest weight of the fundamental
representation. For a representation-theoretic description of the corresponding
local conformal algebra, see \cite{FG,fgrleshouches}. In \cite{fgrleshouches} a
description of the quantum spacetime associated with the $SU(2)$ WZW model is
presented and related to the Fuzzy 3-sphere. Generally, the low-energy
(semi-classical) target space 
manifolds of these conformal field theories are quantum
deformations of the group manifold target space. In fact, in the general case,
there is an intimate connection between the modules of a rational vertex
operator algebra and the modules of a Hopf algebra
associated with the vertex operator algebra. This makes more precise the
relationship between conformal field theory and the theory of quantum groups
within a framework suited to the techniques of noncommutative geometry.

The complete list of concrete vertex operator algebras thus includes the
moonshine vertex operator algebra that we discussed in section 7, vertex
operator algebras based on Heisenberg-Weyl algebras (equivalently Fock spaces)
and even positive-definite lattices, and the vertex operator algebras
associated with the integrable representations of affine Lie algebras, Virasoro
algebras, and $W$-algebras. There are also various generalizations via
twistings and orbifold theories which could be used to describe string
geometries associated with toroidal orbifold sigma-models, for example, and
also the notion of a vertex operator superalgebra \cite{kacwang} leading to
effective target space geometries associated with superconformal field theories
\cite{fgrleshouches}. A more abstract approach is due to Zhu \cite{zhu} who
established a one-to-one correspondence between the set of inequivalent
irreducible (twisted) modules for a given vertex operator algebra, and the set
of inequivalent irreducible modules for an associative algebra associated with
the vertex operator algebra and an automorphism of it. Thus a vertex operator
algebra can be represented as an {\it associative} algebra, a process which we
implicitly carried out in section 5 when we represented the algebra $\alg_X$ as
operators on the Hilbert space ${\cal H}_X$. There is even a geometric
procedure for constructing rational conformal field theories from two given
vertex operator algebras (representing the left and right chiral algebras)
which have only finitely many irreducible modules \cite{huang}.

\end{document}